# INTERLOCKING RESONANCE PATTERNS IN GALAXY DISKS.

J. Font[1,2], J.E. Beckman[1,2,3], M. Querejeta[1,4], B. Epinat[5], P. A. James[6], J. Blasco-herrera[7,8], S. Erroz-Ferrer[1,2], I. Pérez[9]

1. Instituto de Astrofísica de Canarias, c/ Vía Láctea, s/n, E38205, La Laguna, Tenerife, Spain: jfont@iac.es, jeb@iac.es, serroz@iac.es.
2. Departamento de Astrofísica. Universidad de La Laguna, Tenerife, Spain.
3. Consejo Superior de Investigaciones Científicas, Spain.
4 Max-Plack.Institut für Astronomie, Königstuhl 17, D-69117 Heidelberg, Germany: querejeta@mppia.de
5. Laboratoire d'Astrophysique de Marseille, Université d'Aix-Marseille & CNRS, UMR7326, 38 rue F. Juliot-Curie, 13388 Marseille Cedex 13, France: benoit.epinat@oamp.fr.
6 Astrophysics Research Institute, Liverpool John Moores University, Twelve Quays House, Egerton Warf, Birkenhead, CH41 ILD, UK: paj@astro.livjm.ac.uk
7 Department of Astronomy, University of Stockholm, AlbaNova, 10691, Stockholm, Sweden.
8 Instituto de Astrofísica de Andalucia, CSIC, Apdo. 3004, 18080 Granada, Spain: blasco@iaa.es
9 Departamento de Física Teórica y del Cosmos, Universidad de Granada, Spain: isa@ugr.es

ABSTRACT.

We have developed a method for finding dynamical resonances in disk galaxies using the change in sense of the radial component of the in-plane velocity at a resonance radius. Using simulations we show that we would expect to find these changes at corotation radii, with a weaker effect at the ILR and a small effect at the OLR. The method works well with observations at high spectral and angular resolutions, and is suited to the analysis of 2D velocity fields in Hα from Fabry-Perot spectroscopy, (though it is also applicable to fields in 21cm emission from HI, or to CO emission lines). The observations are mainly from the GHASP spectrometer data base, taken on the 1.93m telescope at Haute Provence, plus some data from the GHαFaS spectrometer on the 4.2m WHT at La Palma. We find clear indications of resonance effects in the disk velocity fields of virtually all of the 104 galaxies. The number of resonance radii detected ranges from one to seven, with a median of four. We have derived the resonance curves: $\Omega$, $\Omega \pm \kappa/2$, $\Omega \pm \kappa/4$ against radius for all the galaxies, and used them to derive the ILR, the OLR, and the two 4:1 resonances for each corotation in each galaxy. This led us to discover a pattern in over 70% of the sample: given two pattern speeds, say $\Omega_1$ and $\Omega_2$, the OLR of $\Omega_1$ coincides with the corotation of $\Omega_2$, and the inner 4:1 resonance of $\Omega_2$ coincides with the corotation of $\Omega_1$. Although the second coincidence has been predicted, a prediction of this double coincidence is not found in the literature. This pattern is found once in 42 of the galaxies, twice in a further 26, three times in five, and even four times in one galaxy. We also compute the ratio of corotation radius to bar length where we have good enough image quality, finding a mean value of 1.3, and a shallow increase towards later type galaxies.

*Subject headings*: galaxies: spirals - galaxies: fundamental parameters - galaxies: kinematics and dynamics - galaxies: structure - techniques: interferometric.

I INTRODUCTION

Density wave theory (Lindblad 1961, Lin & Shu 1964) has been the basis for the analysis of the dynamics of disk galaxies. It has the virtue that it can explain the universal presence of well formed spiral arms containing major star forming complexes, although stochastic formation and reformation of arms has been proposed as a rival model (Gerola & Seiden 1978). It also has the virtue that it predicts observable phenomena, both morphological and kinematic, and can therefore be tested. The underlying first order models, based initially on collective stellar dynamics in the potential of either axisymmetric (for unbarred models) or not axisymmetric (for barred models) disks, yield a set of common features which have been widely discussed in both the theoretical and the observational literature. These are the Lindblad resonances, the 4:1 resonances, and corotation, defined using the properties of the rotation curve and the oscillating frequency of a test star under a small perturbation within the gravitational field of the disk. In particular the corotation radius, defined as the radius at which the density wave angular pattern speed coincides with the angular speed of the material in the disk, has been a parameter whose value has been sought via a combination of observations and theoretical modelling. This corotation radius has been quantitatively predicted for barred galaxies. The basic models of Contopoulos (1980) suggested that corotation should be close to the ends of the dominant bar of a galaxy, (and for stability reasons not at a smaller radius). Following from this it has become standard to attempt to determine whether a bar is a fast rotator or a slow rotator by estimating whether its corotation radius is less than 1.4 times the bar length (fast rotators) or greater than this value (slow rotators). As this property has been linked to the presence of dark matter halos, whose dynamical friction has been predicted to slow down the angular speed of the bar on timescales much less that a Hubble time, (Weinberg 1985, Hernquist & Weinberg 1992, Tremaine & Ostriker 1999) the search for a reliable way to determine corotation has taken on cosmological relevance (see also Debattista & Sellwood 2000, and Sellwood & Debattista 2006). Added to this in a recent article Foyle et al. (2011) presented observational evidence that spiral arms may not be long-lived, which would cast some doubt on the applicability of density wave models to galactic structure in general.

Adhering to the conventional view that spiral structure requires an explanation based on at least a fairly stable density wave system, and supported by numerous simulations which show that bars should be stable, albeit evolving, configurations (and indeed supported by the well-established observational result that some two thirds of disk galaxies have bars or oval distortions, Marinova & Jogee 2007) there have been many studies over the years designed to measure the pattern speed of density wave systems associated with galactic bars. One of the best known, due to Tremaine & Weinberg (1984) combines the stellar surface density distribution from photometric imaging with line of sight velocity distribution, via long slit spectra parallel to the bar major axis. It was used successfully on NGC 936 (Merrifield & Kuijken 1995), on five SB0 galaxies by Aguerri et al. (2003), and Corsini (2011) presented results on 18 galaxies using this technique. It is based on clear and well understood principles, but has two drawbacks which make it difficult to extend to large samples of galaxies: it is applicable to stellar populations where interstellar dust does not obscure the stellar emission, so it works less well on late type galaxies, and more important, the spectral measurements require rather long exposures, particularly in the outer disk, thus restricting the numbers of objects observable in practice. As it relies on the continuity of the emitting sources, opinion is divided on whether it can also be applied using measurements of the interstellar gas. Nevertheless, Zimmer, Rand, & McGraw (2004) applied it to CO emission line maps of three galaxies, while Rand & Walllin (2004) applied it to five more. Hernandez et al. (2005) used an Hα emission line map of NGC 4321 (M100), while Fathi et al. (2009) used similar maps to measure pattern speeds in 10 galaxies, (in practice they used the emission line for the velocity distribution, and the stellar continuum distribution derived during the same observations, for the required photometric weighting). Another technically attractive method proposed by Canzian (1993) places corotation where the azimuthal component of the non-circular motion changes from a singlet to a triplet pattern, and calculates the pattern speed from the rotation curve at the point of corotation. It was applied to an HI velocity field of NGC 4321 by Canzian & Allen (1997), and recently Font et al. (2011) showed that the method could be applied using an Hα velocity field, giving comparable results for NGC 4321 but with greater precision due to the higher angular resolution of the data.

These are relatively direct techniques, but less direct methods have also been applied. Comparison of observed HI velocity field with predictions based on the gravitational potential distribution (based on observations of the stellar component in the first instance) was used quite early on (Sanders & Tubbs 1980, England, Gottesman & Hunter 1990, García-Burillo, Combes, & Gérin

1993) to estimate pattern speeds, while Sempere et al. (1995) used a hybrid technique based on the Canzian (1993) method, combined with the use of HI with the gravitational potential distribution. More recently Rautiainen, Salo & Laurikainen (2005, 2008) applied a simulation technique to derive the pattern speeds of 38 barred galaxies. Starting from the gravitational potential distribution from observed near IR imaging, and assuming a single pattern speed they derived the calculated responses of stellar and gaseous disks to this potential, and compared their simulations to observations, finding that the corotation radius (measured in units of bar length) apparently increases from 1.1 to 1.7 from S0 to Sc galaxies. If these results are taken at face value, it could imply that later type galaxies have dark matter halos which are more massive relative to their disks, which is consistent with the increase of M/L ratio to later Hubble types noted as early as 1981 by Tinsley) and confirmed by subsequent authors (see e.g. Broeils 1995). This specific predicted increase in the ratio of corotation radius to bar length is susceptible to observational confirmation which is one motive for studies such as the present article.

The largest sample of derived pattern speeds to date has been presented by Buta & Zhang (2009) who used a potential-density phase-shift method to determine corotation. This is a direct method, in the sense that it does not require the application of simulations, and indeed it does not need any kinematic information as such, relying entirely on the morphological evidence found in H-band images. Buta & Zhang found multiple corotation radii, and for the barred galaxies of their sample used the radius closest to, and just larger than, the bar length, to derive the pattern speed associated with the bar. However they also postulate that in some cases the end of the bar is not at corotation, but at the outer Lindblad resonance, in which case corotation is at a smaller radius than the bar length, and the bar is considered "superfast". In any case they claimed that the presence of resonance patterns in the full sample of galaxies should be considered evidence for the persistence of density wave systems in galaxy disks.

It is interesting to note that predictions of multiple pattern speeds can be found in the literature dating back decades. They appeared in the output of N-body simulations of Sellwood & Sparke (1988), who found bars and spirals in the same galaxies rotating with different pattern speeds; similar behaviour was derived by Rautiainen & Salo (1999). Double-barred galaxies have also been offered as possible evidence for multiple pattern speeds as predicted, for example, in the simulations of Englmeier & Shlosman (2004), Debattista & Shen (2007) and Shen & Debattista (2009). Of particular interest in the context of the present study is the work of Tagger et al. (1987); Sygnet et al. (1988); Patsis et al. (1994); and Masset & Tagger (1997), who considered the possibility of non-linear coupling between bar and spiral waves and modes, work which was cited by Rautiainen & Salo (1999). This group of authors concluded from their model simulations that non-linear coupling could be a more relevant mechanism that the "classical" swing amplifier mechanism (Toomre 1967) to account for the dynamics of disk galaxies beyond the corotation radius of the bar. They find that it can give rise to harmonic and sub-harmonic excitation as well as the stimulation of m=1 spiral waves. This is, as we will see, a theoretical background giving support to the possible presence of multiple pattern speeds, and of interactions among them. The possibility of multiple density wave patterns, each with its own pattern speed dominating within a given range of galactocentric radii, was explored observationally by Meidt et al. (2008a). They developed a more general version of the Tremaine-Weinberg method allowing the possibility of detecting these nested patterns, and tested this first on simulated galaxy data sets. In Meidt et al. (2008b), they applied the method to M51, finding three pattern speeds associated with three corotation radii, suggesting that their technique was an advance on the single pattern speed derived by Zimmer, Rand & McGraw (2004). These more recent measurements suggest strongly that there may be more than one density wave system in a galaxy, and they also suggest that the systems may be coupled such that the outer Lindblad resonance of an inner system may coincide with the corotation of an outer system, as well as other possible couplings between the systems.

Measuring the pattern speed (or speeds) and corotation radius (or radii) is particularly important when investigating the dynamical effects of bars in the radial redistribution of matter in galaxies. There is an abundant literature on this matter which we cannot attempt even to summarize here, but we pick out a few recent salient points to indicate the importance of the subject. The formation and evolution of bars is strongly related to the redistribution of angular momentum; Sellwood (1981) showed originally that the disk harbours sources and the halo sinks of angular momentum. The interactions of the sources and sinks are strongly conditioned by resonances (Athanassoula 2002, Martinez-Valpuesta, Shlosman, & Heller 2006). However the details of these resonant interactions were not at first fully understood, especially in the presence of a significant gas component, since the studies cited were based on models using collisionless approximations. Villa-Vargas, Shlosman & Heller (2010) included gas as well as stars in their model disks. In the interaction with the live dark matter halo, the gas fraction in the model turned out to be qualitatively and quantitatively significant. Gas poor models showed rapidly decelerating bars, while gas rich models produced bars with constant or even slowly accelerating pattern speeds. This seems to contradict the results reported by Rautiainen et al. (2008), who obtained smaller ratios of corotation radius to bar length (i.e. faster bars) in early-type, relatively gas-poor galaxies, than in late-type relatively gas-rich galaxies. Clearly further systematic work is required both on the theoretical and the observational sides.

Responding to the need to obtain more, and if possible more reliable, values for the corotation radii in galaxy disks we have developed a method which can take advantage of the high angular resolution of the data obtained when mapping an entire galaxy in velocity using the emission from its ionized hydrogen in Hα. The technique was first demonstrated in Font et al. (2011). The first step is to derive the rotation curve, and by subtracting off the two dimensional representation of this curve (obtained by rotating it with suitable projection coordinates around the axis of rotation of the galaxy) from the observed velocity field derive a map of the non-circular residual velocities in the disk. As one of the predicted properties of a density wave system at corotation is a change of 180º in phase of the streaming motions in the spiral arms (Kalnajs 1978, Contopoulos 1981), we can look for these phase changes by finding the zero crossings of the velocity vector in the residual velocity map. Plotting the frequency of these zero crossings against galactocentric radius yields, in almost all the galaxies studied, a series of clear maxima, which we attributed in the first instance to corotation radii, although more detailed analysis has revealed a slightly more complex result as we will see below. It is a rather straightforward method in principle, but is not practical to apply with velocity maps where insufficient angular resolution smooths out the effect, which has in general been the case for mapping in HI or CO (these limitations are being superseded with new radiotelescopes and one point of presenting these results is to encourage radio observers to use the method as 1 arcsec or subarcsec resolution becomes available).

We have applied the technique to over 100 galaxies and in this article we present the results obtained. In the second section of the paper we describe briefly the data cubes used in our work, and the data treatment methods. In the third section we apply the method to derive the corotation radii, in the fourth section we use the rotation curves to explore the possibility that corotation in one range of galactocentric radii may coincide with a Lindblad or a 4:1 resonance in a neighbouring range, and in the fifth section we present our conclusions.

II. THE PHASE REVERSALS METHOD

In order to derive the resonant structure of a disk galaxy, we have developed a new technique based on measuring, using observations at high enough resolution in both angle and velocity, the radii at which the non-circular velocity changes phase in the radial direction (Kalnajs 1978). Linear density wave theory (Lin & Shu 1964) includes a prediction that the major axes of the orbits described by the stellar component of a disk galaxy experience a phase shift across the main resonances (ILR, corotation and OLR). This has been confirmed by subsequent authors in simulations which take stars and gas into account (Sanders & Huntley 1976; Huntley, Sanders & Roberts, 1978; Sanders & Tubbs 1980; Huntley 1980; van Albada & Sanders 1982; Teuben & Sanders 1985; and Wada 1994). This is most clearly seen in barred galaxies, and using the terminology of Contopoulos & Mertzinades (1977) $x_1$ periodic orbits which are elongated in the direction of the major axis of the bar, and $x_2$ orbits which are perpendicular to the bar, change their dominance when crossing a resonance as we move outward from the center of the galaxy (Contopoulos & Mertzinades 1977; Contopoulos & Papayannopoulos 1980; Contopoulos 1981; and Contopoulos et al 1989). Although strictly applicable to the linear version of density wave theory, and in its origins in predicting the stellar structure of galaxies, the generic prediction of a radial phase change has been used and applied to velocity maps by several authors in order to determine the corotation radius (Sempere et al. 1995; Rand 1995; and Canzian & Allen 1997). In Font et al. (2011) the present method was introduced briefly and applied to a representative sample of eight galaxies. In essence, the method used in this article remains the same as that presented in Font et al. (2011), but here we give a detailed description, especially as more recently we have introduced modifications which have yielded improved results.

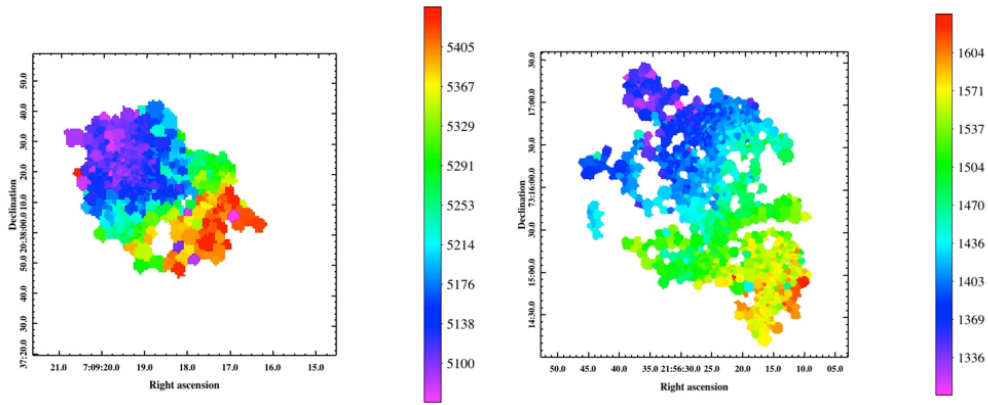

**Figure 1. Left panel**. Velocity field of UGC3709, as measured from Hα emission line observations from GHASP. **Right panel**. Velocity field of UGC 11861. The color bar gives the recession velocity in km·s$^{-1}$.

The input observational material for our technique is a 2D velocity map extracted from data cubes in the Hα emission line, which are produced by a Fabry-Perot interferometer. Figure 1 shows the velocity map for the galaxies UGC 3709 and UGC 11861, which have been picked out as examples in order to illustrate the application of the method. However, our method is elastic enough to accept a velocity field obtained by mapping the galaxy in other emission lines, typically the 21-cm line of HI, and CO molecular emission lines, despite the lower angular resolution generally available in those maps up to now, which reduces the accuracy of the results. Additionally, an optical image of the galaxy with the best resolution available is required for the morphological analysis. Usually this image was taken from the DSS public archive, except for those galaxies available on the Sloan Digital Sky Survey (DR9 Science archive server) for which an RGB image is composed taking r, g and i band images. All the kinematic codes needed to perform our method have been written in IDL and are combined with the available package *kinemetry* (Krajnovic et al. 2006). The complete sequence of steps of the non-circular phase reversal method is:

1. We produce clean maps of the streaming motions by means of an iterative process which yields a separation of circular and non-circular velocity over the full 2D field. We first calculate initial versions of the rotation curve and the residual velocity field using *kinemetry*. Given the input kinematic parameters: the systemic velocity, the position of the center of the galaxy, the position angle of the major axis and the inclination angle of the galactic disk, *kinemetry* performs a harmonic decomposition of the observed velocity field, so that,

$$V_{l.o.s.} = B_0 + \sum_{m=1}^{N} [A_m \cdot sin(m\theta) + B_m \cdot cos(m\theta)]$$

where $V_{l.o.s.}$ is the velocity measured along the line of sight, $A_m$ and $B_m$ are the harmonic coefficients of the *m*-terms, and *θ* is the azimuthal angle with respect to the galaxy major axis. From this, we can identify $B_0$ with the systemic velocity, and the term $B_1(x,y)$ corrected by the inclination angle can be associated with the 2D map of the circular motion, therefore, $V_{l.o.s}$ - $B_1(x,y)$ gives the 2D residual (i.e. non-circular) velocity field. The rotation curve is obtained from the term $B_1$ not as a map but as a function of the galactocentric radius, $B_1(r)$.

Once we have derived the rotation curve and the residual velocity map, we again perform *kinemetry* on the latter to determine any remaining circular component within this map, and then subtract this from the original residual field to produce a corrected map. This procedure is repeated until the values in the circular component of the residual map fall below the observational uncertainty in velocity, yielding, thus, our final corrected residual velocity map. Figure 2 shows the resulting non-circular velocity field for the two galaxies. Subtracting this from the observed velocity field we find an optimized axisymmetric velocity field from which we then extract the final rotation curve, which is plotted in the top-left panel of Figures 4 and 5, for UGC3709 and UGC11861, respectively.

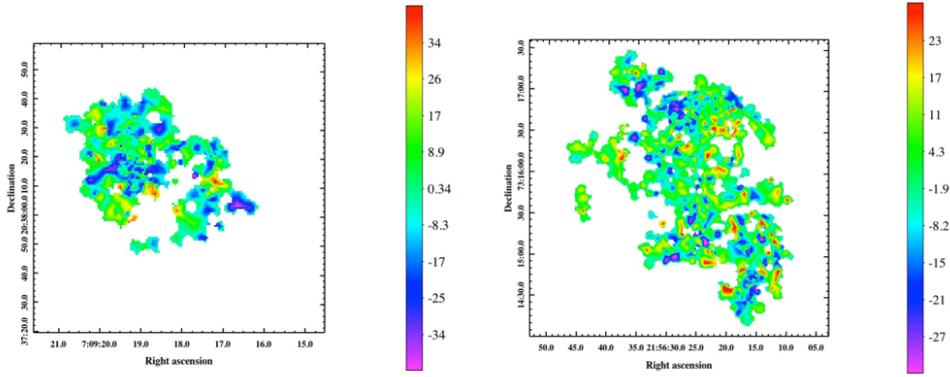

**Figure 2. Left panel.** The residual velocity field for UGC 3709 obtained after subtracting off the 2D rotational velocity field, which is derived after performing the iterative procedure (see text for details), from the observed field. **Right panel**. The residual velocity field for UGC 11861.

2. We set a short slit (~ 8 arcsec) radially oriented and centered on each pixel of the residual velocity map. We apply the following criterion to identify phase reversals or 'zeros' in the velocity of the streaming motion. Setting the slit on the *i*-th pixel which has a radius $r_i$ from the center of the galaxy and with a non-circular velocity $V_{res}^i$, we check two conditions:

   (i)   $V_{res}^{i-1} \cdot V_{res}^{i+1} < 0$
   (ii)  $|V_{res}^{i+1}| > \Delta v, |V_{res}^{i-1}| > \Delta v$

where $V_{res}^{i-1}$ is the non-circular velocity corresponding to the (*i-1*)-th pixel with radius $r_{i-1} < r_i$, located on the radial slit, in a similar way, $V_{res}^{i+1}$ is the non-circular velocity corresponding to the (*i+1*)-th pixel with radius $r_{i+1} > r_i$. *Δv* is the uncertainty in velocity which is the half of the spectral resolution (in km·s$^{-1}$) of the Fabry-Perot. Obviously, condition (i) detects radial reversals of the residual velocity, while condition (ii) is required in order to ensure that those reversals are physically reliable (this ensures that noise is not mistaken for a reversal, but it also ensures that projection effects are effectively eliminated). So, if both of the conditions are satisfied we assume that a real phase reversal in the residual velocity occurs at the *i*-th pixel.

3. We list all the zeros in order of the radial gradients of the residual velocity fields passing through them, starting with the steepest, and define a circle with radius half the angular resolution around each zero. Starting with the zero at the top of the list, we see whether any other zero (which will have a shallower gradient value) lies within the circle. Any of these secondary zeros are rejected as they are not independent of the primary zero in that circle. We proceed down the list across the whole disk rejecting any secondary zeros using this criterion. At the end of the procedure those zeros not rejected in this way are retained. With this binning, we make sure that phase reversals are consistent with the observations and also we avoid overestimating the number of them.

4. At this point, all reliable zeros are positively identified knowing their projected coordinates in the galactic plane. So, we calculate the galactocentric radius of each phase reversal and taking a bin size equal to the angular resolution, a histogram of zeros as a function of the galactocentric radius is built up. In Figure 3 we plot the radial distribution of phase-reversals for the two galaxies. We find that the resulting histogram shows multiple peaks unveiling, thus, the resonant structure of the galaxy, as revealed by the gas kinematics.

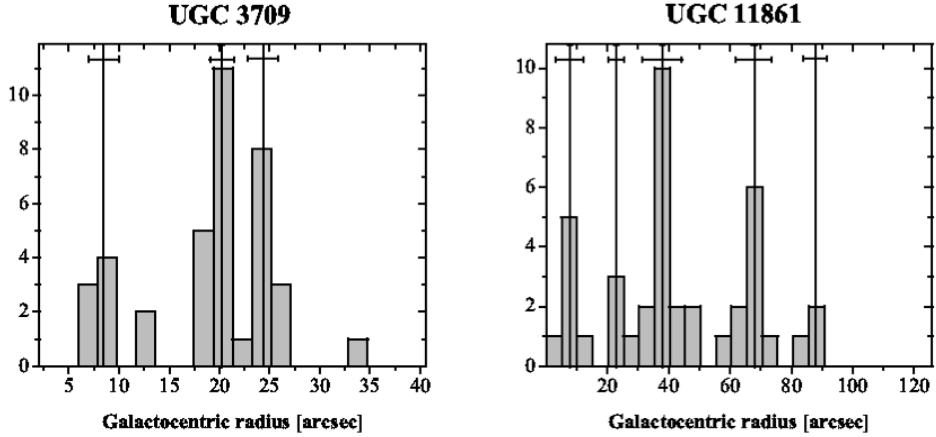

**Figure 3.** Histogram of the number of phase reversals detected in the residual velocity maps of the galaxies, plotted against galactocentric radius for: (a) Left panel UGC 3709 (b) Right panel UGC 11861. The clearly defined peaks in the histograms show the radii of corotation, except the outermost peak for UGC 11861, which is compatible with an outer Lindblad resonance of the third peak, as can be seen in Figure 5.

5. The radial distribution of the measured phase reversals shows peaks at specific radii (see Figure 3), often very marked and with radii easy to determine by fitting each peak with a Gaussian distribution, allowing us to characterize each peak by three parameters: the position, the area and the width of the Gaussian fitting function. As discussed above, we assume that each peak of zeros in this histogram corresponds to a different resonance. Thus, the position of the Gaussian is the galactocentric radius of the resonance which has an associated uncertainty given by the half width of the Gaussian. The area of the fitting function gives an idea of the relative strength of the resonance. We note that the process of identifying and quantifying these resonance peaks does not use the rotation curve of the galaxy, but relies on the separation of the radial and tangential velocity components across the full two-dimensional velocity map. Special attention has been paid to those peaks consisting of only one phase reversal (N=1 peaks). We have used the following criteria in order to distinguish between those N=1 peaks identified as a resonance and those not: (*i*) if the N=1 peak is closer to the nucleus than the resolution limit for that galaxy, then it is rejected as a resonance (*i.e.* the peak must be resolved). (*ii*) If the peak represents an insufficient fraction of the emitting pixels at its radius it is rejected; this is quantified by accepting only those N=1 peaks that lie either within a radius containing 5% of the total number of pixels forming the galaxy Hα image or outside the radius containing 95% of these Hα emitting pixels. All other N=1 peaks are rejected.

6. We next plot the frequency curves corresponding to $\Omega$, $\Omega \pm \kappa/4$ and $\Omega \pm \kappa/2$ as functions of the galactocentric radius using the best fit to the rotation curve previously found (see step 1), and its relevant derivatives. By default, the fit to the rotation curve is performed by means of a potential function of the form:

$$V_{rot} = V_{max} \frac{r/r_{max}}{[1/3 + 2/3 \cdot (r/r_{max})^n]^{3/2n}}$$

being $V_{max}$, $r_{max}$ and $n$ the three fitting parameters, with initial values of 200 km·s$^{-1}$, 50 arcsec and 0.5, respectively. When the fit does not reproduce accurately the experimental rotation curve, a smoothed spline interpolation is applied. The frequency curves for galaxies UGC 3709 and UGC 11861 are plotted in Figures 4 and 5, respectively. In these figures, the positions of the peaks in the histogram (black solid vertical lines) and the associated uncertainty bars (blue dashed vertical lines) are also included. The different panels of Figures 4 and 5 (with the exception of the top-left panel) depict a zoom-in around the value of the pattern speed determined assuming corotation in each peak, in order to identify easily the compatibility between other resonances and the remaining peaks. So, there are as many panels as peaks found in the phase-reversal histogram (Figure 3). This diagram of frequencies enables us to:

a. Calculate the pattern speed and its associated uncertainty for each resonance by determining the value that the curve Ω(r) takes at each resonance radius.
b. Classify as ILR, corotation or OLR, the resonances revealed in the histogram of phase reversals.
c. Establish different coupling relationships between the resonances found for a galaxy.

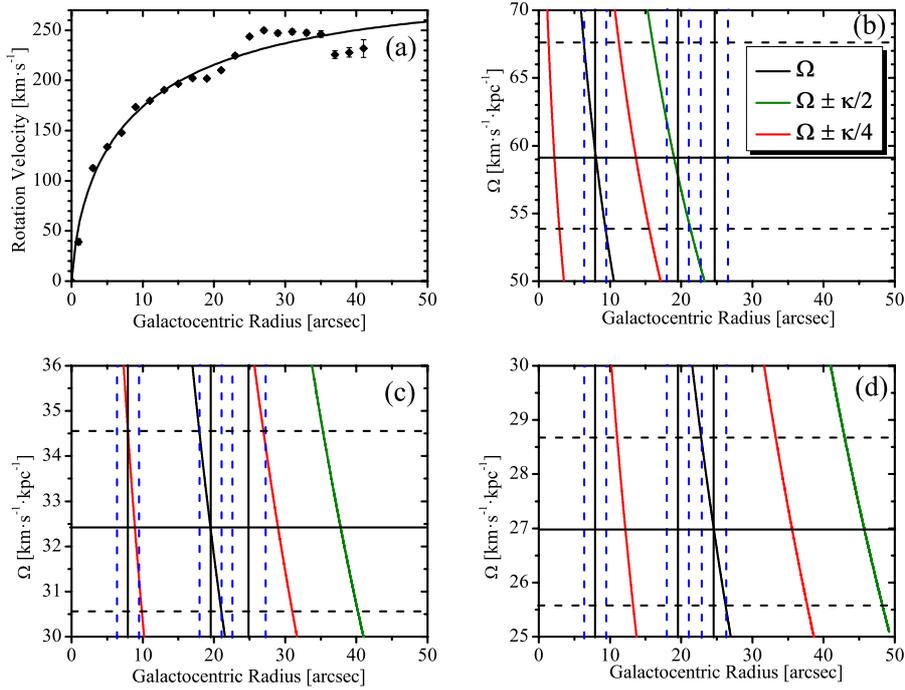

**Figure 4.** (a) Rotation curve for a representative galaxy: UGC 3709 showing the observed values and the best polynomial fit. (b) Plots, against galactocentric radius, of Ω (black curve), Ω ± κ/2 (green curves) and Ω ± κ/4 (red curves). The positions of the three resonance radii found for this galaxy are given by solid black vertical lines, with their uncertainties shown by the accompanying pairs of vertical dashed lines (in blue). The horizontal black line (and its associated uncertainties, shown dashed) gives the pattern speed associated with the innermost resonance peak. The details in this graph show that when corotation is assumed to occur at the position of this peak, its OLR lies very close to the position of the second peak, rather than the ultraharmonic resonances, which do not coincide with the other peak. (c) The same type of plot as in (b), but centered on the second of the three resonance radii and showing that the I4:1 resonace for this corotation coincides (within the uncertainties) with the innermost cortation. (d) The same type of plot as in (b) but centered on the outermost of the three corotations, and showing that neither the inner ultraharmonic resonances nor the Lindblad resonances coincide with the other peaks. The pattern found in this galaxy, where two resonant systems are inter-related such that the OLR of the inner coincides with the corotation of the outer, and the I4:1 resonance of the outer coincides with the corotation radius of the inner is a pattern which we find at least once in the great majority of the galaxies measured.

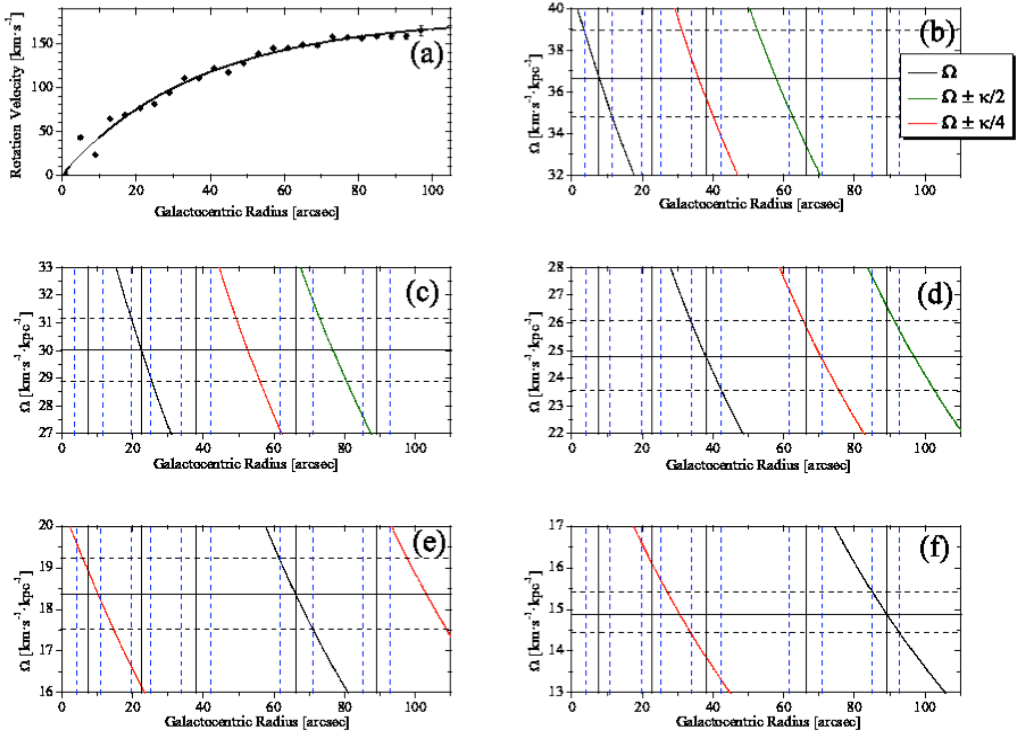

**Figure 5.** The color coding in this figure is the same as in Figure 4. (a) Rotation curve of UGC 11861, showing the observed values and the best fit curve. (b) Plots, against galactocentric radius, of $\Omega$ (black curve), $\Omega \pm \kappa/2$ (green curves) and $\Omega \pm \kappa/4$ (red curves). All five resonance radii are shown as solid black vertical lines, with their corresponding uncertainties as pairs of dashed vertical lines. In this figure we concentrate on the innermost resonance, and can see that if it is assumed to be a corotation, its OLR is compatible with the fourth resonance radius (within the error bars), and the position of the third peak is coincides with the outer ultraharmonic resonance. (c) The same type of plot as in (b) but centered on the second phase-reversal peak, for which none of the other peaks is compatible with its ultraharmonic or Lindblad resonances. (d) The same type of plot as in (b) but centered on the third resonance. The OLR and O4:1 resonance curves coincide with the position of the outermost and fourth peak, respectively. (e) The same type of plot as in (b) but centered on the fourth resonance radius. The I4:1 resonance coincides with the innermost resonance radius, forming the characteristic pattern relation between these two systems. (f) The same type of plot as in (b) but centered on the fifth resonance radius. Although the I4:1 curve is plotted, it shows no coincidences with the position of the other resonances. We conclude that the outermost peak is identified as the OLR associated with the corotation located at the third peak.

## III. TESTING THE PHASE REVERSAL METHOD

In this section we apply the phase reversal method described in the previous section to a modeled velocity field for which the values of all parameters are known beforehand. This enables us to achieve two main goals: firstly to establish the reliability of the method for determining the radii of resonances, and secondly to estimate the error in the resonance radii derived with the method, when the values of those geometrical parameters which for the observed galaxies are subject to observational uncertainties are artificially varied within the model.

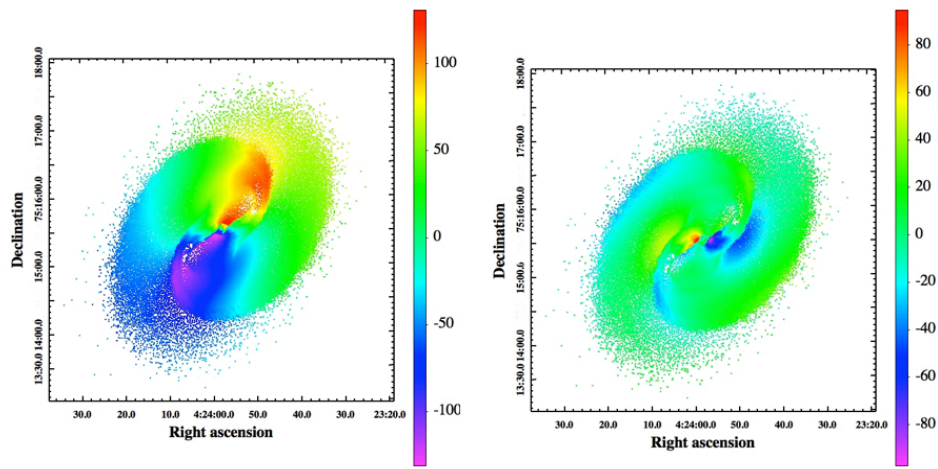

**Figure 6. Left panel.** Velocity field derived from our numerical model, projected as if observed on the sky. **Right panel**. Residual velocity field of the model galaxy, found by subtracting off the 2D rotational velocity field from the complete field.

The modeled map of the velocity field, which is shown in Figure 6 (left panel), is obtained from a 3D N-body hydrodynamical simulation used to explore the barred galaxy NGC 4123 (see Perez 2008 for details). The geometrical parameters are taken from the kinematical study of Weiner et al. (2001). The values of the pattern speed and the coefficient between the corotation radius and the bar radius, $\mathcal{R}$, are obtained from the best-fit model of Weiner, Sellwood & Williams (2001). All relevant parameters together with the corresponding values, which have been used to produce our model of velocity field, are summarized in Table 1.

After subtracting the rotation curve from the velocity map, we apply the phase reversal method to the derived map of non-circular velocity (Figure 6 right panel). The resulting histogram of zeros, plotted in panel (a) of Figure 8, shows four different peaks, which are identified as four resonances. Using the rotation curve, plotted in Figure 7 left panel, we can produce the diagram of the angular velocity, $\Omega$, the angular velocity plus and minus half the epicyclic frequency, $\Omega+\kappa/2$, $\Omega-\kappa/2$, and the angular velocity plus and minus one fourth of the epicyclic frequency, $\Omega+\kappa/4$, $\Omega-\kappa/4$ plotted as function of radius r, all of them are plotted in Figure 7 right panel, where we have also shown the four resonance radii. In this figure, the solid vertical lines show the positions of the centers of the peaks of the phase reversal distribution, while dotted vertical lines show the error bars associated with each peak (the error bars are calculated as the sum of the half width of the Gaussian fit to the peak, and the angular resolution of the map, taken in quadrature). The value of the pattern speed for each resonance peak is then obtained numerically by determining the crossing point of the curve $\Omega(r)$ (solid black curve in Figure 7 right panel) and the vertical line which indicates the position of the resonance. The crossing points of the $\Omega(r)$ curve and error bar vertical lines are used to estimate the uncertainty in the pattern speed. As can be seen in right panel of Figure 7, we have assumed that the corotation occurs at the third peak: $r_3 = 53.4 \pm 4.2$ arcsec, and the corresponding pattern is $\Omega_P = 19.7 \pm 0.7$ km·s$^{-1}$·kpc$^{-1}$, which is in very good agreement with the value derived directly from the model (see Table 1). Taking the radial length of the bar given in Table 1, we find a value of $\mathcal{R}=1.37 \pm 0.11$ for the ratio $r_{CR}/r_{bar}$, which is in excellent agreement with the value $\mathcal{R}=1.35$, which has been fixed as an input parameter in our numerical model of velocities. In addition, assuming the above pattern speed value, we can see in Figure 7, that the remaining three peaks, located at $r_1$, $r_2$ and $r_4$ are compatible with iILR, oILR and OLR (inner inner Linblad, outer inner Lindblad, and outer Lindblad resonances) respectively, as the $\Omega(r) \pm \kappa(r)/2$ curves (green dotted curves in Figure 7, right panel) pass through the error box defined for each of these radii (the vertical sides of the error box are given by error bars of the peak radii, and the horizontal ones by the pattern speed error bars).

Table 1. Parameters of the modeled velocity field

| Parameter | Value | Reference |
|---|---|---|
| Position Angle | 57º | 1 |
| Inclination | 45º | 1 |
| Bar radius | 5.6 Kpc | 1 |
| Distance | 22.4 Mpc | 1 |
| $r_{CR}/r_{bar}$ | 1.35 | 2 |
| Pattern Speed | 20 km·s$^{-1}$·kpc$^{-1}$ | 2 |
| Angular resolution | 1.2 arcsec | 3 |
| Spectral resolution | 10 km·s$^{-1}$ | 3 |

References: (1) Weiner et al. 2001 (2) Weiner, Sellwood & Williams. 2001b (3) Perez 2008.

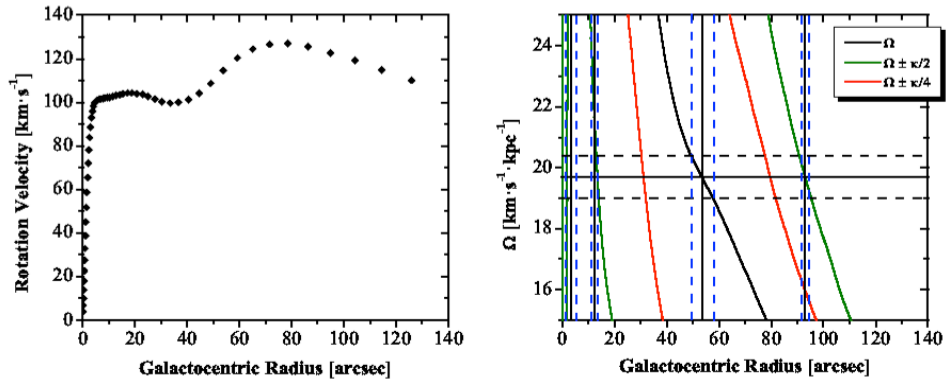

**Figure 7. Left panel**. Rotation curve for the model used to make comparisons with our observational results. We have not made any attempt to smooth out the rather abrupt change of gradient at 100 km s$^{-1}$ given by this model. **Right panel**. Plots, against galactocentric radius, of $\Omega$ ( black curve), $\Omega \pm \kappa/2$ (green curves), $\Omega \pm \kappa/4$ ( red curves) for the model galaxy. The derived positions of the resonance radii are shown as vertical solid black lines, with their corresponding uncertainties shown as pairs of vertical dashed lines. The diagram is centered on the third of the four resonances, and shows that in this case the OLR coincides with the fourth resonance, while the ILR coincides with the second resonance.

We conclude that the phase reversal method identifies and correctly locates phase reversals in the streaming motion, which are produced by the main resonances of the galaxy: ILR, Corotation and OLR, in agreement with the original simulations of Contopoulos & Papayannopoulos (1980). The phase reversal histogram of Figure 8 panel (a), shows two important features: (1) Corotation occurs at the position of the main peak (larger area), although the iILR peak is also quite strong. (2) The second peak corresponding to the oILR resonance is the weakest and we draw the conclusion that it could well be difficult to be identified in a real galaxy for which a more complex, and noisy, histogram is expected. Finally we note that the iILR resonance takes place at a very small radius (peak 1), so it could be unresolved if the angular resolution of the observations is insufficient. We have used the diagram of frequencies for two main purposes: (1) To assign a resonance to each peak of the phase reversal histogram, taking into account the associated uncertainties. (2) To determine the pattern speed and its uncertainty at corotation.

III.1 *Sensitivity of the phase reversal method to parameter uncertainties*

In order to estimate the sensitivity of our method with respect to the main geometrical parameters, i.e. position angle, inclination and position of the center, we have used the modeled velocity field to quantify the effects on the peaks of the histogram of zeros in non-circular velocity when these parameters are deliberately modified. Table 2 shows the values of the parameters, which have been taken in our method, along with the resulting values for the corotation radius and the pattern speed. In the first row we give the original values of these parameters.

Table 2. Variation of geometrical parameters for the test model

| PA (°) (1) | $i$ (°) (2) | $(x,y)_c$ (pixel) (3) | Corotation Radius (kpc) (4) | Pattern Speed (km·s$^{-1}$·kpc$^{-1}$) (5) |
|---|---|---|---|---|
| 57 | 45 | (256,256) | 7.8 ± 0.6 | 19.7 ± 0.7 |
| 49$^{(*)}$ | 45 | (256,256) | 7.4 ± 0.6 | 20.0 ± 0.7 |
| 65 | 45 | (256,256) | 8.1 ± 0.6 | 19.2 ± 0.7 |
| 57 | 38$^{(*)}$ | (256,256) | 7.5 ± 0.6 | 20.0 ± 0.7 |
| 57 | 52 | (256,256) | 8.1 ± 0.6 | 19.2 ± 0.7 |
| 57 | 45 | (253,253)$^{(*)}$ | 7.8 ± 0.6 | 19.5 ± 0.7 |

**Notes.** Columns (1), (2) and (3) give the value of the position angle, inclination angle and the position of the center, respectively, that have been assumed in our method for the modeled velocity field. The corotation radius and the pattern speed along with their uncertainties calculated with the phase-reversal method are given in Columns (4) & (5). $^{(*)}$ The second peak corresponding to the oILR does not appear in the histogram. The first and fourth peaks do not match the iILR and OLR, respectively.

First, we modify the position angle by ±15% with respect to its original value while keeping the remaining parameters fixed (see rows two and three of Table 2), the resulting values of the corotation radius and the pattern speed are given in columns (4) and (5) of Table 2, respectively. When the position angle is overestimated, the calculated radius of the corotation is shifted to higher values, and the pattern speed is slightly reduced. The effect is reversed in sign when lower values of the position angle are taken: the pattern speed is increased and corotation is shifted inwards. In both parameter configurations, the variation of the corotation radius is ≈ 5%, while for the pattern speed it is ≈ 2.5%. In particular, when the position angle value is fixed at 49º then the second peak in the histogram of zeros, which corresponds to the oILR, cannot be positively identified. These features can be seen in Figure 8, panels (c) and (d), which shows the histogram of the phase reversals obtained with the values of the parameters appearing in rows 2 and 3 of Table 2. In these plots the new resonances are indicated with a solid vertical line, and its uncertainty as a horizontal solid bar, while the position of the original resonances are marked with a dashed vertical line, in order to evidence the relative shift in radial position between original and new resonances. From panels (c) and (d) of Figure 8, we also can see that the radial position of the iILR does not change significantly, while the OLR radius is shifted in the opposite direction that the corotation does. Typical uncertainties in the kinematically derived position angle are less than 5º (Epinat et al. 2008).

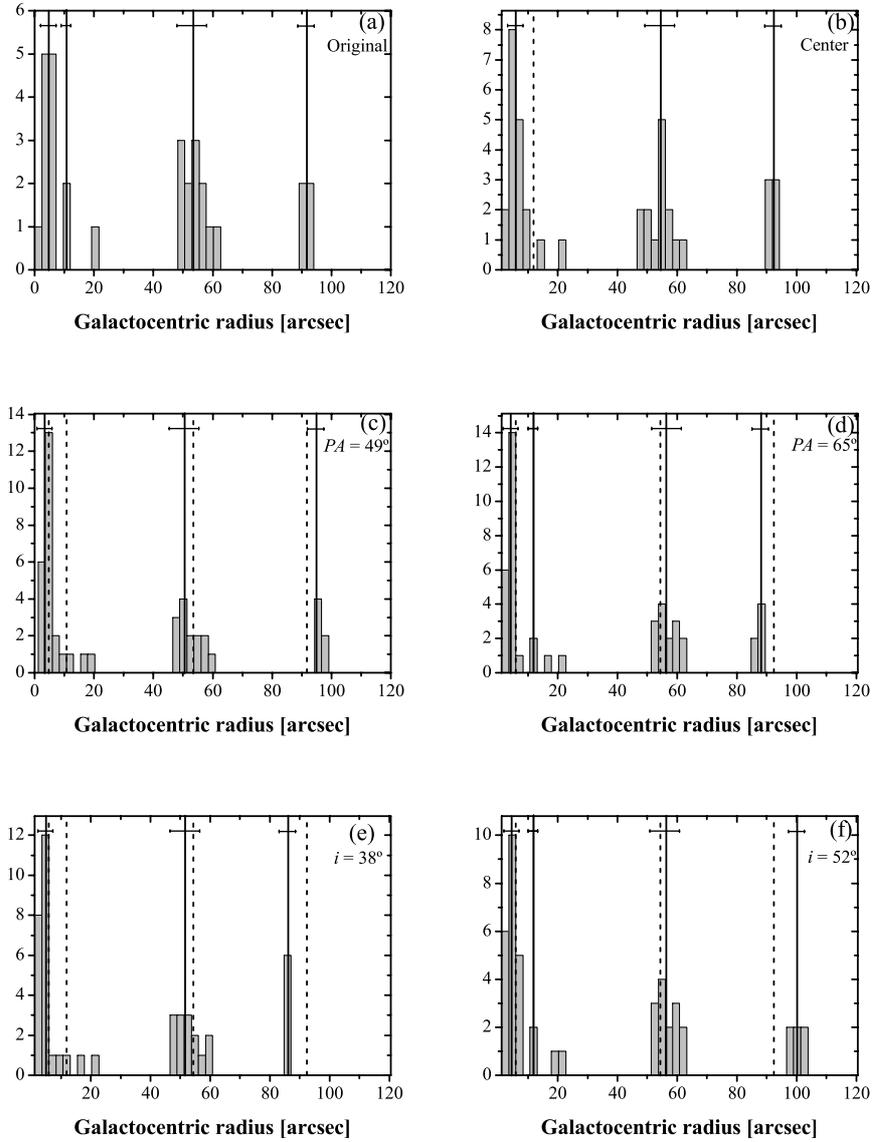

**Figure 8.** (a) Histogram of the number of phase reversals detected in the residual velocity maps derived from the numerical model, plotted against galactocentric radius, taking the original values of the geometrical parameters: center position = (256,256), position angle = 57° and inclination = 45° (see Table 1, row 1). The vertical solid line marks the radial position of the resonance, and the horizontal solid segment gives the uncertainty associated with the resonance radius. (b) The same type of histogram as in (a) but modifying the position of the center (see Table 2, row 6). The new corotation lies in the same radial position as in (a). (c) The same type of histogram as in (a) but modifying the position angle (see Table 2, row 2). The resonance radii obtained with the original parameters (panel a) is now included and plotted as a vertical dashed line, showing the effect of the variation of this parameter on each resonance radius. (d) The same type of histogram as in (c) but increasing the value of the position angle (Table 2, row 3). The new resonance radii are shifted with respect to the original position in the opposite way to those in (c). (e) The same type of histogram as in (a) but changing the inclination angle (see Table 2, row 4). The new resonance radii (solid lines) are shifted inwards when compared with the original one (dashed line). (f) The same type of histogram as in (e) but increasing the value of the inclination (Table 2, row 5). The new resonances are shifted with respect to the original position in the opposite way to those in (e).

When we study the effect of the variation of the inclination angle, we find the same behavior as for the variation of the position angle. A change of ±15% in the inclination, which is of he order of the variance in the GHASP sample (Epinat et al. 2008), produces an offset of ≈ 5% of the corotation and the pattern speed is modified by ≈ 2.5%, as shown in rows 4 & 5 of Table 2. Here, the iILR radius is not affected significantly by the variation of the inclination angle, contrarily the OLR, which is now shifted outwards, as well as the corotation, are more sensitive to the adopted value of the inclination, showing a change of ≈ 10%. These features can be seen in the corresponding histograms, which are plotted in panels (e) and (f) of Figure 8. We can also check that the oILR in the histogram of phase reversals (see panel (e) of Figure 8) cannot be identified when the inclination is underestimated ($i = 38°$).

Finally, the offset in the position of the center consistent with the angular resolution of the map, apparently seems to introduce no effect on the resonance radius and the pattern speed, as we see in Table 2, row six. However, the oILR resonance peak is not reproduced, as can be seen in Figure 8 (panel b), and the compatibility between peaks 1 & 4 and iILR & OLR, respectively, is lost. In this case, the radial position of the new resonances (solid vertical line) coincides with the original positions, hiding its dashed vertical line in the plot.

From the results obtained in this study with the modeled velocity field we may conclude that the phase reversal method is well behaved under the variation of the main parameters (15%), yielding errors of ≈ 5% in the radial position for the corotation, less than 10% for the OLR, and ≈ 2.5% in the angular rate, while the iILR is the resonance that shows a position more stable under these variations. The peak resonance in the histogram of phase reversals corresponding to the oILR has not been positively identified in three cases; overestimation of the position angle, underestimation of the inclination and offset of the center of the galaxy, in all these cases we find only one phase reversal in the histogram (N=1 peak), which could be associated with this resonance, but it has been rejected as it does not satisfy the N=1 peak criteria described in step 5 of section II.

IV. THE OBSERVATIONAL DATA.

The data for most of the galaxies studied in the present article, 101 in number, were obtained from the Gassendi Hα survey of Spirals data base[1]. We have rejected all those galaxies from the GHASP sample having the following parameter properties: (*i*) galaxies with inclinations higher than 85°. (*ii*) galaxies with inclinations lower than 5° and (*iii*) galaxies where the detected Hα emission covers less than 50% of the continuum image. In all three cases we were not able to derive reliable rotation curves for the galaxies. Application of these criteria led us to select only one half of the objects in the GHASP data base. GHASP is a survey of 203 galaxies conducted at the Haute Provence observatory using a Fabry-Perot instrument on the 1.93 m reflector. The observations were taken during a series of runs between 1998 and 2004. Details of the survey can be found in Epinat et al. (2008). Most of the observations were made with a version of the instrument with a field of view of 4x4 arcmin$^2$ and the pixel size is 0.96 arcsec, but a fraction were taken with an updated version having a field of 6x6 arcmin$^2$ and the pixel size is 0.68 arcsec. The version of the instrument used is indicated in the last column of Table 3, where the pixel scale of the data is given. The range of seeing limited image resolution is between 1 and 6 arcsec. The remaining four galaxies (UGC3013, UGC5303, UGC6118 and UGC7420) were observed with the GHαFaS (Galaxies Hα Fabry-Perot System) on the 4.2m William Herschel telescope at the Roque de los Muchachos observatory, during observing runs between 2009 and 2011. Details of the instrument and its use can be found in Hernandez et al. (2008). The field of view of this instrument is 3.4x3.4 arcmin$^2$, and its pixel size is ∼ 0.2 arcsec, with the exception of UGC3013 for which the pixel size is double. Typical seeing limited image resolution is of order 1.2 arcsec. The values of the main parameters, such as morphological type, inclination angle, position angles, distance and seeing, for each galaxy of the present survey, are given in Table 3.

---

[1] web site: http://fabryperot.oamp.fr/FabryPerot/jsp/tiles/content/html/info_ghasp.jsp

Table 3. Basic Parameters of the Galaxies

| Object Name | | Morphology | $i$ | PA | Distance | Seeing | Pixel Scale |
| UGC | NGC | | (º) | (º) | (Mpc) | (arcsec) | (arcsec/pix) |
| (1) | (2) | (3) | (4) | (5) | (6) | (7) | (8) |
|---|---|---|---|---|---|---|---|
| 508 | 266 | SB(rs)ab | 25 | 123 | 63.8 | 3.8 | 0.68 |
| 763 | 428 | SAB(s)m | 54 | 117 | 12.7 | 4.0 | 0.69 |
| 1256 | 672 | SB(s)cd | 76 | 73 | 7.2 | 3.2 | 0.96 |
| 1317 | 674 | SAB(r)c | 73 | 106 | 42.2 | 3.4 | 0.69 |
| 1437 | 753 | SAB(rs)bc | 47 | 307 | 66.8 | 3.8 | 0.69 |
| 1736 | 864 | SAB(rs)c | 35 | 27 | 17.6 | 4.7 | 0.69 |
| 1913 | 925 | SAB(s)c | 48 | 288 | 9.3 | 2.9 | 0.69 |
| 2080 | -- | SAB(rs)cd | 25 | 336 | 13.7 | 2.6 | 0.95 |
| 2141 | 1012 | S0 | 74 | 191 | 12.2 | 2.3 | 0.68 |
| 2193 | 1058 | SA(rs)c | 6 | 305 | 9.8 | 3.8 | 0.68 |
| 2855 | -- | SABc | 68 | 100 | 17.5 | 3.4 | 0.97 |
| 3013 | 1530 | SB(rs)b | 55 | 195 | 37.0 | 1.3 | 0.39 |
| 3273 | -- | Sm | 82 | 42 | 12.2 | 7.4 | 0.68 |
| 3463 | -- | SAB(s)bc | 63 | 110 | 38.6 | 7.2 | 0.69 |
| 3574 | -- | SA(s)cd | 19 | 99 | 21.8 | 4.4 | 0.95 |
| 3685 | -- | SB(rs)b | 12 | 298 | 26.3 | 2.5 | 0.68 |
| 3691 | -- | SAcd | 64 | 284 | 28.9 | 4.6 | 0.68 |
| 3709 | 2342 | S (pec) | 55 | 232 | 70.7 | 3.3 | 0.69 |
| 3734 | 2344 | SA(rs)c | 43 | 139 | 15.9 | 2.9 | 0.68 |
| 3740 | 2276 | SAB(rs)c | 48 | 247 | 17.1 | 3.4 | 0.68 |
| 3809 | 2336 | SAB(r)bc | 58 | 357 | 32.9 | 3.1 | 0.97 |
| 3826 | -- | SAB(s)d | 20 | 254 | 25.7 | 3.2 | 0.68 |
| 3876 | -- | SA(s)d | 59 | 358 | 14.5 | 3.4 | 0.67 |
| 3915 | -- | SB | 47 | 30 | 63.6 | 3.4 | 0.69 |
| 4165 | 2500 | SB(rs)d | 41 | 265 | 11 | 2.5 | 0.68 |
| 4273 | 2543 | SB(s)b | 60 | 212 | 35.4 | 3.7 | 0.69 |
| 4284 | 2541 | SA(s)cd | 59 | 176 | 9.8 | 3.8 | 0.96 |
| 4325 | 2552 | SA(s)m | 63 | 57 | 10.9 | 3.7 | 0.95 |
| 4422 | 2595 | SAB(rs)c | 25 | 36 | 58.1 | 2.7 | 0.69 |
| 4555 | 2649 | SAB(rs)bc | 38 | 90 | 58 | 6.0 | 0.69 |
| 4936 | 2805 | SAB(rs)d | 13 | 294 | 25.6 | 4.0 | 0.98 |
| 5175 | 2977 | Sb | 56 | 143 | 44.1 | 6.8 | 0.69 |
| 5228 | -- | SB(s)c | 72 | 120 | 24.7 | 2.8 | 0.69 |
| 5251 | 3003 | Sbc | 73 | 260 | 21.5 | 3.1 | 0.66 |
| 5253 | 2985 | SA(rs)ab | 40 | 356 | 21.1 | 5.8 | 0.96 |
| 5303 | 3041 | SAB(rs)c | 36 | 273 | 26.4 | 1.8 | 0.19 |
| 5319 | 3061 | SB(rs)c | 30 | 345 | 35.8 | 2.4 | 0.68 |
| 5414 | 3104 | IAB(s)m | 71 | 219 | 10 | 2.9 | 0.97 |
| 5510 | 3162 | SAB(rs)bc | 31 | 200 | 18.6 | 3.5 | 0.69 |
| 5532 | 3147 | SA(rs)bc | 32 | 147 | 41.1 | 3.3 | 0.68 |
| 5786 | 3310 | SAB pec | 53 | 153 | 14.2 | 4.0 | 0.68 |
| 5840 | 3344 | SAB(r)bc | 25 | 333 | 6.9 | 3.2 | 0.68 |
| 5842 | 3346 | SB(rs)cd | 47 | 292 | 15.2 | 3.5 | 0.68 |
| 5982 | 3430 | SAB(rs)c | 55 | 28 | 20.8 | 4.3 | 0.95 |
| 6118 | 3504 | SAB(s)ab | 19 | 330 | 19.8 | 1.2 | 0.19 |
| 6277 | 3596 | S0 | 17 | 76 | 16.9 | 2.9 | 0.67 |
| 6521 | 3719 | SA pec | 46 | 20 | 78.6 | 3.7 | 0.69 |
| 6523 | 3720 | SAa | 24 | 353 | 80 | 3.7 | 0.69 |
| 6537 | 3726 | SAB(r)c | 47 | 200 | 14.3 | 5.8 | 0.96 |
| 6702 | 3840 | Sa | 38 | 256 | 99.8 | 3.7 | 0.68 |
| 6778 | 3893 | SAB(rs)c | 49 | 343 | 15.5 | 4.2 | 0.96 |
| 7021 | 4045 | SAB(r)a | 56 | 266 | 26.8 | 4.3 | 0.68 |
| 7045 | 4062 | SA(s)c | 68 | 99 | 11.4 | 2.1 | 0.68 |
| 7154 | 4145 | SAB(rs)d | 65 | 275 | 16.2 | 4.2 | 0.68 |
| 7323 | 4242 | SAB(s)dm | 51 | 38 | 8.1 | 6.0 | 0.97 |
| 7420 | 4303 | SAB(rs)bc | 29 | 135 | 16.1 | 1.5 | 0.19 |
| 7766 | 4559 | SAB(rs)cd | 69 | 323 | 13 | 2.0 | 0.68 |
| 7831 | 4605 | SBc pec | 56 | 290 | 5.2 | 2.2 | 0.69 |
| 7853 | 4618 | SB(rs)m | 58 | 217 | 8.9 | 2.6 | 0.68 |
| 7861 | 4625 | SAB pec | 47 | 297 | 10.2 | 3.3 | 0.69 |
| 7876 | 4635 | SAB(s)d | 53 | 344 | 14.5 | 3.0 | 0.68 |
| 7901 | 4651 | SA(rs)c | 53 | 254 | 20.7 | 3.8 | 0.68 |
| 7985 | 4713 | SAB(rs)d | 49 | 276 | 13.7 | 5.4 | 0.69 |
| 8334 | 5055 | SA(rs)bc | 66 | 100 | 9.8 | 2.3 | 0.69 |
| 8403 | 5112 | SB(rs)cd | 57 | 121 | 19.1 | 4.4 | 0.68 |
| 8490 | 5204 | SA(s)m | 40 | 167 | 4.7 | 3.8 | 0.94 |
| 8709 | 5297 | SAB(s)c | 76 | 330 | 35 | 3.1 | 0.69 |
| 8852 | 5376 | SAB(r)b | 52 | 63 | 30.6 | 2.8 | 0.69 |
| 8937 | 5430 | SB(s)b | 32 | 185 | 49 | 4.4 | 0.68 |
| 9179 | 5585 | SAB(s)d | 36 | 49 | 5.7 | 3.1 | 0.68 |
| 9248 | 5622 | Sb | 58 | 261 | 54.9 | 4.7 | 0.68 |
| 9358 | 5678 | SAB(rs)b | 54 | 182 | 29.1 | 4.2 | 0.68 |
| 9363 | 5668 | SA(s)d | 18 | 147 | 22.3 | 2.8 | 0.69 |
| 9366 | 5676 | SA(rs)bc | 62 | 225 | 37.7 | 3.2 | 0.68 |
| 9465 | 5727 | SABdm | 65 | 127 | 26.4 | 3.4 | 0.68 |
| 9736 | 5874 | SAB(rs)c | 51 | 219 | 45.4 | 4.8 | 0.69 |
| 9753 | 5879 | SA(rs)bc | 69 | 3 | 12.4 | 3.9 | 0.68 |
| 9866 | 5949 | SA(r)bc | 56 | 148 | 7.4 | 6.9 | 0.7 |
| 9943 | 5970 | SB(r)c | 54 | 266 | 28 | 3.9 | 0.69 |
| 9969 | 5985 | SAB(r)b | 61 | 16 | 36 | 3.3 | 0.96 |
| 10075 | 6015 | SA(s)cd | 62 | 210 | 14.7 | 3.2 | 0.68 |
| 10359 | 6140 | SB(s)cd pec | 44 | 284 | 16 | 2.4 | 0.68 |
| 10445 | -- | Scd | 47 | 110 | 16.9 | 3.0 | 0.68 |
| 10470 | 6217 | SB(rs)bc | 34 | 287 | 21.2 | 3.3 | 0.68 |
| 10502 | -- | SA(rs)c | 50 | 99 | 61.2 | 3.6 | 0.68 |
| 10521 | 6207 | SA(s)c | 59 | 20 | 18 | 3.0 | 0.69 |
| 10546 | 6236 | SAB(s)cd | 42 | 182 | 20.4 | 3.4 | 0.69 |
| 10564 | 6248 | SBd | 77 | 149 | 18.4 | 3.0 | 0.68 |
| 10652 | 6283 | S | 21 | 225 | 18.2 | 3.2 | 0.69 |
| 10757 | -- | Scd | 44 | 56 | 19.5 | 2.4 | 0.69 |
| 10897 | 6412 | SA(s)c | 31 | 115 | 20.5 | 3.8 | 0.95 |
| 11012 | 6503 | SA(s)cd | 72 | 299 | 5.3 | 2.6 | 0.68 |
| 11124 | -- | SB(s)cd | 51 | 182 | 23.7 | 2.7 | 0.68 |
| 11218 | 6643 | SA(rs)c | 58 | 42 | 22.8 | 3.6 | 0.96 |
| 11283 | -- | SB(s)bm | 34 | 120 | 31.3 | 3.3 | 0.96 |
| 11407 | 6764 | SB(s)bc | 64 | 65 | 35.8 | 2.6 | 0.69 |
| 11466 | -- | S | 66 | 226 | 14.2 | 2.3 | 0.69 |
| 11557 | -- | SAB(s)dm | 29 | 276 | 19.7 | 3.1 | 0.68 |
| 11861 | -- | SABdm | 43 | 218 | 25.1 | 6.1 | 0.69 |
| 11872 | 7177 | SAB(r)b | 47 | 86 | 18.1 | 4.7 | 0.68 |
| 11914 | 7217 | SA(r)ab | 33 | 266 | 15 | 4.3 | 0.69 |
| 12276 | 7440 | SB(r)a | 33 | 322 | 77.8 | 2.7 | 0.69 |
| 12343 | 7479 | SB(s)c | 52 | 203 | 26.9 | 5.5 | 0.67 |
| 12754 | 7741 | SB(s)cd | 53 | 342 | 8.9 | 3.0 | 0.96 |

**Notes.** Columns (1) & (2) provide the names of the galaxies according to the UGC and NGC system. Column (3) describes the morphology of each galaxy. The inclination angle, the position angle and the distance are written in columns (4), (5) and (6), respectively. Column (7) & (8) give the angular resolution and the pixel scale for each galaxy observed. All values of the parameters come from the GHASP data base (Epinat et al. 2008), except those from UGC5303 and UGC6118 which come from S4G data base (Sheth et al. 2010). Parameters of UGC 3013 and UGC7420 are obtained applying the tilted ring model to the velocity field. Spectral resolution for all galaxies is ~16 km·s$^{-1}$, apart from UGC 3013, UGC5303, UGC6118 and UGC7420 for which the spectral resolution is 8.3 km·s$^{-1}$.

The data reduction for spectrographs using FP etalons has a number of common features which can be briefly summarized as follows. A lamp emitting sharp lines at precisely known wavelengths is required in order to perform the phase calibration and the wavelength calibration. The data from a Fabry-Perot takes the basic form of a data cube, containing a series of channel maps, or interferograms. Observing the calibration lamp with the FP instrument produces, thus, a calibration cube from which a monochromatic focal surface, consisting on a paraboloid of revolution about the optic axis, is extracted and used for the phase calibration. The wavelength calibration is straightforward given the wavelength of the selected emission line of the lamp. The two calibrations combined allow us to transform the H$\alpha$ cube of the observed galaxy into a set of monochromatic channel maps. Each channel map gives the surface brightness of the object (the galaxy) in a well defined narrow wavelength range. This full set of channel maps is termed a calibrated data cube, which gives the spectrum of the H$\alpha$ emission line for each pixel of the map. From these data cubes we can derive a map in H$\alpha$ surface brightness determining the integrated area of the emission peaks across the field, sometimes referred to as zeroth moment map, a map in velocity corresponding to the velocities of the peaks of the emission lines across the full field, which is the first moment map, and a map in velocity dispersion, corresponding to the line widths across the field, also known as the second moment map. Details of how this was performed in GHASP can be found in Garrido et al. (2002), while the equivalent details for GH$\alpha$FaS are given in Hernandez et al. (2008). All the work in the present paper is based on the first moment maps of the galaxies observed.

Table 4. Resonance radii

| Object UGC | N | P | $r_1$ | $r_2$ | $r_3$ | $r_4$ (arcsec) | $r_5$ | $r_6$ | $r_7$ |
|---|---|---|---|---|---|---|---|---|---|
| (1) | (2) | (3) | (4) | (5) | (6) | (7) | (8) | (9) | (10) |
| 508 | 5 | 2 | **7.98±1.8**(ILR:4) | 25.6±1.8(CR:1)[1] | 37.3±2.4(CR:2)[2] | 51.8±2.3(CR:3)[1] | 65.9±4.1(CR:4)[2] | -- | -- |
| 763 | 4 | 1 | 40.6±3.7(CR:1)[1] | 51.9±1.9(CR:2) | 98.6±1.9(CR:3)[1] | 129.1±1.9(CR:4) | -- | -- | -- |
| 1256 | 5 | 0 | 10.9±1.9(CR:1) | 36.6±3.0(CR:2) | 58.7±1.7(CR:3) | 90.7±4.4(CR:4) | 113.7±6.3(CR:5) | -- | -- |
| 1317 | 4 | 2 | 10.5±1.5(CR:1)[1] | 26.3±5.1(CR:2)[1,2] | 48.2±1.5(CR:3)[2] | 66.1±2.3(CR:4) | -- | -- | -- |
| 1437 | 4 | 1 | **4.1±1.9**(ILR:2) | 31.3±3.8(CR:1)[1] | 44.2±3.5(CR:2)[1] | 57.2±2.7(CR:3) | -- | -- | -- |
| 1736 | 5 | 0 | 5.8±2.2(CR:1) | 29.7±3.6(CR:2) | 56.1±2.1(CR:3) | 84.4±2.3(CR:4) | 102.1±2.1(CR:5) | -- | -- |
| 1913 | 4 | 0 | 36.1±1.4(CR:1) | 47.6±1.3(CR:2) | 96.0±1.4(CR:3) | 118.5±3.8(CR:4) | -- | -- | -- |
| 2080 | 2 | 0 | 32.1±1.3(CR:1) | **74.8±3.4**(OLR:1) | -- | -- | -- | -- | -- |
| 2141 | 5 | 0 | 13.5±1.8(CR:1) | 38.3±1.9(CR:2) | 47.9±1.1(CR:3) | 61.6±1.8(CR:4) | 70.2±2.2(CR:5) | -- | -- |
| 2193 | 3 | 1 | 26.6±4.6(CR:1)[1] | 47.6±1.7(CR:2)[1] | 55.6±1.7(CR:3) | -- | -- | -- | -- |
| 2855 | 4 | 1 | 11.5±1.7(CR:1) | 41.2±6.4(CR:2)[1] | **70.4±6.3**(OLR:1) | 95.5±5.9(CR:3)[1] | -- | -- | -- |
| 3013 | 7 | 2 | **9.0±2.4**(ILR:3) | 43.6±1.5(CR:1)[1] | 53.2±3.7(CR:2)[2] | 67.9±2.6(CR:3)[1] | 76.5±3.2(CR:4) | 86.9±2.8(CR:5)[2] | **100.1±2.7**(OLR:3) |
| 3273 | 4 | 0 | 7.4±3.2(CR:1) | 31.9±3.7(CR:2) | 73.9±4.4(CR:3) | 99.4±6.3(CR:4) | -- | -- | -- |
| 3463 | 4 | 3 | 8.9±3.8(CR:1)[1] | 21.7±4.4(CR:2)[1,2] | 45.5±7.6(CR:3)[2,3] | 68.7±5.1(CR:4)[3] | -- | -- | -- |
| 3574 | 3 | 2 | 5.4±1.9(CR:1)[1] | 16.8±3.0(CR:2)[1,2] | 42.6±7.4(CR:3)[2] | -- | -- | -- | -- |
| 3685 | 4 | 0 | 16.8±1.3(CR:1) | 27.7±1.7(CR:2) | 49.5±1.4(CR:3) | 55.7±1.9(CR:4) | -- | -- | -- |
| 3691 | 4 | 1 | 10.3±2.3(CR:1) | 24.3±2.4(CR:2)[1] | 42.0±2.0(CR:3) | 54.2±2.3(CR:4)[1] | -- | -- | -- |
| 3709 | 3 | 1 | 7.9±1.5(CR:1)[1] | 19.6±1.5(CR:2)[1] | 24.6±1.7(CR:3) | -- | -- | -- | -- |
| 3734 | 4 | 2 | 17.6±1.3(CR:1)[1] | 34.8±1.9(CR:2)[1,2] | 41.9±1.7(CR:3) | 64.3±1.7(CR:4)[2] | -- | -- | -- |
| 3740 | 5 | 0 | 47.5±1.8(CR:1) | 61.3±2.2(CR:2) | 69.8±1.8(CR:3) | 81.5±1.5(CR:4) | 89.6±1.5(CR:5) | -- | -- |
| 3809 | 7 | 2 | 35.4±1.8(CR:1) | 62.0±4.9(CR:2)[1] | 82.5±2.9(CR:3)[2] | 91.9±2.1(CR:4) | 107.3±3.1(CR:5)[1] | 119.9±3.6(CR:6) | 128.4±2.5(CR:7)[2] |
| 3826 | 2 | 0 | 13.9±1.8(CR:1) | **39.2±1.4**(OLR:1) | -- | -- | -- | -- | -- |
| 3876 | 2 | 1 | 33.9±2.2(CR:1)[1] | 59.1±3.0(CR:2)[1] | -- | -- | -- | -- | -- |
| 3915 | 1 | 0 | 15.3±2.3(CR:1) | -- | -- | -- | -- | -- | -- |

| | | | | | | | | |
|---|---|---|---|---|---|---|---|---|
| 4165 | 5 | 1 | 16.4±1.2(CR:1) | 25.6±1.6(CR:2)[1] | 48.9±2.0(CR:3)[1] | 65.3±1.2(CR:4) | 75.4±1.2(CR:5) | -- | -- |
| 4273 | 4 | 0 | **7.7±1.8**(ILR:3) | 31.8±4.1(CR:1) | 47.4±3.2(CR:2) | 79.6±1.8(CR:3) | -- | -- | -- |
| 4284 | 5 | 2 | 10.8±4.8(CR:1)[1] | 31.4±6.4(CR:2)[2] | 52.9±1.9(CR:3)[1] | 68.1±3.9(CR:4) | 93.8±6.5(CR:5)[2] | -- | -- |
| 4325 | 3 | 1 | 16.6±1.7(CR:1)[1] | 49.4±2.5(CR:2)[1] | 67.2±2.4(CR:3) | -- | -- | -- | -- |
| 4422 | 4 | 1 | **8.9±1.2**(ILR:3) | 24.8±1.2(CR:1)[1] | 36.5±2.4(CR:2) | 44.3±1.5(CR:3)[1] | -- | -- | -- |
| 4555 | 3 | 1 | 13.0±2.6(CR:1)[1] | 28.0±2.6(CR:2)[1] | 35.3±2.6(CR:3) | -- | -- | -- | -- |
| 4936 | 7 | 1 | 16.9±1.8(CR:1) | 34.8±1.8(CR:2)[1] | 70.8±2.1(CR:3) | 82.6±1.8(CR:4)[1] | 102.2±1.8(CR:5) | 112.3±1.8(CR:6) | **137.7±1.7**(OLR:3) |
| 5175 | 3 | 1 | 6.7±2.9(CR:1) | 15.6±4.4(CR:2)[1] | 35.3±3.6(CR:3)[1] | -- | -- | -- | -- |
| 5228 | 4 | 0 | 16.6±1.3(CR:1) | 30.4±1.3(CR:2) | 50.1±1.6(CR:3) | **59.7±1.3**(OLR:2) | -- | -- | -- |
| 5251 | 6 | 2 | 12.9±1.6(CR:1)[1] | 47.4±2.3(CR:2)[1,2] | 70.2±2.0(CR:3) | 87.4±2.1(CR:4) | 102.6±2.6(CR:5)[2] | 110.8±2.3(CR:6) | -- |
| 5253 | 4 | 2 | 43.2±2.6(CR:1)[1] | 58.8±3.7(CR:2)[2] | 70.4±3.1(CR:3)[1] | 82.9±3.4(CR:4)[2] | -- | -- | -- |
| 5303 | 4 | 2 | 16.0±4.6(CR:1)[1] | 35.2±6.2(CR:2)[1,2] | 48.7±2.5(CR:3) | 71.7±2.4(CR:4)[2] | -- | -- | -- |
| 5319 | 5 | 2 | 9.3±1.1(CR:1)[1] | 16.2±1.2(CR:2)[2] | 21.8±1.2(CR:3)[1] | 25.8±1.2(CR:4) | 33.1±2.8(CR:5)[2] | -- | -- |
| 5414 | 6 | 0 | 19.3±3.1(CR:1) | 33.0±2.1(CR:2) | 51.1±1.8(CR:3) | 62.0±3.4(CR:4) | 77.0±1.3(CR:5) | 92.3±2.3(CR:6) | -- |
| 5510 | 6 | 1 | 13.6±1.6(CR:1)[1] | 30.8±1.6(CR:2)[1] | 36.7±1.6(CR:3) | 46.7±2.4(CR:4) | 54.4±1.6(CR:5) | **66.5±1.6**(OLR:2) | -- |
| 5532 | 5 | 4 | 7.0±1.6(CR:1)[1] | 16.3±3.3(CR:2)[1,2] | 34.7±4.8(CR:3)[2,3] | 54.3±1.7(CR:4)[3,4] | 76.7±1.8(CR:4)[4] | -- | -- |
| 5786 | 5 | 1 | **20.6±2.1**(ILR:1) | 29.0±3.5(CR:1)[1] | 43.6±8.7(CR:2) | 63.0±1.9(CR:3) | 80.1±6.0(CR:4)[1] | -- | -- |
| 5840 | 5 | 1 | 31.6±4.2(CR:1)[1] | 60.9±5.4(CR:2)[1] | 79.7±1.6(CR:3) | 93.1±1.6(CR:4) | **120.6±1.6**(OLR:2) | -- | -- |
| 5842 | 3 | 1 | 15.5±2.0(CR:1) | 23.4±1.5(CR:2)[1] | 66.4±6.1(CR:3)[1] | -- | -- | -- | -- |
| 5982 | 5 | 1 | **11.7±1.8**(ILR:4) | 52.6±3.5(CR:1)[1] | 68.3±3.5(CR:2) | 80.0±2.8(CR:3) | 96.2±4.6(CR:4)[1] | -- | -- |
| 6118 | 6 | 2 | **9.9±2.2**(CR:1,ILR:4)[1] | 19.2±2.5(CR:2)[1,2] | 29.1±2.8(CR:3) | 37.6±2.2(CR:4) | 43.7±1.7(CR:5)[2] | **66.6±1.6**(OLR:4) | -- |
| 6277 | 4 | 1 | 8.3±1.3(CR:1)[1] | 23.6±1.4(CR:2)[1] | 33.2±1.4(CR:3) | **47.5±1.4**(OLR:2) | -- | -- | -- |
| 6521 | 4 | 2 | 11.0±2.1(CR:1)[1] | 21.0±1.6(CR:2)[1,2] | 27.9±1.8(CR:3) | 41.4±3.7(CR:4)[2] | -- | -- | -- |
| 6523 | 2 | 1 | 4.6±2.6(CR:1)[1] | 10.9±2.1(CR:2)[1] | -- | -- | -- | -- | -- |
| 6537 | 5 | 2 | 35.8±2.5(CR:1)[1] | 59.8±2.5(CR:2)[2] | 105.8±3.2(CR:3) | 122.0±3.1(CR:4) | **143.7±2.5**(OLR:3) | -- | -- |
| 6702 | 3 | 2 | 4.5±1.7(CR:1)[1] | 10.5±1.7(CR:2)[1,2] | 20.8±1.9(CR:3)[2] | -- | -- | -- | -- |
| 6778 | 5 | 2 | 13.0±6.2(CR:1)[1] | 34.9±4.2(CR:2)[1,2] | 63.0±3.0(CR:3) | 75.2±2.0(CR:4)[2] | 88.8±4.5(CR:5) | -- | -- |
| 7021 | 3 | 1 | 7.6±1.9(CR:1)[1] | 18.5±2.0(CR:2)[1] | 39.4±1.9(CR:3) | -- | -- | -- | -- |
| 7045 | 3 | 1 | 43.9±1.1(CR:1)[1] | 54.2±2.4(CR:2) | 85.4±1.2(CR:3)[1] | -- | -- | -- | -- |

| | | | | | | | | |
|---|---|---|---|---|---|---|---|---|
| 7154 | 5 | 1 | 18.1±2.1(CR:1)[1] | 46.3±2.1(CR:2)[1] | 63.4±2.7(CR:3) | 135.3±2.1(CR:4) | 187.0±5.3(CR:5) | -- | -- |
| 7323 | 5 | 2 | 13.8±2.7(CR:1)[1] | 44.7±2.8(CR:2)[1,2] | 92.4±7.4(CR:3) | 117.7±2.7(CR:4) | 130.1±3.1(CR:5) | -- | -- |
| 7420 | 6 | 2 | 4.6±1.0(CR:1) | 26.6±1.0(CR:2)[1] | 36.1±3.1(CR:3)[2] | 49.9±1.1(CR:4) | 60.4±2.3(CR:5)[1] | 68.0±1.0(CR:6)[2] | -- |
| 7766 | 5 | 1 | 15.8±1.0(CR:1)[1] | 37.3±1.9(CR:2)[1] | 51.6±1.6(CR:3) | 60.0±4.1(CR:4) | 89.6±1.0(CR:5) | -- | -- |
| 7831 | 5 | 0 | 4.7±1.0(CR:1) | 17.8±1.0(CR:2) | 41.3±3.6(CR:3) | 53.2±1.0(CR:4) | 63.0±1.0(CR:5) | -- | -- |
| 7853 | 3 | 0 | 41.7±1.1(CR:1) | 54.5±1.7(CR:2) | 87.1±3.5(CR:3) | -- | -- | -- | -- |
| 7861 | 3 | 0 | 15.2±1.5(CR:1) | 31.1±2.6(CR:2) | 37.1±1.5(CR:3) | -- | -- | -- | -- |
| 7876 | 3 | 0 | 19.6±1.5(CR:1) | 28.7±1.9(CR:2) | 43.9±2.1(CR:3) | -- | -- | -- | -- |
| 7901 | 4 | 1 | **15.6±1.6**(ILR:3) | 48.6±5.8(CR:1)[1] | 58.0±1.7(CR:2) | 80.4±2.9(CR:3)[1] | -- | -- | -- |
| 7985 | 5 | 3 | 12.6±2.3(CR:1)[1] | 27.6±2.3(CR:2)[2] | 38.3±5.3(CR:3)[1,3] | 54.5±3.5(CR:4)[2] | 72.5±8.1(CR:5)[3] | -- | -- |
| 8334 | 7 | 1 | 12.6±1.0(CR:1)[1] | 28.8±1.0(CR:2)[1] | 47.0±1.0(CR:3) | 71.7±1.0(CR:4) | **89.7±1.0**(OLR:3) | **129.5±1.0**(OLR:4) | 205.1±1.0(CR:5) |
| 8403 | 6 | 0 | 30.8±2.0(CR:1) | 44.5±2.1(CR:2) | 51.7±2.8(CR:3) | 70.4±2.2(CR:4) | 86.6±5.4(CR:5) | 105.9±3.8(CR:6) | -- |
| 8490 | 5 | 0 | 9.4±2.5(CR:1) | 19.3±2.1(CR:2) | 31.6±3.2(CR:3) | 49.5±2.9(CR:4) | **67.7±2.8**(OLR:2) | -- | -- |
| 8709 | 5 | 0 | 17.2±1.9(CR:1) | 27.9±1.4(CR:2) | 50.0±1.4(CR:3) | 73.9±1.4(CR:4) | 92.7±4.7(CR:5) | -- | -- |
| 8852 | 3 | 1 | 10.0±3.1(CR:1)[1] | 26.8±1.3(CR:2)[1] | 32.7±1.3(CR:3) | -- | -- | -- | -- |
| 8937 | 4 | 2 | 15.2±3.0(CR:1)[1] | 25.2±2.1(CR:2)[1] | 33.3±3.0(CR:3)[2] | 47.5±2.6(CR:4)[2] | -- | -- | -- |
| 9179 | 6 | 2 | 19.5±1.4(CR:1)[1] | 34.4±1.5(CR:2) | 48.1±3.0(CR:3)[2] | 74.3±3.4(CR:4)[1] | 100.8±1.4(CR:5) | 144.7±1.4(CR:6)[2] | -- |
| 9248 | 4 | 2 | 11.6±2.1(CR:1)[1] | 17.4±2.1(CR:2)[2] | 22.3±2.4(CR:3)[1] | 30.4±2.2(CR:4)[2] | -- | -- | -- |
| 9358 | 3 | 1 | 14.7±1.9(CR:1)[1] | 30.4±2.4(CR:2)[1] | 39.3±3.1(CR:3) | -- | -- | -- | -- |
| 9363 | 3 | 0 | 36.7±1.2(CR:1) | 47.4±1.2(CR:2) | 62.0±3.6(CR:3) | -- | -- | -- | -- |
| 9366 | 4 | 2 | 11.6±3.1(CR:1)[1] | 21.3±1.6(CR:2)[1,2] | 36.9±5.6(CR:3)[2] | 47.2±2.6(CR:4) | -- | -- | -- |
| 9465 | 3 | 1 | 21.5±2.0(CR:1)[1] | 28.1±1.5(CR:2) | 47.7±1.5(CR:3)[1] | -- | -- | -- | -- |
| 9736 | 2 | 1 | 27.5±2.8(CR:1)[1] | 56.1±3.2(CR:2)[1] | -- | -- | -- | -- | -- |
| 9753 | 2 | 1 | 12.2±3.8(CR:1)[1] | 30.6±1.9(CR:2)[1] | -- | -- | -- | -- | -- |
| 9866 | 1 | 0 | 31.2±4.9(CR:1) | -- | -- | -- | -- | -- | -- |
| 9943 | 5 | 2 | 19.1±2.8(CR:1)[1] | 34.4±4.2(CR:2)[1,2] | 52.2±2.0(CR:3) | 59.7±1.7(CR:4)[2] | 69.7±1.7(CR:5) | -- | -- |
| 9969 | 6 | 3 | 27.7±2.8(CR:1)[1] | 40.1±2.3(CR:2)[2] | 57.3±5.8(CR:3)[1] | 67.2±1.6(CR:4)[3] | 82.3±7.3(CR:5)[2] | 105.5±4.3(CR:6)[3] | -- |
| 10075 | 5 | 2 | 9.7±1.5(CR:1)[1] | 35.6±2.0(CR:2)[1,2] | 54.2±1.5(CR:3) | 93.2±5.5(CR:4)[2] | 127.5±1.5(CR:5) | -- | -- |
| 10359 | 3 | 0 | 16.6±1.4(CR:1) | 32.7±3.6(CR:2) | 58.3±2.3(CR:3) | -- | -- | -- | -- |

| (1) | (2) | (3) | (4) | (5) | (6) | (7) | (8) | (9) | (10) |
|---|---|---|---|---|---|---|---|---|---|
| 10445 | 3 | 0 | 6.0±2.2(CR:1) | 28.5±1.5(CR:2) | **55.3±3.0(OLR:2)** | -- | -- | -- | -- |
| 10470 | 5 | 2 | 11.3±1.4(CR:1)[1] | 35.6±2.0(CR:2)[1,2] | 43.9±1.5(CR:3) | 57.8±3.4(CR:4) | 65.7±1.5(CR:5)[2] | -- | -- |
| 10502 | 3 | 1 | 34.1±3.1(CR:1)[1] | 48.8±1.7(CR:2) | 68.4±4.4(CR:3)[1] | -- | -- | -- | -- |
| 10521 | 6 | 2 | 12.6±1.4(CR:1)[1] | 24.3±1.4(CR:2)[2] | 38.0±1.4(CR:3)[1] | 49.5±1.7(CR:4) | 55.4±1.7(CR:5)[2] | 64.1±3.6(CR:6) | -- |
| 10546 | 4 | 2 | 28.1±1.5(CR:1)[1] | 38.9±3.3(CR:2)[2] | 47.5±1.5(CR:3)[1] | 55.7±1.5(CR:4)[2] | -- | -- | -- |
| 10564 | 3 | 0 | 8.8±1.4(CR:1) | 45.6±1.5(CR:2) | 81.2±2.6(CR:3) | -- | -- | -- | -- |
| 10652 | 2 | 0 | 12.4±1.4(CR:1) | 22.9±1.4(CR:2) | -- | -- | -- | -- | -- |
| 10757 | 2 | 0 | 12.4±1.3(CR:1) | 22.5±1.0(CR:2) | -- | -- | -- | -- | -- |
| 10897 | 5 | 0 | 7.5±1.9(CR:1) | 19.0±2.0(CR:2) | 26.3±2.8(CR:3) | 36.4±2.1(CR:4) | **42.6±1.8(OLR:2)** | -- | -- |
| 11012 | 4 | 1 | 21.8±1.8(CR:1) | 37.2±1.1(CR:2)[1] | 56.8±1.6(CR:3) | 89.3±1.5(CR:4)[1] | -- | -- | -- |
| 11124 | 4 | 0 | 11.1±2.8(CR:1) | 25.4±1.4(CR:2) | 48.3±2.7(CR:3) | 59.5±1.3(CR:4) | -- | -- | -- |
| 11218 | 6 | 3 | 9.8±1.7(CR:1)[1] | 22.5±1.7(CR:2)[2] | 32.0±4.1(CR:3)[1,3] | 48.6±4.0(CR:4)[2] | 61.8±3.1(CR:5)[3] | 76.4±1.8(CR:6) | -- |
| 11283 | 5 | 1 | 4.0±1.4(CR:1) | 12.6±1.5(CR:2)[1] | 24.1±1.5(CR:3)[1] | 30.7±1.8(CR:4) | **38.4±2.6(OLR:3)** | -- | -- |
| 11407 | 5 | 3 | 3.3±1.7(CR:1)[1] | 9.5±2.2(CR:2)[2] | 20.3±1.5(CR:3)[1,3] | 31.4±1.8(CR:4)[2] | 64.9±2.3(CR:5)[3] | -- | -- |
| 11466 | 3 | 1 | 20.2±4.9(CR:1)[1] | 32.8±2.9(CR:2) | 52.2±1.0(CR:3)[1] | -- | -- | -- | -- |
| 11557 | 5 | 0 | 9.1±1.5(CR:1) | 19.0±1.4(CR:2) | 32.4±2.6(CR:3) | 49.0±3.2(CR:4) | 56.6±1.7(CR:5) | -- | -- |
| 11861 | 5 | 1 | 7.6±3.6(CR:1)[1] | 22.3±2.7(CR:2) | 37.8±3.7(CR:3) | 67.3±3.7(CR:4)[1] | **88.6±3.1(OLR:3)** | -- | -- |
| 11872 | 4 | 2 | 5.8±2.1(CR:1)[1] | 19.9±2.8(CR:2)[1,2] | 30.8±2.4(CR:3)[2] | **57.9±2.1(OLR:3)** | -- | -- | -- |
| 11914 | 5 | 3 | 17.0±5.8(CR:1)[1] | 31.4±4.1(CR:2)[1,2] | 48.8±2.2(CR:3)[2,3] | 67.4±2.1(CR:4) | 78.5±4.1(CR:5)[3] | -- | -- |
| 12276 | 3 | 1 | 13.8±1.3(CR:1)[1] | 21.8±1.3(CR:2) | 27.4±1.3(CR:3)[1] | -- | -- | -- | -- |
| 12343 | 5 | 0 | 12.7±2.5(CR:1) | 36.3±3.9(CR:2) | 52.4±3.3(CR:3) | 75.1±3.9(CR:4) | 91.1±4.3(CR:5) | -- | -- |
| 12754 | 6 | 1 | 28.6±1.4(CR:1)[1] | 46.5±1.4(CR:2) | 69.6±7.3(CR:3) | 83.7±1.4(CR:4)[1] | 95.4±3.9(CR:5) | 106.3±1.7(CR:6) | -- |

**Notes.** The name of the galaxy is given in Column (1), the number of peaks in the radial distribution of phase reversals appears in Column (2). Column (3) gives the number of coupling patterns found. The resonance radii from the inside outwards of each galaxy sorted outwards starting from the innermost one are given in Columns (4) to (10). In brackets, each radius is identified with the type of resonance with which is associated (CR=corotation, ILR= Inner Lindblad Resonance, OLR = Outer Lindblad Resonance); for instance, in galaxy UGC508, CR:1 to CR:4 indicate that these are corotation radii for the pattern speeds $\Omega_1$ to $\Omega_4$ of Table 5, and ILR:4 indicates that this is the ILR for the $\Omega_4$ pattern speed, so ILR:4 and CR:4 are two resonances of the same pattern speed. Identical superscripts relating to pairs of resonance radii indicate coupled corotations. ILR & OLR superscripts label those resonances which are found to be Inner and Outer Lindblad Resonances, respectively; their radii are shown in bold face characters.

As outlined above in the section describing the phase reversal method, we use the first moment map to derive the rotation curve of the galaxy, using standard procedures, and then produce a map of the residual velocity by subtracting off the rotational velocity in 2 dimensions. The residual velocity map is used to find the radial positions of the 180° phase changes in the radial velocity component. These positions are then initially assigned the status of resonance radii. Figure 9 (available only online) displays four panels for each galaxy of our sample, from left to right we plot the velocity map of the galaxy; the derived rotation curve together with the best fit; the final non-circular velocity map; and the radial distribution of phase reversals in which the position of the peaks positively identifies as resonances are marked with vertical lines and their uncertainties with horizontal lines. We will see when considering the individual galaxies that the great majority of these turn out indeed to be corotations. Resonance radii, labeled $r_i$, for each galaxy are provided in Table 4, in which the number of peaks is also given in column (2), as well as the number of coupling patterns (column 3). Those pairs of resonance radii which has been found to be interlocked (a coupling pattern), show identical numbered superscripts. As the majority of the resonances are identified as corotation, the radii of these resonences in Table 4 are labeled with a "CR" along with a number indicating the pattern speed with which are associated. These pattern speeds are given in columns (2) to (8) of Table 5. A small fraction of the resonances were identified as Inner Lindblad Resonances, as can be seen in Table 4, where these resonances are bold faced and marked with an "ILR" and a number indicating the pattern speed with which they are associated. We also found a small number of cases where a peak (normally the outermost peak) represents only an Outer Lindblad Resonance (the bold faced peak radius, then, is labeled with an "OLR" in Table 4, the following number obeys to the same criterion than for the case of the other resonances), as well as the more common cases where the resonance peaks harbor both a corotation and the Outer Lindblad Resonance corresponding to another corotation. Note that the special N=1 peak criteria, needed in order to identify a peak with a single phase reversal as a resonance (see step 5 of section II) have been taken into account for all galaxies of our sample. In Table 5 we give the value for the pattern speeds, together with the upper and lower uncertainty, corresponding to those peaks, which have been identified as corotations, for all galaxies with the exception of those having a solid body rotation curve**.** In those galaxies for which the pattern speed has not been calculated, the number that goes with the type of resonance, which in used to label its resonance radii in Table 4, indicates only the relative position of the resonances sorted from inside outwards. At this point it will be best for the reader to look at our descriptions of the results for the individual galaxies, given below.

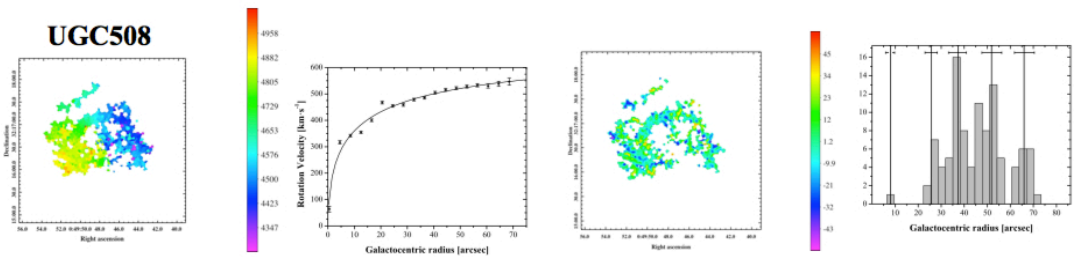

**Figure 9 (only online figure)**. General description of all the panels. From left to right: Velocity field derived from the emission line map in Hα obtained using the Fabry-Perot interferometers named in the text (GHASP or GHαFaS). Rotation curve corresponding to the velocity field shown in left panel, with the observational points and the best fit curve both shown in the figure. Residual velocity field obtained by subtracting off the rotated and projected rotation curve in 2D from the total velocity map shown in left panel. Histogram showing the numbers of phase reversals in the radial velocity component of map (c), as a function of galactocentric radius. The peaks in the histogram give the resonance radii (in a few cases the innermost peak corresponds to the ILR of one of the outer systems, see text); the derived radii are shown as vertical lines, and their uncertainties by horizontal error bars.

Table 5. Pattern Speeds.

| Object UGC | $\Omega_1$ | $\Omega_2$ | $\Omega_3$ | $\Omega_4$ (km·s⁻¹·kpc⁻¹) | $\Omega_5$ | $\Omega_6$ | $\Omega_7$ |
|---|---|---|---|---|---|---|---|
| (1) | (2) | (3) | (4) | (5) | (6) | (7) | (8) |
| 508 | $58.1^{+3.5}_{-3.1}$ | $43.0^{+2.4}_{-2.2}$ | $32.6^{+1.3}_{-1.2}$ | $26.5^{+1.5}_{-1.4}$ | -- | -- | -- |
| 763 | $30.4^{+1.7}_{-1.5}$ | $26.2\pm0.6$ | $16.9\pm0.2$ | $13.8\pm0.2$ | -- | -- | -- |
| 1256 | -- | -- | -- | -- | -- | -- | -- |
| 1317 | $65.1^{+6.1}_{-5.1}$ | $34.7^{+6.2}_{-4.6}$ | $20.8\pm0.6$ | $15.5\pm0.5$ | -- | -- | -- |
| 1437 | $22.2^{+3.4}_{-3.0}$ | $13.3^{+2.0}_{-1.5}$ | $9.6\pm0.4$ | -- | -- | -- | -- |
| 1736 | -- | -- | -- | -- | -- | -- | -- |
| 1913 | -- | -- | -- | -- | -- | -- | -- |
| 2080 | $46.1^{+1.3}_{-1.2}$ | -- | -- | -- | -- | -- | -- |
| 2141 | -- | -- | -- | -- | -- | -- | -- |
| 2193 | $91.6^{+20.2}_{-13.9}$ | $52.0\pm1.8$ | $44.6\pm1.4$ | -- | -- | -- | -- |
| 2855 | $55.5^{+0.2}_{-0.3}$ | $47.5^{+2.7}_{-3.0}$ | $24.2^{+1.9}_{-1.7}$ | -- | -- | -- | -- |
| 3013 | $27.2\pm1.0$ | $22.1^{+1.7}_{-1.5}$ | $17.0\pm0.7$ | $14.9\pm0.7$ | $13.0\pm0.4$ | $11.2\pm0.3$ | -- |
| 3273 | -- | -- | -- | -- | -- | -- | -- |
| 3463 | $58.0^{+17.9}_{-10.5}$ | $33.7^{+5.6}_{-4.2}$ | $19.1^{+3.1}_{-2.4}$ | $13.3^{+1.0}_{-0.8}$ | -- | -- | -- |
| 3574 | $123.3^{+28.1}_{-17.6}$ | $65.9^{+8.3}_{-6.4}$ | $35.5^{+5.1}_{-3.9}$ | -- | -- | -- | -- |
| 3685 | -- | -- | -- | -- | -- | -- | -- |
| 3691 | $42.4^{+4.1}_{-3.4}$ | $28.0^{+1.7}_{-1.5}$ | $19.4\pm0.7$ | $15.9\pm0.5$ | -- | -- | -- |
| 3709 | $59.6^{+7.4}_{-5.9}$ | $32.3^{+2.1}_{-1.8}$ | $27.0^{+1.6}_{-1.5}$ | -- | -- | -- | -- |
| 3734 | $60.0^{+3.7}_{-3.3}$ | $34.4^{+1.7}_{-1.6}$ | $29.2\pm1.0$ | $19.7\pm0.5$ | -- | -- | -- |
| 3740 | -- | -- | -- | -- | -- | -- | -- |
| 3809 | $37.5^{+1.3}_{-1.2}$ | $24.6^{+1.8}_{-1.6}$ | $19.1\pm0.6$ | $17.2\pm0.4$ | $14.8\pm0.4$ | $13.2\pm0.4$ | $12.3\pm0.4$ |
| 3826 | $23.2\pm0.4$ | -- | -- | -- | -- | -- | -- |
| 3876 | $38.5^{+1.7}_{-1.6}$ | $24.9^{+1.2}_{-1.1}$ | -- | -- | -- | -- | -- |
| 3915 | $42.0^{+6.7}_{-5.1}$ | -- | -- | -- | -- | -- | -- |
| 4165 | $57.1^{+2.2}_{-2.1}$ | $44.0^{+1.8}_{-1.7}$ | $27.3^{+1.0}_{-0.9}$ | $21.1\pm0.4$ | $18.5\pm0.3$ | -- | -- |
| 4273 | $30.5^{+3.3}_{-2.6}$ | $22.4^{+1.3}_{-1.1}$ | $14.7\pm0.3$ | -- | -- | -- | -- |
| 4284 | $58.2^{+12.3}_{-7.3}$ | $38.4^{+3.8}_{-3.0}$ | $30.3\pm0.5$ | $26.7\pm0.8$ | $22.6\pm0.8$ | -- | -- |
| 4325 | $42.4^{+1.3}_{-1.2}$ | $24.1\pm1.0$ | $18.5\pm0.6$ | -- | -- | -- | -- |
| 4422 | $50.2^{+2.4}_{-2.2}$ | $35.1^{+2.4}_{-2.1}$ | $29.2^{+1.0}_{-0.9}$ | -- | -- | -- | -- |
| 4555 | $41.1^{+6.3}_{-4.9}$ | $22.3^{+2.0}_{-1.8}$ | $18.0^{+1.4}_{-1.2}$ | -- | -- | -- | -- |
| 4936 | $54.7^{+2.3}_{-2.1}$ | $38.8^{+1.2}_{-1.1}$ | $24.1^{+0.6}_{-0.5}$ | $21.3\pm0.4$ | $17.8\pm0.3$ | $16.3\pm0.2$ | -- |
| 5175 | $69.2^{+8.2}_{-7.7}$ | $48.5^{+9.2}_{-7.3}$ | $25.2^{+2.8}_{-2.4}$ | -- | -- | -- | -- |
| 5228 | $47.5^{+2.0}_{-1.9}$ | $32.3^{+1.1}_{-1.0}$ | $21.0^{+0.7}_{-0.6}$ | -- | -- | -- | -- |
| 5251 | $37.5^{+1.3}_{-1.2}$ | $21.3\pm0.7$ | $16.1\pm0.4$ | $13.4\pm0.3$ | $11.7\pm0.3$ | $10.8\pm0.2$ | -- |
| 5253 | $54.4^{+3.2}_{-2.9}$ | $40.2^{+2.8}_{-2.5}$ | $33.4^{+1.6}_{-1.5}$ | $28.1^{+1.3}_{-1.2}$ | -- | -- | -- |
| 5303 | $108.3^{+30.7}_{-19.3}$ | $56.9^{+10.4}_{-7.7}$ | $42.5^{+2.1}_{-1.9}$ | $29.6\pm1.0$ | -- | -- | -- |
| 5319 | $70.6^{+4.2}_{-3.8}$ | $51.3^{+2.5}_{-2.3}$ | $41.7^{+1.7}_{-1.6}$ | $36.6^{+1.4}_{-1.3}$ | $29.8^{+2.4}_{-2.1}$ | -- | -- |
| 5414 | -- | -- | -- | -- | -- | -- | -- |
| 5510 | $89.9^{+7.3}_{-6.2}$ | $50.9^{+2.0}_{-1.9}$ | $44.5^{+1.5}_{-1.4}$ | $36.8^{+1.6}_{-1.5}$ | $32.4\pm0.8$ | -- | -- |
| 5532 | $231.8^{+44.0}_{-32.3}$ | $115.9^{+26.6}_{-18.8}$ | $54.1^{+9.2}_{-7.1}$ | $32.6\pm1.2$ | $21.6\pm0.6$ | -- | -- |
| 5786 | $26.4^{+5.0}_{-2.9}$ | $20.0^{+2.2}_{-1.2}$ | $16.6\pm0.4$ | $13.4^{+1.0}_{-0.8}$ | -- | -- | -- |
| 5840 | $135.8^{+12.1}_{-10.4}$ | $84.6^{+6.7}_{-5.8}$ | $67.0\pm1.2$ | $57.9\pm0.9$ | -- | -- | -- |
| 5842 | $43.6\pm1.2$ | $39.0\pm0.8$ | $21.8^{+1.7}_{-1.5}$ | -- | -- | -- | -- |
| 5982 | $34.4^{+2.0}_{-1.8}$ | $27.7^{+1.2}_{-1.1}$ | $24.3^{+0.8}_{-0.7}$ | $20.7^{+0.9}_{-0.8}$ | -- | -- | -- |
| 6118 | $230.5^{+104.7}_{-59.9}$ | $101.7^{+14.7}_{-10.8}$ | $79.9^{+1.9}_{-3.5}$ | $67.1^{+3.5}_{-3.3}$ | $58.9\pm1.9$ | -- | -- |
| 6277 | $191.1^{+13.6}_{-12.0}$ | $102.4^{+4.7}_{-4.4}$ | $77.3^{+2.9}_{-2.7}$ | -- | -- | -- | -- |
| 6521 | $53.9^{+9.0}_{-7.0}$ | $30.4^{+2.4}_{-2.1}$ | $22.7^{+1.6}_{-1.4}$ | $14.6^{+1.7}_{-1.4}$ | -- | -- | -- |
| 6523 | $52.6^{+24.5}_{-16.3}$ | $23.0^{+6.3}_{-4.4}$ | -- | -- | -- | -- | -- |
| 6537 | 44.0±1.0 | 35.4±0.8 | 23.7±0.6 | 20.9±0.5 | -- | -- | -- |
| 6702 | $75.2^{+35.2}_{-17.9}$ | $36.4^{+6.3}_{-5.1}$ | $19.0^{+1.8}_{-1.5}$ | -- | -- | -- | -- |
| 6778 | $126.1^{+66.4}_{-29.4}$ | $63.4^{+6.2}_{-5.1}$ | $41.0^{+1.5}_{-1.4}$ | $35.8^{+0.8}_{-0.7}$ | $31.5^{+1.3}_{-1.2}$ | -- | -- |
| 7021 | $126.9^{+7.6}_{-8.5}$ | $78.9^{+7.5}_{-6.8}$ | $34.4^{+2.2}_{-2.0}$ | -- | -- | -- | -- |
| 7045 | 54.1±1.0 | 46.4±1.6 | 32.3±0.4 | -- | -- | -- | -- |
| 7154 | $41.5^{+2.9}_{-2.4}$ | 24.1±0.6 | 19.8±0.5 | 12.0±0.1 | 9.5±0.2 | -- | -- |
| 7323 | $51.7^{+5.3}_{-4.1}$ | 28.5±1.0 | 18.6±0.9 | 16.0±0.2 | 15.0±0.2 | -- | -- |
| 7420 | $135.0^{+10.0}_{-8.3}$ | $60.5^{+1.5}_{-1.4}$ | $49.6^{+3.2}_{-2.9}$ | $39.0^{+0.7}_{-0.6}$ | 33.6±1.0 | 30.4±0.4 | -- |
| 7766 | $68.8^{+2.5}_{-2.3}$ | $39.4^{+1.5}_{-1.4}$ | $30.8^{+0.8}_{-0.7}$ | $27.3^{+1.6}_{-1.5}$ | 19.4±0.2 | -- | -- |
| 7831 | 89.9±0.2 | 85.0±0.6 | 68.1±2.1 | 58.9±0.8 | 52.3±0.6 | -- | -- |
| 7853 | -- | -- | -- | -- | -- | -- | -- |
| 7861 | $36.8^{+1.2}_{-1.1}$ | $28.4^{+1.0}_{-0.9}$ | $26.5^{+0.5}_{-0.4}$ | -- | -- | -- | -- |
| 7876 | $43.3^{+1.4}_{-1.3}$ | $36.4^{+1.3}_{-1.2}$ | 28.6±0.9 | -- | -- | -- | -- |
| 7901 | $44.2^{+5.9}_{-4.7}$ | 37.0±1.1 | 26.2±1.0 | -- | -- | -- | -- |
| 7985 | $75.6^{+7.1}_{-6.0}$ | $48.4^{+4.4}_{-2.6}$ | $38.3^{+4.4}_{-3.7}$ | $28.9^{+1.7}_{-1.5}$ | $22.4^{+2.6}_{-2.1}$ | -- | -- |
| 8334 | $218.5^{+11.0}_{-9.9}$ | $124.0^{+3.3}_{-3.1}$ | 84.4±1.4 | 58.8±0.7 | 21.4±0.1 | -- | -- |
| 8403 | -- | -- | -- | -- | -- | -- | -- |
| 8490 | $99.5^{+1.2}_{-1.4}$ | $93.0^{+1.6}_{-1.7}$ | 82.7±2.8 | $67.5^{+2.4}_{-2.3}$ | -- | -- | -- |
| 8709 | 36.9±0.5 | 33.2±0.6 | 24.1±0.5 | 16.7±0.4 | 12.8±0.8 | -- | -- |
| 8852 | $79.9^{+9.8}_{-8.7}$ | $44.5^{+1.8}_{-1.7}$ | $37.4^{+1.4}_{-1.3}$ | -- | -- | -- | -- |
| 8937 | $79.5^{+17.1}_{-12.4}$ | $48.6^{+4.0}_{-4.0}$ | $35.9^{+4.0}_{-3.4}$ | $23.8^{+1.7}_{-1.5}$ | -- | -- | -- |
| 9179 | $82.1^{+2.4}_{-2.2}$ | 65.1±1.3 | 55.8±1.7 | 44.7±1.1 | 37.7±0.3 | 30.3±0.2 | -- |
| 9248 | $46.5^{+6.1}_{-5.2}$ | $33.8^{+3.9}_{-3.4}$ | $26.6^{+3.2}_{-2.7}$ | $19.0^{+1.7}_{-1.5}$ | -- | -- | -- |
| 9358 | $92.4^{+10.3}_{-8.4}$ | $50.0^{+3.9}_{-3.4}$ | $39.4^{+3.2}_{-2.8}$ | -- | -- | -- | -- |
| 9363 | 27.7±0.6 | 22.7±0.5 | 18.1±0.9 | -- | -- | -- | -- |
| 9366 | $90.9^{+23.3}_{-15.3}$ | $55.9^{+3.7}_{-3.3}$ | $34.1^{+5.6}_{-4.3}$ | $26.9^{+1.6}_{-1.3}$ | -- | -- | -- |
| 9465 | 26.7±1.2 | 22.9±0.8 | 14.7±0.5 | -- | -- | -- | -- |
| 9736 | $25.3^{+2.2}_{-1.8}$ | $14.5^{+0.7}_{-0.6}$ | -- | -- | -- | -- | -- |
| 9753 | $152.9^{+26.5}_{-22.2}$ | $74.6^{+5.0}_{-4.4}$ | -- | -- | -- | -- | -- |
| 9866 | $84.9^{+7.8}_{-6.4}$ | -- | -- | -- | -- | -- | -- |
| 9943 | $64.5^{+7.9}_{-6.5}$ | $39.5^{+4.9}_{-4.0}$ | $26.4^{+1.1}_{-1.0}$ | 23.0±0.7 | 19.5±0.5 | -- | -- |
| 9969 | $50.2^{+2.5}_{-2.4}$ | $40.4^{+1.6}_{-1.5}$ | $30.6^{+2.9}_{-2.5}$ | 26.4±0.6 | $21.5^{+2.2}_{-1.9}$ | 16.3±0.8 | -- |
| 10075 | $87.3^{+4.8}_{-4.2}$ | $48.0^{+1.7}_{-1.6}$ | 36.8±0.7 | $24.6^{+1.2}_{-1.1}$ | 18.9±0.2 | -- | -- |
| 10359 | 36.7±0.6 | $30.4^{+1.3}_{-1.2}$ | 23.3±0.5 | -- | -- | -- | -- |
| 10445 | 32.7±1.6 | 14.8±1.0 | -- | -- | -- | -- | -- |
| 10470 | $76.3^{+4.3}_{-3.9}$ | $37.8^{+1.7}_{-1.6}$ | 31.8±0.9 | $24.9^{+1.4}_{-1.3}$ | 22.0±0.5 | -- | -- |
| 10502 | $14.8^{+1.1}_{-1.0}$ | $11.0^{+0.4}_{-0.3}$ | $7.9^{+0.6}_{-0.5}$ | -- | -- | -- | -- |
| 10521 | $66.4^{+3.6}_{-3.2}$ | $46.2^{+1.7}_{-1.6}$ | 34.0±0.9 | 27.7±0.8 | 25.2±0.7 | 22.3±1.0 | -- |
| 10546 | $36.3^{+1.5}_{-1.4}$ | $27.4^{+2.3}_{-2.1}$ | $22.4^{+0.8}_{-0.7}$ | 18.8±0.6 | -- | -- | -- |
| 10564 | -- | -- | -- | -- | -- | -- | -- |
| 10652 | $93.1^{+2.6}_{-2.9}$ | $68.0^{+3.4}_{-3.3}$ | -- | -- | -- | -- | -- |
| 10757 | $45.4^{+2.4}_{-2.1}$ | 33.8±0.8 | -- | -- | -- | -- | -- |
| 10897 | -- | -- | -- | -- | -- | -- | -- |
| 11012 | 107.7±2.0 | 90.7±1.2 | 71.8±1.3 | -- | -- | -- | -- |
| 11124 | 20.3±0.7 | 17.4±0.2 | 14.4±0.3 | 13.2±0.1 | -- | -- | -- |
| 11218 | $98.8^{+9.5}_{-7.9}$ | $60.6^{+3.3}_{-3.0}$ | $46.9^{+5.1}_{-4.2}$ | $33.2^{+2.6}_{-2.2}$ | $26.8^{+1.3}_{-1.2}$ | 21.8±0.5 | -- |
| 11283 | $81.9^{+0.7}_{-1.2}$ | $67.0^{+3.6}_{-3.7}$ | $40.5^{+2.8}_{-2.5}$ | $30.7^{+2.3}_{-2.1}$ | -- | -- | -- |
| 11407 | $43.8^{+9.9}_{-5.4}$ | $30.7^{+3.0}_{-2.3}$ | $22.6^{+0.8}_{-0.7}$ | $18.6\pm0.5$ | $12.9\pm0.3$ | -- | -- |
| 11466 | $58.1^{+9.2}_{-6.5}$ | $44.3^{+2.5}_{-2.2}$ | $33.6\pm0.4$ | -- | -- | -- | -- |
| 11557 | -- | -- | -- | -- | -- | -- | -- |
| 11861 | $36.7^{+1.9}_{-1.8}$ | $30.1^{+1.1}_{-1.0}$ | $24.8^{+1.1}_{-1.0}$ | $18.2^{+0.7}_{-0.6}$ | -- | -- | -- |
| 11872 | $208.4^{+48.3}_{-32.3}$ | $94.5^{+11.9}_{-9.6}$ | $65.2^{+4.9}_{-4.3}$ | -- | -- | -- | -- |
| 11914 | $222.8^{+94.8}_{-55.4}$ | $123.7^{+17.9}_{-14.2}$ | $79.3^{+3.6}_{-3.3}$ | $57.2^{+1.9}_{-1.8}$ | $48.7^{+2.9}_{-2.7}$ | -- | -- |
| 12276 | $16.6\pm1.0$ | $11.9\pm0.6$ | $9.7\pm0.4$ | -- | -- | -- | -- |
| 12343 | $25.6\pm1.2$ | $22.3\pm0.9$ | $18.4\pm1.0$ | -- | -- | -- | -- |
| 12754 | $59.7^{+1.5}_{-1.4}$ | $46.1^{+0.8}_{-0.7}$ | $36.6^{+2.4}_{-2.1}$ | $32.8\pm0.3$ | $30.3^{+0.8}_{-0.7}$ | $28.4\pm0.3$ | -- |

**Notes.** Column (1) gives the name of the galaxy. Pattern speed values corresponding to the corotation radii of Table 4, together with the associated upper and lower uncertainties are given in columns (2) to (8); for each galaxy $\Omega_i$ is the pattern speed associated with the corotation CR:i appearing in Table 4. No values are shown for those galaxies showing a rigid body-like rotation curve.

V. NOTES ON INDIVIDUAL GALAXIES

We use the following terminology throughout the descriptions of the resonances found for the individual galaxies.

We use the conventional terms ILR and OLR for the Inner and outer Lindblad resonances respectively, and where there are two Inner Lindblad resonances we call the inner and outer of these, respectively, the iILR and the oILR. We use I4:1 and O4:1 for the inner 4:1 and the outer 4:1 ultraharmonic resonances respectively. When a resonance corresponding to a given corotation lies within the error box of another corotation we say that the resonance is "compatible with" the corotation peak, when it is within one third of the error box from the corotation we say that it is "close to" the corotation peak, and when it coincides with the corotation we say that it lies "on" the peak. The corotation peaks are numbered successively starting from the innermost peak, peak 1. The strongest peak is the one with the largest number of observed velocity zero-crossings (largest area of the Gaussian fit). When we have assigned a given peak X as a corotation, and have found by inspecting the frequency curves that none of its Lindblad resonances or ultraharmonic resonances coincide with any of the other peaks, we describe peak X as showing "no coincidence" or "isolated". This description is maintained, for clarity, even if peak X coincides with Lindblad or ultraharmonic resonances of other peaks. One of the most striking findings of our work is a pattern of interlocking resonances which is found in the majority of the objects studied, and in some galaxies is found more than once. This pattern is defined as follows. Given two pattern speeds $\Omega_1$ and $\Omega_2$ corresponding to two corotations, CR1 and CR2, the OLR of $\Omega_1$ falls on, or close to, CR2, while the I4:1 resonance of $\Omega_2$ falls on, or close to, CR1. Where this occurs, we have included the descriptor (xP) in the header of the text referring to the galaxy, where x is the number of times the pattern is repeated. The quantitative details for all galaxies, essentially the resonance radii and the pattern speeds, are given in Tables 4 and 5. In Table 4 each radius is labeled with the type of resonance with which has been identified, along with the pattern speed which the resonance belongs to (so, if a radius is labeled as CR:3 this means that it is the corotation radius of the pattern speed $\Omega_3$). We should point out that our initial assignments of the resonance peaks were as corotations and that the majority of these assignments were verified internally via their couplings patterns with other resonances. In the descriptions of individual objects, if a resonance is described otherwise, one can assume that we have assigned it corotation status on the basis of the evidence described.

1. UGC 508. NGC 266.                  (2P)

This has a strong bar with arms breaking perpendicularly from the ends. There are five clear resonance peaks. The resolved Peak 1 lies close to the galaxy center. Inspection of the bar suggests that peak 3, which is the strongest, should be at its corotation radius. The computed resonance radii for peak 1 show no coincidences with the other peaks. Peak 2 has its O4:1 resonance very close to peak 3, and its OLR very close to peak 4. Peak 3 has its O4:1 resonance close to peak 4, and its OLR is compatible with peak 5. Peak 4 has its I4:1 resonance just compatible with peak 2. Peak 5 has its I4:1 resonance on peak 2, and its ILR close to peak 1. We note that for this galaxy the OLR´s corresponding to two of its finally assigned corotation peaks (peaks 2 and 3) lie on or close to two peaks further out (peaks 4 and 5, respectively), while the I4:1 resonances of these latter lie on or close to the respective inner peaks. This pattern, where there is a reciprocal relation between two peaks, with the OLR of one coinciding with the other at larger radius, and the I4:1 of the second coinciding with the first, is found in almost two thirds of the galaxies studied. If we assume peak 1 is a corotation its OLR and its O4:1 resonance do not coincide with any of the other peaks. However it does lie at the ILR of the resonance whose corotation is at peak 5, so we infer that it does not correspond to a corotation but simply to the ILR of peak 5, as shown in Table 4, where this resonance is labeled as ILR:4, indicating that it corresponds to an ILR of the pattern speed $\Omega_4$ (in Table 5) which has its corotation labeled as CR:4 (in Table 4).

2. UGC 763. NGC 428.              (1P)

This is an asymmetric and not cleanly designed galaxy. It shows four resonance peaks, of which peak 4 is significantly the strongest. Peak 1 taken as a corotation has its OLR on peak 3. Peaks 2 and show no coincidences with any other peak. Peak 3 taken as a corotation has its I4:1 resonance lying close to peak 1. Buta & Zhang (2009) reported a corotation radius of 54.8 arcsec, which is very close to the position of peak 2 at $r_2 = 51.9 \pm 1.9$ arcsec.

3. UGC 1256. NGC 672.

This is virtually edge-on and shows an almost rigid body rotation curve in its inner portion. The number of resonance peaks is 5, but the difficulty of assigning realistic values to any corotation radius is clear, since peaks 1 to 4 show O4:1 resonances compatible with peak 5, but peak 5 shows no coincidences with any of the inner peaks.

4. UGC 1317. NGC 674.              (2P)

This is flocculent and highly inclined. It shows four resonance peaks, of which peak 2 is strongest, but peak 4 is also strong. Peak 1 has its OLR on peak 2, which in its turn has the I4:1 close to peak 1, and its OLR is close to peak 3. Peak 3 has its I4:1 on peak 2 and shows its O4:1 resonance on peak 4. The outer peak shows no compatibility with any other peak.

5. UGC 1437. NGC 753.              (1P)

This is a small spiral with no obvious bar. It shows four peaks of which the second is the strongest. Peak 1 has no coincidences. Peak 2 has its OLR compatible with peak 3. Peak 3 has its ILR compatible with peak 1, its I4:1 resonance close to peak 2, and its O4:1 resonance on peak 4. Peak 4 has its I4:1 resonance very close to peak 3. It is reasonable to infer that peak 1 in this case does not represent a corotation, but is the kinematic response to the ILR of peak 3.

6. UGC 1736. NGC 864.

This is a clearly barred galaxy, and not very inclined. The best fit to the inner part (r ≤ 40 arcsec) of the rotation curve has a linear "rigid body" appearance, although points close to the center may imply a rotation offset. It shows four peaks, of which peak 2 is the strongest, and an obvious candidate for the corotation associated with the bar. Peak 1 has its O4:1 resonance just compatible with peak 3. Peak 2, but neither peak 3 nor peak 4, shows resonant coincidences. Multiple corotation radii at 6.5, 33.0 and 57.3 arcsec, are found by Buta & Zhang (2009) applying the "potential-density" phase-shift method.

These results are in very good agreement with our peaks, 1, 2 and 3 respectively, where $r_1 = 5.8\pm2.2$ arcsec, $r_2 = 29.7\pm3.6$ arcsec and $r_3 = 56.1\pm2.1$ arcsec.

7. UGC 1913. NGC 925.

This galaxy is highly inclined, with a very asymmetric bar-like structure. Its rotation curve is almost completely linear, i.e. it shows rigid body-like behavior, though this is almost certainly an effect of the inclination foreshortening. It shows 4 weak peaks and no resonance coincidences.

8. UGC 2080

This galaxy shows clean structure but it is nearly face on with the implied difficulties of making good in-plane velocity determinations. It has 2 weak peaks, of which peak 2 is slightly stronger than peak 1. The only coupling between both peaks is that the position of peak 2 is compatible with OLR of peak 1.

9. UGC 2141. NGC 1012.

This highly inclined galaxy shows 5 peaks, of which peak 2 is the strongest. However its rotation curve is another example of a pseudo-rigid body curve, caused by the different dust penetration optical depths along the lines of sight in the disk. No resonance coincidences are found under these circumstances

10. UGC 2193. NGC1058.            (1P)

This galaxy is nearly face on. It has three weak peaks of which peak 1 is the strongest. The rotation curve is not particularly well determined. Peak 1 and peak 2 are coupled as the OLR of the innermost peak is on peak 2, which in its turn has its I4:1 resonance close to peak 1. Peak 3 shows no resonance coincidences. Peak 1 located a radial distance of 26.6±4.6 arcsec coincides, within our error limits, with the value found by Buta & Zhang (2009) as the corotation radius of this galaxy ($r_{CR} = 30.5$ arcsec).

11. UGC 2855                       (1P)

This is clearly a barred galaxy. It shows 4 resonance peaks, of which the third and forth peaks are the strongest ones. Peak 2 has its O4:1 resonance on peak 3, and peak 4 is compatible with its OLR. Peak 3 has its O4:1 resonance on peak 4. Peak 4 has its I4:1 resonance close to peak 2.

12. UGC 3013. NGC 1530.            (2P)

This is a very strongly barred galaxy. It shows 7 peaks. Peaks 1 and 6 show no coincidences with any other peak. Peak 2 has its O4:1 resonance close to peak 3, its OLR close to peak 4. Peak 3 has its O4:1 resonance close to peak 4, its OLR on peak 5. Peak 4 has its ILR compatible with peak 1, its I4:1 and O4:1 resonances very close to peaks 2 and 6 respectively, and its OLR just compatible with peak 7. The ILR of peak 5 is also compatible with peak 1, peak 5 has its I4:1 resonance on peak 3, and its O4:1 resonance on peak 7. And, peak 7 has its I4:1 resonance on peak 5. We can infer from the resonant structure found for this galaxy that peaks 1 and 7 are not corotations but ILR and OLR respectively, associated with peak 4, which can be taken as the corotation of the bar. Using H$\alpha$ Fabry-Perot data from NGC 1530, Regan et al. (1996) derived the positions of the ILR, CR and OLR of the bar at 2.1 kpc, 12.6 kpc and 19±4 kpc, respectively (values corrected for distance), which are in agreement (within uncertainties) with our peaks 1, 4 and 7 located at $r_1 = r_{ILR} = 1.6\pm0.5$ kpc, $r_4 = r_{CR} = 12.2\pm0.5$ kpc and $r_7 = r_{OLR} = 18.0\pm0.5$ kpc. The pattern speed of the bar is also in good agreement; while these authors reported a value between 18.2 and 20.2 ± 2.0 km·s$^{-1}$·kpc$^{-1}$ (corrected for distance and inclination), we find a pattern speed associated with peak 4 of $\Omega_4 = 17.0\pm0.7$ km·s$^{-1}$·kpc$^{-1}$. Other authors (Downes et al. 1996, Reynaud & Downes 1998) obtain different values of the ILR and CR radius when they analyze CO maps of this galaxy. They find the ILR at 1.5 kpc and the CR at 14 kpc, which could be well reproduced if peak 5 is assumed as the bar corotation radius (peak 1 is compatible as the ILR of this peak), so $r_5 = r_{CR}$ = 13.7±0.6 kpc and $r_1 = r_{ILR}$ = 1.6±0.5 kpc. In addition, the corotation radius at 77 arcsec = 13.8 kpc derived by Vera-Villamizar et al. (2003) matches perfectly with the position of peak 5. The corresponding pattern speed for peak 5 is $\Omega_5$ = 15.0±0.7 km·s$^{-1}$·kpc$^{-1}$., which is in good agreement with $\Omega_P$ = 14.3 km·s$^{-1}$·kpc$^{-1}$ derived by Reynaud & Downes (1998). The beamwidth of the CO observations was 22 arcsec (compared with the angular resolution of 4 arcsec reported by Regan et al. 1996, and our resolution of 1.2 arcsec), which can explain the differences between the results.

13. UGC 3273

This is a virtually edge-on galaxy, with an almost rigid body rotation curve. It has four peaks of which peak 1 is the strongest, but no coincidences, as the data do not give true in-plane velocities

14. UGC 3463                    (1P)

This galaxy shows very little structure. It has three resonance peaks. Peak 1 has its OLR compatible with peak 2, while peak 2 has its I4:1 close to peak 1 and its O4:1 resonance very close to the position of peak 3. Peak 3 shows no resonances or coincidences with the others.

15. UGC 3574                    (2P)

This is a face on galaxy for which we find 3 peaks. Peak 1 has its OLR compatible with peak 2. Peak 2 has its I4:1 resonance close to peak 1, and its OLR close to peak 3. Peak 3 has its I4:1 resonance close to peak 2.

16. UGC 3685

This is a neat, barred object. There are 4 peaks of which peak 4 is the strongest. The rotation curve is almost linear, i.e. rigid body, so there are no resonance coincidences between any peaks. Peak 2 is probably at the bar corotation radius

17. UGC 3691                    (1P)

This face on galaxy has a clear bar and well defined arms. There are 4 peaks, all quite weak, with peak 3 somewhat stronger than the rest. Peak 1 has its O4:1 resonance compatible with peak2 . Peak 2 has its I4:1 resonance just compatible with peak 1, its O4:1 resonance quite close to peak 3, and its OLR on peak 4. Peak 3 shows no coincidences. Peak 4 has its I4:1 resonance on peak 2.

18. UGC 3709. NGC 2342.         (1P)

This is angularly small, with a quite strong spiral structure, but not especially regular, and has a possible small bar. It shows three peaks, of which peak 2 is the strongest. Peak 1 has its OLR on peak 2. Peak 2 has its I4:1 resonance close to peak 1, and its O4:1 resonance just compatible with peak 3. Peak 3 shows no resonance coincidences.

19. UGC 3734. NGC 2344.         (2P)

Although there is not much structure in the image, this galaxy shows 4 corotation peaks. All are quite weak, and peak 2 is just the strongest. Peak 1 has its OLR on peak 2. Peak 2 has its I4:1 resonance quite near peak 1, and its OLR on peak 4. Peak 3 has its O4:1 resonance compatible with peak 4. Peak 4 has its I4:1 resonance compatible with peak 2 (although also quite near peak 3)

20. UGC 3740. NGC 2276. (Arp 25)

This object has a 'rotation curve' which raises linearly, a rigid rotator curve. There are five peaks on the outer galaxy, but the rotation curve implies that there will be no coincidences for any peak.

21. UGC 3809. NGC 2336.                  (2P)

This angularly large galaxy is barred. It has seven resonance peaks. Peak 2 is slightly stronger than the rest. Peak 1 has its O4:1 resonance compatible with peak 2. Peak 2 has its I4:1 resonance compatible with peak 1, its O4:1 resonance between peaks 3 and 4 (compatible with either), and its OLR on peak 5. Peak 3 has its O4:1 resonance compatible with peak 5, and its OLR just compatible with peak 7. Peak 4 has its I4:1 resonance compatible with peak 2 and its O4:1 resonance compatible with peak 6. Peak 5 has its I4:1 resonance on peak 2. Peak 6 has its I4:1 resonance almost on peak 3. Peak 7 has its I4:1 resonance on peak 3.

22. UGC 3826

This galaxy is almost face on, which makes the in-plane velocities difficult to measure with precision. The rotation curve of the inner galaxy is that of a rigid rotator. There are two weak peaks, of which peak 1 has its OLR close to peak 2, so we assume that this galaxy shows one single pattern speed located at the position of peak 1, and that peak 2 is simply at the OLR of this corotation.

23. UGC 3876                              (1P)

This galaxy shows two peaks. Peak 2 lies close to the OLR of peak 1. The I4:1 resonance of peak 2 is just compatible with peak 1.

24. UGC 3915                              (1P)

This angularly small galaxy shows three peaks, of which peak 2 is the strongest. Peak 1 has no compatibility with any other peak. Peak 2 has its oILR resonance very close to peak 1, and its OLR close to peak 3. Peak 3 has its I4:1 resonance close to peak 2 and also its oILR is just compatible with peak 1.

25. UGC 4165. NGC 2500.                   (1P)

This barred galaxy shows five corotation peaks, of which peak 4 is the strongest. Peak 1 does not show any coincidence with any other peak. Peak 2 has its OLR just compatible with peak 3. Peak 3 has its I4:1 resonance compatible with peak 2, and its O4:1 resonance compatible with peak 4. There are no coincidences for peaks 4 and 5.

26. UGC 4273. NGC 2543.

This has a neat two-armed structure, with almost certainly a nuclear bar as well as the large bar. Peaks 2 and 4 are the strongest. Peak 1 and peak 3 show no coincidences. Peak 2 has its O4:1 resonance close to peak 3. Peak 4 has its ILR close to peak 1 and its I4:1 almost compatible with peak 3. Using a method based on a Fourier analysis of azimuthal profiles, Vera-Villamizar et al. (2003) identify two corotation resonances at 27 and 46 arcsec. The first is close to our peak 2 ($r_2$ = 31.8±4.1 arcsec) and the other falls in the position of peak 3 within the uncertainty bars ($r_3$ = 47.4±3.2 arcsec).

27. UGC4284. NGC 2541.                    (1P)

The image of this quite flocculent galaxy does not show very clear structure. There are, however, six resonance peaks, all weak, with peak 3 a little stronger. Peak 1 has its O4:1 resonance just compatible with peak 2, and its OLR compatible with peak 4. Peaks 2, 3 and 6 show no coincidences. Peak 4 has its I4:1 resonance on peak 1 . Peak 5 has its I4:1 resonance just compatible with peak 1.

28. UGC 4325. NGC 2552.                   (1P)

This galaxy is quite inclined, and shows three peaks, of which peak 3 is the strongest. Peak 1 has its OLR on peak 2. Peak 2 has its I4:1 resonance on peak 1. Peak 3 shows no coincidences.

29. UGC 4422. NGC 2595.             (1P)

This is a cleanly presented barred galaxy, with a cleanly displayed smaller bar across the nucleus. There are four peaks of which peak 2 is the strongest. Peaks 1 and 3 show no resonance coincidences. Peak 2 has its OLR on peak 4, and its O4:1 resonance close to peak 3. Peak 4 has its ILR on peak 1, and its I4:1 resonance compatible with peak 2. As peak 1 coincides with the position of the ILR associated with corotation at peak 4, and none of the frequency curves are compatible with any of the other peaks when corotation is assumed at peak 1, we may conclude that peak 1 is a pure ILR.

30. UGC 4555. NGC 2649.             (1P)

This small galaxy has a faint inner ring just inside the first of three peaks, of which peak 3 is the strongest. Peak 1 has its OLR close to peak 2. Peak 2 has its I4:1 resonance close to peak 1, and its O4:1 resonance close to peak 3. Peak 3 shows no further coincidences.

31. UGC 4936. NGC 2805.             (1P)

The inner disk is regular but this galaxy has a ragged disturbed appearance in the outer disk. It shows seven peaks, of which the third peak is the strongest. Peak 1 has its O4:1 resonance very close to peak 2. Peak 2 has its OLR close to peak 4. Peak 3 has its I4:1 resonance just compatible with peak 2, and its O4:1 resonance close to peak 5. Peak 4 has its I4:1 resonance compatible with peak 2. Peak 5 has its O4:1 resonance just compatible with peak 7. Peak 6 shows no coincident resonances with any other peak and peak 7 has its I4:1 resonance just compatible with peak 4.

32. UGC 5175. NGC 2977.             (1P)

This is small quite flocculent object with three resonance peaks. Peak 1 has its O4:1 resonance on peak 2. Peak 2 has its OLR quite close to peak 3. Peak 3 has its I4:1 resonance quite close to peak 2.

33. UGC 5228

This galaxy is highly inclined, and shows 4 peaks. Peak 1 has its O4:1 resonance just compatible with peak 2. Peak 2 has its O4:1 resonance just compatible with peak 3, and its OLR just compatible with peak 4. Peak 3 has its O4:1 resonance on peak 4. Peak 4 shows no other resonance coincidences, so it can be assumed to be a pure OLR. The galaxy's high inclination impedes an adequate survey of its internal kinematics.

34. UGC 5251. NGC 3003.             (2P)

This is a regular spiral, but quite inclined galaxy, with a clear small nuclear bar, and a probable outer bar. Peaks 5 and 3 are the strongest. Peak 1has its OLR quite near peak 2. Peak 2 has its I4:1 resonance compatible with peak 1, its O4:1 resonance compatible with peak 3, and its OLR is compatible with peak 5. Peak 3 has its O4:1 resonance on peak 4. Peak 4 shows no resonance coincidences. Peak 5 has its I4:1 resonance compatible with peak 2. Peak 6 shows no coincidences.

35. UGC 5253. NGC 2985.             (2P)

This galaxy is flocculent and with quite a low inclination. It shows four peaks of which peak 3 is the strongest. Peak 1 has its O4:1 resonance on peak 2, and its OLR on peak 3. Peak 2 has its I4:1 resonance compatible with peak 1, its O4:1 resonance compatible with peak 3, and its OLR compatible with peak 4.Peak 3 has its I4:1 resonance compatible with peak 1, and its O4:1 resonance compatible with peak 4. Peak 4 has its I4:1 resonance on peak 2.

36. UGC 5303. NGC 3041.             (2P)

This is a low inclination galaxy showing a multi-arm structure. Its radial distribution of phase-reversals has four peaks. Peak 1 has its OLR close to peak 2, which has its I4:1 resonance close to peak 1, its

O4:1 resonance is close to peak 3 and its OLR is compatible with peak 4. Peak 3 has its I4:1 resonance compatible with peak 2 and its O4:1 resonance on peak 4. Last peak has its ILR compatible with peak 1 and its I4:1 resonance close to peak 2.

37. UGC 5319. NGC 3061. (2P)

This is a galaxy with strong spiral structure, at low inclination, with a bar. Peak 2 is the main one. Peak 1 has its O4:1 resonance quite close to peak 2, and its OLR is compatible with both peaks 3 and 4. Peak 2 has itsO4.1 resonance on peak 4, and its OLR on peak 5. Peak 3 has its I4:1 resonance on peak 1, and its O4:1 resonance on peak 5. Peak 4 has its I4:1 resonance compatible with peak 1 and its O4:1 resonance on peak 5. Peak 5 has its I4:1 resonance on peak 2.

38. UGC 5414. NGC 3104.

This galaxy is clearly edge-on. It shows six peaks of which peak 4 is the strongest. Peak 1 has its O4:1 resonance on peak 5. Peak 2 has its O4:1 resonance just compatible with peak 6. Peaks 3, 4, 5, and 6 show no coincidences. This is as we have come to expect from these edge-on systems where the rotation curve has a rigid body form, and is clearly strongly affected by dust extinction.

39.UGC 5510. NGC 3162. (1P)

This is angularly a small object with a short bar. There are six peaks, or which peaks 3 and 4 are the strongest. Peak 1 has its OLR on peak 2. Peak 2 has its I4:1 resonance close to peak 1, its O4:1 resonance on peak 4, and its OLR just compatible with peak 6. Peak 3 has its O4:1 resonance on peak 5. Peak 4 has its O4:1 resonance compatible with peak 6. Peak 5 has its I4:1 resonance quite close to peak 2. Peak 6 has its I4:1 resonance on peak 3.

40. UGC 5532. NGC 3147. (4P)

This object is angularly small, with a bar, and shows 5 peaks, of which peaks 2 and 3 are the strongest. Peak 1 has its OLR close to peak 2. Peak 2 has its I4:1 resonance close to peak 1, its OLR resonance quite compatible with peak 3. Peak 3 has its oILR compatible with peak 1, its I4:1 resonance just compatible with peak 2, and its OLR resonance on peak 4. Peak 4 has its oILR on peak 2, I4:1 resonance close to peak 3 and its OLR just compatible with peak 5. The last peak has its I4:1 resonance close to peak 4. This galaxy shows a remarkable set of 4 coupling patterns. Peak 2 with radius $r_2$ = 16.3±3.3 arcsec and its associated pattern speed, $\Omega_2 = 115.9^{+26.6}_{-18.8}$ km·s$^{-1}$·kpc$^{-1}$, are in excellent agreement with the values derived by Cassola et al. (2008); $r_{CR}$ = 15.2±3.0 arcsec and $\Omega_P$ = 115.6±23.1 km·s$^{-1}$·kpc$^{-1}$.

41. UGC 5786. NGC 3310. (1P)

This is probably a post merger object (Arp 217) with starburst galaxy. It shows five peaks. The outermost peak is the strongest one. There is a strong mass concentration at the center, which may produce the inner peak 1, which has its OLR on peak 3 (and may well be due to an ILR rather than a corotation). Peak 2 has its ILR on peak 1, and its OLR close to peak 5. Peak 3 has its O4:1 resonance compatible with peak 5. Peak 4 shows no coincidences with any other peak. Peak 5 has its I4:1 resonance on peak 2.

42. UGC 5840. NGC 3344. (1P)

This is a large and regular spiral galaxy, with a small central bar. There are five peaks, although its face-on presentation implies that the precision of the results must be reduced. Peak 1 has its OLR close to peak 2, which has its ultraharmonic resonances, I4:1 and O4:1, close to peaks 1 and 3, respectively, and its OLR is just compatible with peak 5. Peak 3 turns out to be an isolated corotation. Peak 4 has its O4:1 resonance compatible with peak 5. Neither of the two frequency curves with corotation at peak 5 is compatible with any other peak, so this peak, 5, is considered to be simply the OLR of peak 2.

43. UGC 5842. NGC 3346.            (1P)

This is a regular spiral galaxy with a shortish bar. It has three peaks. Peak 1 is close to peak 2, but significantly stronger, while peak 3 is broad and strong. Peak 1 has its OLR compatible with peak 3. Peak 2 has its OLR on peak 3. Peak 3 has its I4:1 resonance on peak 2.

44. UGC 5982. NGC 3430.            (1P)

This galaxy has a possible nuclear bar. It shows five peaks, or which peak 5 is the strongest. Peak 1 shows no resonance coincidences with any other peak, and it could well simply be the ILR corresponding to peak 5 rather than a corotation. Peak 2 has its O4:1 resonance compatible with peak 4, and its OLR compatible with peak 5. Peak 2 has its O4:1 resonance close to peak 5. Peak 3 has its O4:1 resonance close to peak 5. Peak 4 shows no resonance coincidences with other peaks. Peak 5 has its ILR compatible with peak 1, and its 4:1 resonance on peak 2.

45. UGC 6118. NGC 3504.            (2P)

This is a barred and ringed galaxy, with a strong large bar, and a possible nuclear bar. There are 6 peaks, of which peaks 1 and 2 is the strongest. Peak 1 has its OLR compatible with peak 2. Peak 2 has its I4:1 resonance just compatible with peak 1, its O4:1 resonance close to peak 3, and its OLR on peak 5. Peak 3 has its O4:1 resonance close to peak 5. Peak 4 has its OLR very close to peak 6 and its ILR just compatible peak 1. Peak 5 has its ILR close to peak 1 and its I4:1 resonance compatible with peak 3. Finally, the isolated peak 6 doesn't show any coincidence with any other peak, so it may be considered as OLR of peak 4 rather than corotation. Peak 1 plays a dual role; it is a corotation as it is linked with peak 2 in their coupling pattern, and it is also an ILR of peak 4.

46. UGC 6277. NGC 3596.            (1P)

This is galaxy with clear spiral structure, but its presentation is nearly face on with consequent reduction in the precision of the measurements, indeed of the rotation curve itself. It has four peaks of which peak 2 is the strongest. Peak 1 has its OLR resonance compatible with peak 2. Peak 2 has its I4:1 resonance compatible with peak 1, its O4:1 resonance compatible with peak 3, and its OLR just compatible with peak 4. Peak 3 has its O4:1 resonance close to peak 4. Peak 4 remains an isolated resonance and can be considered an OLR rather than a corotation, which, however, is different from the result of Buta & Zhang (2009). These authors identify four different corotation radii at 12.3, 32.2, 49.8 and 71.1 arcsec. The innermost corotation is quite close to our peak 1 ($r_1$ = 8.3±1.3 arcsec), the second corotation is in very good agreement with our peak 3 ($r_3$ = 33.2±1.4 arcsec), the next corotation occurs at a radial position close to our OLR at peak 4 ($r_4$ = 47.5±1.4 arcsec), and their last resonance is beyond the limiting radius of our data.

47. UGC 6521. NGC 3719.            (2P)

This small angular size galaxy shows a regular spiral structure, with four peaks of which peak 4 is by far the strongest. Peak 1 has its OLR on peak 2. Peak 2 has its I4:1 resonance compatible with peak 1, its O4:1 resonance compatible with peak 3, and its OLR compatible with peak 4. Peak 3 has its OLR compatible with peak 4. Peak 4 has its ILR on peak 1, and its I4:1 resonance on peak 3.

48. UGC 6523. NGC 3720.            (1P)

This has a very small angular size, and very weak spiral arms. It has only two peaks, with peak 2 the main one. Peak 1 has its OLR just compatible with peak 2, and peak 2 has its I4:1 resonance just compatible with peak 1.

49. UGC 6537. NGC 3726.            (2P)

This galaxy has a weak bar, and shows five peaks of which peak 3 is the strongest. Peak 1 has its OLR compatible with peak 3. Peak 2 has its OLR compatible with peak 4. Peak 3 has its I4:1 resonance on peak 1, and its O4:1 resonance on peak 5. Peak 4 has its I4:1 resonance compatible with peak 2. Peak 5 shows no coincidences and it is assumed to be a pure OLR associated corotation at peak 3. Rautiainen et al. (2008) locate the corotation resonance at 83.5±13.9 arcsec, which does not coincide with any of the resonances found for this galaxy in this work (see Table 4).

50. UGC 6702. NGC 3840.                     (2P)

This is a small object, with a possible bar. It shows three peaks with similar strengths. Peak 1 has its OLR close to peak 2. Peak 2 has its I4:1 resonance close to peak 1 and its OLR resonance close to peak3. Peak 3 has its I4:1 resonance close to peak 2 and shows an ILR radius which is compatible with peak 1.

51.UGC 6778. NGC 3893.                      (2P)

This is a strong spiral structure galaxy showing a histogram with 5 peaks. Peaks 2 and 3 are both the strongest. Peak 1 has its OLR close to peak 2. Peak 2 has its I4:1 resonance close to peak 1, its O4:1 resonance close to peak 3, and its OLR on peak 4. Error bars associated with the position of peak 2 enable to peaks 3 and 4 to have its I4:1 resonance close to this peak . Peak 5 has its ILR resonance close to peak 1. So, peak 1 not only is a corotation as it is associated with peak 2 in a coupling pattern, but it is also an ILR of the corotation with radius at peak 5. Peaks 1 and 3 ($r_1$ = 13.0±6.2 arcsec and $r_3$ = 63.0±3.0 arcsec) are in good agreement with the two corotations found by Buta & Zhang (2009) at 20.8 and 61.0 arcsec. In addition, Kranz et al. (2003), by means of dynamical simulations, determine a corotation radius of 66.7±6.0 arcsec, which confirms peak 3 as a corotation resonance.

52. UGC 7021. NGC 4045.                     (1P)

This is a galaxy with quite complex structure and an oval distortion in its center region. It shows three peaks, of which peak 3 is the strongest. Peak 1 has its OLR compatible with peak 2, and peak 2 has its I4:1 resonance compatible with peak 1. Peak 3 shows no coincidences.

53. UGC 7045. NGC 4062.                     (1P)

This is a normal spiral with no special structure and does not have a strong bar. There are three corotation peaks, of which peak 2 is the strongest. Peak 1 has its OLR just compatible with peak 3. Peak 2 has its O4:1 resonance just compatible with peak 3. Peak 3 has its I4:1 resonance on peak 1.

54. UGC 7154. NGC 4145.                     (2P)

This is a two-armed spiral galaxy with a short well-defined bar. It shows 6 peaks, of which peak 6 is the strongest. Peak 1 has its OLR compatible with peak 5. Peak 2 has its O4:1 resonance compatible with peak 3, and its OLR close to peak 6. Peak 3 shows no coincidences. Peak 4 has its I4:1 resonance just compatible with peak 1.  Peak 5 has its I4:1 resonance compatible with peak 1. Peak 6 has its I4:1 resonance compatible with peak 2. Corotation at peak 2 ($r_2$ = 46.3±2.1 arcsec) is also derived by Buta & Zhang (2009); according to these authors, there are two corotations at 46.6 and 122.4 arcsec. Their outer resonance does not coincide with any of the other peaks.

55. UGC 7323. NGC 4242.                     (2P)

This galaxy shows possible signs of being a post-merger object. The bar is clean to one side of the nucleus but disturbed on the other. There are five peaks, of which peak 3 is the broadest. Peak 1 has its OLR on peak 2. Peak 2 has its I4:1 resonance on peak 1, its O4:1 resonance is just compatible with peak 3, and its OLR is compatible with peak 4. Peak 3 and peak 5 are isolated resonances showing no compatibilities with other peaks for its corresponding pattern speed. Finally, peak 4 has its I4:1 resonance close to peak 2. None of the corotations identified with our method (see Table 4) occur where Buta & Zhang (2009) place this resonance ($r_{CR}$ = 30.4 arcsec).

56. UGC 7420. NGC 4303 (M61)          (2P)

This is a galaxy with strong spiral structure, at quite low inclination, with a nuclear bar and a main bar. The histogram of phase-reversals shows six peaks of which peak 4 is the strongest and could be associated with the corotation of the spiral arms. Peaks 1 and 4 have no coincidences with any other peak. Peak 2 has its OLR close to peak 5. The O4:1 and OLR resonances of peak 3 are compatible with peaks 4 and 6, respectively. Peak 5 has its I4:1 resonance compatible with peak 2, and its O4:1 on peak 6. Peak 6 has its I4:1 close to peak 3. This galaxy has been studied by a number of authors. Schinnerer et al. (2002) reported a corotation radius between 36 and 48 arcsec, assuming a factor 1.2 between corotation and bar radius, and a pattern speed for the bar between 40 and 53 km·s$^{-1}$·kpc$^{-1}$.. This result is close to the position of the peak 4 at $r_4$ = 49.9±1.0 arcsec, which has a pattern speed of $39.0^{+0.7}_{-0.6}$ km·s$^{-1}$·kpc$^{-1}$. is supported by Koda & Sufue (2006) and Egusa et al. (2009), but Rautiainen et al (2004) modeling optical images, derive a corotation radius of 89.1±8.5 arcsec, where we do not detect H$\alpha$ emission.

57. UGC 7766. NGC 4559.          (1P)

This galaxy is highly inclined; it contains a possible short bar, but this is not easy to discern in the images. It is quite flocculent and shows signs of disturbance due to interaction. There are five peaks, of which peak 4 appears to be the strongest. Peak 1 has its OLR close to peak 2 and peak 2 has its I4:1 resonance close to peak 1, and its O4:1 resonance on peak 4. Peaks 3 and 5 show no coincidences. Peak 4 has its O4:1 resonance compatible with peak 5.

58. UGC 7831. NGC 4605.          (1P)

This is a flocculent highly inclined galaxy. Although we find six peaks in our resonance diagram, only one coupling pattern between peaks 4 and 6 is detected.

59. UGC 7853. NGC 4618. (Arp 23)

This object is highly asymmetric, showing a disturbed morphology, especially in the outer disk, a clear sign of interaction. It has a bar. The rotation curve is almost that of a rigid body. There are three resonance peaks, of which peak 3 is the strongest. None of the peaks show resonance coincidences. Buta & Zhang (2009) determine a corotation radius for this galaxy of 83.0 arcsec, which very close to the position of peak 3 ($r_3$ = 87.1±3.5 arcsec).

60. UGC 7861. NGC 4625.

This is a weak strongly asymmetric object, possibly a post-merger. It has three peaks, of which peak 2 is clearly the strongest. Peak 1 has its O4:1 resonance compatible with peak 3. There are no further coincidences

61. UGC 7876. NGC 4635.

This is a small poorly structured, flocculent galaxy, showing three resonance peaks. Peak 1 may be associated with a weak bar. Peak 1 has its O4:1 resonance compatible with peak 3. There are no further coincidences.

62. UGC 7901. NGC 4651.          (1P)

Although it is classified as an Arp object (Arp 189), the galaxy shows a regular spiral structure with a possible weak bar. It shows 4 peaks, of which peak 2 is the strongest, and also broad. Peak 1 does not show resonance coincidences, and may well simply be due to the ILR of peak 4. Peak 2 has its OLR on peak 4. Peak 3 has its O4:1 resonance compatible with peak 4. Peak 4 has its ILR on peak 1, and its I4:1 resonance compatible with peak 2. There is a clear discrepancy between our results (see Table 4) and those given in Buta & Zhang (2009). According to these authors, this galaxy has four corotations, which are located at 6.1±0.5, 36.9, 51.8 and 88.9±2.5 arcsec. Only the two outermost resonances are quite close to our peaks 3 and 4. We attribute this difference in the positions of the corotations between the two works to the fact that this galaxy is a post-merger object.

63. UGC 7985. NGC 4713.                 (3P)

This is an angularly small object, fairly flocculent, with a short bar. It shows five peaks of which peak 3 is the strongest, followed by peak 5. Peak 1 has its O4:1 resonance compatible with peak 2, and its OLR compatible with peak 3. Peak 2 has its I4:1 resonance compatible with peak 1, its O4:1 resonance is compatible with peak 3, and its OLR compatible with peak 4. Peak 3 has its I4:1 resonance compatible with peak 1, its O4:1 resonance compatible with peak 4, and its OLR on peak 5. Peak 4 has its I4:1 resonance on peak 2 and its O4:1 resonance compatible with peak 5. Peak 5has its I4:1 resonance on peak 3.

64. UGC 8334. NGC 5055.                 (1P)

This is a very flocculent grand design galaxy for which we obtain a complex distribution of phase reversals consisting on a total of 7 very narrow peaks. The OLR of peak 1 lies on the position of peak 2. Peak 2 has its I4:1 resonance on peak 1, and its O4:1 resonance close to peak 3. Peak 3 has its O4:1 resonance just compatible with peak 4 and its OLR on peak 5. Peak 4 has its OLR close to peak 6. The remaining peaks show no coincidences, however, the I4:1 resonance of peak 7 is very near to the position of peak 6. So, peaks 5 and 6 are assumed to be the OLR of peaks 3 and 4, respectively.

65. UGC 8403. NGC 5112.

This is a barred galaxy but the structure is highly asymmetric and flocculent. There are six resonance peaks, of which peak 4 is the strongest, but peak 2 is the best candidate for the bar corotation. Peak 1 has its O4:1 resonance compatible with peak 4. Peak 2 shows no coincidences. Peak 3 has its O4:1 resonance almost compatible with peak 6. Peaks 4, 5, and 6 show no coincidences.

66. UGC 8490. NGC 5204.

This is a very flocculent galaxy. It does show 5 weak peaks, of which peak 5 is the strongest, followed by peak 1. Peak 1 has its O4:1 resonance compatible with peak 4., and its OLR compatible with peak 5. Peak 2 also has its 4:1 resonance compatible with peak 4, and its OLR compatible with peak 5. Neither peak 3, nor peak 4, nor peak 5, show coincidences. Although we have interpreted peaks 1 and 2 as if they were corotations, given the extreme flocculence, and the very small number of residual velocity zero crossings in each peak, this is almost certainly an over-interpretation.

67. UGC 8709. NGC 5297.

This galaxy is almost edge-on. There are five peaks of which the innermost and the outermost are the strongest. Peak 1 does not show any resonance coincidences. Peak 2 has its O4:1 resonance very close to peak 3. Peak 3 does not show coincidences. Peak 4 has its O4:1 resonance on peak 5. Peak 5 does not show coincidences. The rotation curve is smooth and has a normal appearance, and we attribute the absence of coincidences to the very high inclination of the galaxy.

68. UGC 8852. NGC 5376.                 (1P)

This galaxy has a short bar, and presents three peaks of which peak 3 is the strongest. The OLR of peak 1 coincides with peak 2, and the I4:1 resonance of peak 2 coincides with peak 1. The 4:1 resonance of peak 3 is just compatible with peak 1. Peak 1 is probably at the corotation radius of the bar.

69. UGC 8937. NGC 5430.            (2P)

This object has a strong, dominant bar, but is quite asymmetric and could be post-merger. It has 4 peaks of which peak 4 is by far the strongest. Peak 1 has its OLR on peak 2. Peak 2 has its I4:1 resonance on peak 1, its O4:1 resonance on peak 3. Peak 3 has its I4:1 resonance compatible with peak 2, and its OLR compatible with peak 4. Peak 4 has its I4:1 resonance on peak 3, and its oILR compatible with peak 1 (iILR cannot be resolved according to the angular resolution of this data cube).

70. UGC 9179. NGC 5585.            (2P)

This galaxy does not have a strongly ordered structure, though there is a weak bar. It shows six corotation peaks, of which either peak 1 or peak 2 could be at the bar corotation. Peak 4 is the strongest. Peak 1 has its O4:1 resonance compatible with peak 3, and its OLR close to peak 4. Peak 2 has its O4:1 resonance compatible with peak 4. Peak 3 has its O4:1 resonance compatible with peak 5, and its OLR close to peak 6. Peak 4 has its I4:1 resonance on peak 1, and its O4:1 resonance on peak 6. Peak 5 has its I4:1 resonance close to peak 2. Peak 6 has its 4:1 resonance just compatible with peak 3.

71. UGC 9248. NGC 5622.            (2P)

This is angularly small, and with strong spiral structure, but with no evident bar. It shows four resonance peaks, of which peaks 1 and 3 are the strongest. Peak 1 has its O4:1 resonance on peak 2, and its OLR very close to peak 3. Peak 2 has its I4:1 resonance compatible with peak 1, its O4:1 resonance compatible with peak 3, and its OLR compatible with peak 4. Peak 3 has its I4:1 resonance compatible with peak 1, and its O4:1 resonance compatible with peak 4. Peak 4 has its I4:1 resonance compatible with peak 3.

72. UGC 9358. NGC 5678.            (1P)

This is a moderately inclined galaxy with rather flocculent structure. It has three corotation peaks, peak 3 being the strongest. Peak 1 has its OLR coincident with peak 2. Peak 2 has its I4:1 resonance compatible with peak 1, and its O4:1 resonance on peak 3. Peak 3 does not show coincidences.

73. UGC 9363. NGC 5668.            (1P)

This has low inclination and is fairly flocculent, though symmetrical. There are three peaks, of which peak 2 is somewhat stronger than the others. Peak 1 has its O4:1 resonance compatible with peak 2, and its OLR compatible with peak 3. Peak 2 has its I4:1 resonance on peak 1, and its O4:1 resonance on peak 3. Peak 3 has its I4:1 resonance compatible with peak 1.

74. UGC 9366. NGC 5676.

This is a regular spiral galaxy with not too high inclination. There are 4 peaks of which peak 3 is the strongest. Peak 1 does not show coincidences, and is probably not a corotation peak, but the ILR corresponding to peak 4. Peak 2 has its O4:1 resonance almost on peak 3, and its OLR compatible with peak 4. Peak 3 has its O4:1 resonance compatible with peak 4. Peak 4 has its ILR compatible with peak 1, and its I4:1 resonance compatible with peak 3. Kranz et al. (2003) place corotation at a radial distance of $68.8^{+6.2}_{-12.4}$ arcsec, which is out of range of the H$\alpha$ maps analyzed in this work.

75. UGC 9465. NGC 5727.            (1P)

This object has small angular size, and is quite highly inclined. There are three corotation peaks, of which peaks 1 and 2 are stronger than peak 3. Peak 1 has its OLR compatible with peak 3. Peak 2 shows no coincidences. Peak 3 has its I4:1 resonance compatible with peak 1.

76. UGC 9736. NGC 5874.            (1P)

This galaxy has a strong spiral structure but with no evidence of a major bar. There are two corotation peaks of which peak 1 is clearly the stronger. Peak 1 has its OLR on peak 2. Peak 2 has its I4:1 resonance close to peak 1. ILR compatible with peak 2 is under the limit of the angular resolution.

77. UGC 9753. NGC 5879.          (1P)

This has quite a high inclination but there is a discernible bar. It shows two peaks, of which peak 2 is the strongest, and the most likely to be at the bar corotation radius. Peak 1 has its OLR close to peak 2. Peak 2 has its I4:1 resonance close to peak 1.

78. UGC 9866. NGC 5949.

This is a small, very flocculent object, quite highly inclined and shows only one weak peak.

79. UGC 9943. NGC 5970.          (2P)

This galaxy is regularly shaped and has a clear bar. There are 5 resonance peaks, of which peak 3 is the strongest, but peak 1 is strong, and appears to be the best candidate for the bar corotation. Peak 1 has its OLR close to peak 2. Peak 2 has its I4:1 resonance on peak 1, its O4:1 resonance compatible with peak 3, and its OLR very close to peak 4. Peak 3 has its I4:1 resonance very close to peak 2, and its O4:1 resonance very close to peak 5. Peak 4 has its I4:1 resonance quite close to peak 2. Peak 5 shows no resonant coincidences.

80. UGC 9969. NGC 5985.          (3P)

This is a somewhat flocculent galaxy, but very massive with its plateau rotational velocity of some 300 km/s. It shows 6 peaks in all. Peak 4 is the strongest followed by peaks 3 and 5. Peak 1 has its OLR close to peak 3. Peak 2 has its O4:1 resonance close to peak 3, and its OLR compatible with peak 5. Peak 3 has its I4:1 resonance compatible with peak 1, and its O4:1 resonance close to peak. Peak 4 has its O4:1 resonance close to peak 5, and its OLR on peak 6. Peak 5 has its I4:1 resonance just compatible with peak 2, and its O4:1 resonance on peak 6. Peak 6 has its I4:1 resonance compatible with peak 4.

81. UGC 10075. NGC 6015.          (1P)

This galaxy is quite neatly regular spiral, but without a bar. It shows three peaks, of which peak 3 is the strongest. Peak 1 has its O4:1 resonance compatible with peak 2, and its OLR just compatible with peak 3. Peak 2 has its O4:1 resonance compatible with peak 3. Peak 3 has its I4:1 resonance just compatible with peak1..

82. UGC 10359. NGC 6140.

This is a lopsided "peculiar" galaxy. It shows 3 peaks of which peak 2 is the strongest. All the peaks obtained for this galaxy are isolated.

83. UGC 10445

This galaxy has a strangely irregular morphology. It shows three peaks, one near the center and two further out. Peak 3 is the strongest. Peaks 1 and 3 have no resonances compatible with the other peaks. Peak 2 has its OLR compatible with peak 3.

84. UGC 10470. NGC 6217.          (2P)

This galaxy (Arp 185) has a strong bar with a somewhat complex morphology. It shows five peaks, of which peak 4 is the strongest, and the best candidate for the bar corotation radius. Peak 1 has its O4:1 resonance just compatible with peak 2, and its OLR compatible with peak 3. Peak 2 has its I4:1 resonance compatible with peak 1, and its O4:1 resonance just compatible with peak 4, and its OLR on peak 5. Peak 3 has its I4:1 resonance compatible with peak 1, and its O4:1 resonance compatible with peak 4. Peak 4 has its I4:1 resonance compatible with peak 2. Peak 5 has its I4:1 resonance compatible with peak 2.

85. UGC 10502        (1P)

This is a very regular spiral galaxy, but with no obvious bar. It shows three peaks, of which peak 1 is the strongest, followed by peak 3. Peak 1 has its O4:1 resonance on peak 2, and its OLR is compatible with peak 3. Peak 2 has its O4:1 resonance compatible with peak 3. Peak 3 has its I4:1 resonance just compatible with peak 1.

86. UGC 10521. NGC 6207.        (2P)

This is a fairly inclined object, showing six corotation peaks. Peak 4 is the strongest, followed by peak 2. Peak 1 has its O4:1 resonance on peak 2, and its OLR just compatible with peak 3. Peak 2 has its O4:1 resonance just compatible with peak 3, and its OLR compatible with peak 5. Peak 3 has its I4:1 resonance just compatible with peak 1, and its O4:1 resonance just compatible with peak 5. Peak 4 has its I4:1 resonance close to peak 2. Peak 5 has its I4:1 resonance just compatible with peak 2. Peak 6 has no coincidence with any other peak.

87. UGC 10546. NGC 6236.        (2P)

This is a low inclination object with a possible rather short bar. It shows 4 peaks, none of them very strong. Peak 1 has its O4:1 resonance on peak 2 and its OLR close to peak 3. Peak 2 has its O4:1 resonance compatible with peak 3, and its OLR just compatible with peak 4. Peak 3 has its I4:1 resonance on peak 1. Peak 4 has its I4:1 resonance close to peak 2.

88. UGC 10564. NGC 6248.

This is an edge-on galaxy which shows an apparently rigid body rotation curve. It has 5 weak peaks, of which the strongest is peak 3. There are no coincidences between the resonant systems.

89. UGC 10652. NGC 6283.

This is a small galaxy without much structure. It shows two weak peaks. Peak 1 has its O4:1 resonance on peak2, but peak 2 shows no coincidences with peak1.

90. UGC 10757

This is an angularly small galaxy and shows only two weak peaks. Peak 1 has its O4:1 resonance compatible with peak 2, but peak 2 shows no coincidences with peak 1.

91. UGC 10897. NGC 6412.

This galaxy has a small central bar, and possibly a weak large bar. The rotation curve for the inner region ($r \leq 20$ arcsec) responds to a solid body, so it is difficult to determine the pattern speed associated with each peak. There are five corotation peaks, of which peak 3 is the strongest. Peak 1 could be the corotation of the nuclear bar. Peak 1 has its O4:1 resonance on peak 4, and its OLR close to peak 5. Peak 2 has its O4:1 resonance on peak 4, and its OLR on peak 5. Peak 3 has its O4:1 resonance close to peak 4, and its OLR close to peak 5. Peak 4 has its O4:1 resonance close to peak 5. Peak 5 shows no coincidences with other peaks, so we infer that this peak is not a corotation but the OLR of peak 2.

92. UGC 11012. NGC 6503.        (1P)

This is a fully edge-on galaxy. It does show four well separated peaks. Peak 1 has its O4:1 resonance just compatible with peak 3. Peak 2 has its OLR just compatible with peak 4. Peak 3 has its O4:1 resonance just compatible with peak 4. Peak 4 has its I4:1 resonance compatible with peak 2.

93. UGC 11124

This galaxy has arms with a very low pitch angle, almost forming an S-shape with the bar, and also has a short inner bar-like structure which is almost aligned with the main bar. It shows four peaks of which peak 3 is the strongest. Peak 1 has its O4:1 resonance compatible with peak 3. Peaks 2, 3, and 4 do not show coincidences.

94. UGC 11218. NGC 6643.            (3P)

This galaxy is well structured. It shows six peaks. The strongest are peaks 3 and 4. There is no obvious bar in the image. Peak 1 has its O4:1 resonance slightly compatible with peak 2, and its OLR compatible with peak 3. Peak 2 has its O4:1 resonance compatible with peak 3, and its OLR close to peak 4. Peak 3 has its I4:1 resonance just compatible with peak 1, its O4:1 resonance on peak 4, and its OLR on peak 5. Peak 4 has its I4:1 and O4:1 resonances compatible with peaks 2 and 5, respectively. Peak 5 and 6 has its I4:1 resonance compatible with peak 3 and 4, respectively. Peak 5 of the resonant structure found for this galaxy located at $r_5 = 61.8 \pm 3.1$ arcsec is in good agreement with the corotation radius of $58.3^{+13.5}_{-4.5}$ arcsec, derived by Kranz et al. (2003) using numerical simulations and NIR images.

95. UGC 11283            (1P)

This galaxy has a dominant bar, with rather clumpy structure. There are five peaks, of which peak 3 is the strongest. Peak 1 has its OLR compatible with peak 3. Peak 2 has its OLR on peak 3. Peak 3 has its I4:1 resonance just compatible with peaks 1 and 2, its O4:1 resonance close to peak 4, and its OLR just compatible with peak 5. Peak 4 has its I4:1 resonance compatible with peak 3, and its O4:1 resonance compatible with peak 5. Peak 5 has its I4:1 resonance on peak 4. So, peak 5 is assumed to be the OLR of peak 3 rather than a corotation.

96. UGC 11407. NGC 6764.            (3P)

This galaxy has a long strong bar. It shows five resonance peaks. Peak 5 is the best candidate for the bar corotation, but peak 2 is also strong. Peak 1 has its O4:1 resonance on peak 2 and its OLR close to peak 3. Peak 2 has its I4:1 resonance compatible with peak 1, its O4:1 resonance compatible with peak 3, and its OLR compatible with peak 4. Peak 3 has its I4:1 resonance on peak 1, and its OLR just compatible with peak 5. Peak 4 has its I4:1 resonance compatible with peak 1, and its O4:1 resonance close to peak 5. Peak 5 has its I4:1 resonance close to peak 3.

97. UGC 11466            (1P)

The radial phase reversal distribution of this galaxy shows three peaks of which peak 1 is the strongest. Peak 1 has its O4:1 and OLR compatible with peaks 2 and 3, respectively. Peak 2 has its O4:1 just compatible with peak 3 and peak 3 has its I4:1 resonance on peak 1.

98. UGC 11557

This galaxy has an almost "rigid body" rotation curve, which does not produce resonance coincidences. It shows five peaks, of which peak 2 is the strongest. Peak 2 is suggested as marking the corotation of the bar. There are no coincidences between any pair of peaks.

99. UGC 11861            (1P)

In this weakly spiral structured galaxy, with a central bar-like structure, there are five resonance peaks, of which peak 2 is the weakest. Peak 1 has its O4:1 resonance close to peak 3 and its OLR just compatible with peak 4. Peaks 2 and 5 turn out to be an isolated resonance as none of them show any coincidence with any other peak. Peak 3 has its O4:1 resonance close to peak 4, and its OLR is just compatible with peak 5. Peak 4 has its I4:1 resonance close to peak 1. There is a small bar for which peak 2 could be associated with its corotation, while peak 5 is not a corotation but an OLR of peak 3.

100. UGC 11872. NGC 7177.            (2P)

This is a well-structured galaxy, showing 4 peaks. Peak 1 is the strongest, and the candidate for the bar corotation. Peak 1 has its OLR compatible with peak 2. Peak 2 has its I4:1 resonance compatible with peak 1, and its OLR compatible with peak 3. Peak 3 has its I4:1 resonance compatible with peak 2 and its OLR compatible with peak 4, which shows no coincidences with any other innermost peak, assuming, thus, there is no corotation at this position.

101. UGC 11914. NGC 7217.            (2P)

This is a flocculent galaxy, showing five peaks. Peak 1 has its OLR on peak 2. Peak 2 has its I4:1 resonance very close to peak 1, and its OLR close to peak 3. Peak 3 has its I4:1 resonance on peak 2, its O4:1 resonance is very close to peak 4, its ILR is compatible with peak 1, and its OLR is on peak 5. Peak 4 has its ILR just compatible with peak 1. Peak 5 has its I4:1 resonance close to peak 3. Here, peaks 1 and 5 are not only a corotation as they are interlocked with peaks 2 and 3, respectively, but also are the ILR and OLR of peak 3. This result is compatible not only with the corotation radii determined by Buta & Zhang (2009), but also with the resonant structure reported by Combes et al. (2004). Buta & Zhang (2009) identify two corotations at 14.5±1.6 and 43.1 arcsec, which match our peaks 1 and 3 located at $r_1$ = 17.0±5.8 arcsec and $r_3$ = 48.8±2.2 arcsec. In the complete dynamical analysis performed by Combes et al. (2004) with this galaxy, the corotation resonance is placed at a radius of $r_{CR}$ = 52.6±10.4 arcsec with a pattern speed $\Omega_P$ = 83.5±16.6 km·s$^{-1}$·kpc$^{-1}$. Both values are in very good agreement with peak 3 and its associated pattern speed $\Omega_3$ = 79.3$^{+3.6}_{-3.3}$ km·s$^{-1}$·kpc$^{-1}$. As described above, when corotation is assumed to occur at the position of peak 3, the corresponding ILR and OLR are compatible with peaks 1 and 5, respectively, which are fully compatible with the radial position of these resonances found by Combes et al. (2004), who placed the ILR at $r_{ILR}$ = 14.3±2.8 arcsec and the OLR in the range $r_{OLR}$ = (51.7 − 71.4)$^{+14.4}_{-10.2}$ arcsec.

102. UGC 12276. NGC 7440.            (1P)

This galaxy shows three peaks, of which peak 3 is the best candidate for the bar corotation. Peak 1 has its O4:1 resonance on peak 2, and its OLR on peak 3. Peak 2 has its O4:1 resonance compatible with peak 3. Peak 3 has its I4:1 resonance on peak 1.

103. UGC 12343. NGC 7479.

This galaxy has a very strong but twisted bar, and its structure suggests that it has suffered a recent merger. Peaks 1 and 2 are situated at sharp changes in the bar brightness and peak 3 where the bar starts to twist. Peak 3 has it O4:1 resonance compatible with peak 5. We derive the pattern speed only for peaks 3, 4 and 5, because the best fit to the rotation curve derived from our data, shows a linear behavior for the central region (r ≲ 70 arcsec). This is a galaxy extensively studied by other authors as shown in Table 6 in which we summarize the values of the corotation radius and the pattern speed (when available) derived by these authors; additionally the type of images as well as the method they have applied are also given in the table. It seems that there is a consensus (in eight of the 13 studies) to place the corotation resonance at a radius between 44 and 57.7 arcsec with a mean value of 52.5 arcsec, which turns out to be in very good agreement with our peak 3 ($r_3$ = 52.4±3.3 arcsec). The remaining results push the corotation further out, in particular Elemgreen & Elmegreen (1985), del Rio & Cepa (1998) and Fathi et al. (2009) obtain a corotation radius between 74 and 106 arcsec, which is fully compatible with our peak 5 ($r_5$ = 91.1±4.3 arcsec). The corotation radius derived by Wilke et al. (2000) is close to the position of our peak 4 at 75.1±3.9 arcsec. Values for the pattern speed are scarce, as can be seen in Table 6, in which only five authors determine this parameter. Those authors who place corotation close to 52.5 arcsec give a pattern speed between 27 and 30 km·s$^{-1}$·kpc$^{-1}$, while the other authors coincide with a value of 18 km·s$^{-1}$·kpc$^{-1}$. These values are in good agreement with the pattern speed associated with corotations at peaks 3 and 5 ($\Omega_3$ = 25.6±1.5 km·s$^{-1}$·kpc$^{-1}$ and $\Omega_5$ = 18.4±1.0 km·s$^{-1}$·kpc$^{-1}$).

Table 6. Corotation radius and pattern speed for UGC 12343 from literature

| Reference | Corotation Radius (arcsec) | Pattern speed (km·s$^{-1}$·kpc$^{-1}$) | Method | Type of data |
|---|---|---|---|---|
| (1) | (2) | (3) | (4) | (5) |
| Elmegreen & Elmegreen, 1985 | 74-106 | -- | Morphological inspection | Optical, NIR |
| Duval & Monet, 1985 | 56 | 27 | Mass model | Long slit |
| Beckman & Cepa, 1990 | 56 | -- | Band crossing | Optical |
| Quillen et al. 1995 | 45 | -- | Bar radius | Optical, NIR, H$\alpha$, CO |
| Sempere et al. 1995 | 57 | 30 | Numerical simulations | CO |
| Puerari & Dottori, 1997 | 55 | -- | Fourier analysis of azimuthal profiles | Optical |
| Laine et al. 1998 | 50±5 | 27±2 | SPH simulations | Optical, CO |
| del Rio & Cepa, 1998 | 85 | -- | Band crossing | Optical |
| Aguerri et al. 2000 | 63 | -- | Fourier analysis of azimuthal profiles | Optical |
| Wilke et al., 2000 | 70 | 18 | Kinematical modeling | NIR |
| Vera-Villamizar et al. 2003 | 18 / 44 | -- | Fourier analysis of azimuthal profiles | Optical |
| Buta & Zhang, 2009 | 6.3 / 57.7 | -- | Potential-density phase-shift | NIR |
| Fathi et al. 2009 | $94^{+6}_{-26}$ | 18±3 | Tremaine-Weinberg | H$\alpha$ |

**Notes.** Column (1) shows the reference of the works that have determined the corotation radius and pattern speed values, which are given in columns (2) & (3) respectively. In columns (4) & (5) are found, respectively, the method applied in order to derive these values and the type of data used. Pattern speed values are corrected by distance and inclination.

104. UGC 12754. NGC 7741. (1P)

This galaxy shows six peaks, of which the strongest, peak 3, appears to lie at a suitable length to be at the bar corotation, while peak 1 is where the bar brightness shows a sharp change. Peak 6 is also strong. Peak 1 has its OLR on peak 4. Peak 2 has its O4:1 resonance on peak 4. Peak 3 has its I4:1 resonance compatible with peak 1, and its O4:1 resonance compatible with peak 6. Peak 4 has its I4:1 resonance close to peak 1. Peak 5 is an isolated peak, and peak 6 shown no coincidences with any other peak. Peak 6 at 106.3±1.7 arcsec was measured as a corotation radius using the Tremaine-Weinberg method by Fathi et al. (2009), $r_{CR} = 109^{?}_{-17}$ arcsec, and peak 2 ($r_2$ = 45.1±1.4 arcsec) is in agreement within the error bars with the corotation radius of 51.9±6.8 arcsec determined by Buta & Zhang (2009) applying the potential-density phase-shift method. Duval et al. (1991) find a different value for the corotation radius, $r_{CR}$ = 99 arcsec, which is close to our peak 5 at 95.4±3.9 arcsec. Although Fathi et al. (2009) and Duval et al. (1991) report compatible values of the corotation radius, the pattern speed they derive differs significantly (after correcting by distance and inclination, $\Omega_P = 25^{+8}_{-6}$ km·s$^{-1}$·kpc$^{-1}$ and $\Omega_P$ = 53 km·s$^{-1}$·kpc$^{-1}$, respectively). Only the pattern speed calculated by means of the Tremaine-Weinberg method is in good agreement with the angular rate associated with peak 6, $\Omega_6$ = 28.4±0.3 km·s$^{-1}$·kpc$^{-1}$.

To complete this study we would have preferred to include measurements of the ratio of the bar length to the corresponding corotation radius for all the galaxies. The problem here is the absence of images of suitable photometric quality to make reliably accurate measurements of the bar length. Images at wavelengths in the visible of galaxies with widespread star formation are normally subject to the distorting effects of dust extinction. Previous near IR images such as those from 2MASS are of only moderate quality. When more of the images from the "Spitzer warm" imaging survey S4G are in the public domain it may be possible to make more reliable measurements of bar lengths (which always contain a subjective element) for a significant fraction of the galaxies in the present article, and we intend to do this. In the meantime we list in Table 7 the corotation radius to bar length ratio, $\mathcal{R}$, for only those galaxies for which we could find the radial bar length in the literature. Due to the multi-spin nature of the galaxies, we have defined two criteria in order to systematically pick out that resonance, which is identified as the corotation of the bar: (*i*) The bar corotation occurs not far from the tip of the bar, in other words the radius of the resonance is close to the bar length. (*ii*) Among all peaks identified as resonances in the histogram of phase reversals which satisfy the criterion (*i*), the strongest peak (defined after comparing the areas under the histogram forming the peak) is taken as the bar corotation. After applying these criteria to the galaxies listed in Table 7, we have found that all values are over 1 taking into account the uncertainties, locating, thus, the corotation just at or beyond the end of the bar, as predicted theoretically by Contopoulos (1980). This ratio is also reported by other authors only for a small fraction of the galaxies listed in Table 7. Aguerri et al (1998) obtained $\mathcal{R}$=1.22 for UGC 3013, which is slightly higher than the value we derive. For UGC 5532, Casasola et al. (2008) find $\mathcal{R}$=2, which is in agreement with our result. The ratio for UGC 5840 we have determined is also in agreement with the value obtained by Meidt et al. (2009), $\mathcal{R}$=1.4. Rautiainen et al. (2008) obtain the following values: $\mathcal{R}$=1.19±0.27, $\mathcal{R}$=1.95±0.55 and $\mathcal{R}$=1.70±0.45 for UGC6118, UGC6537 and UGC7240, respectively, the values for the latter two galaxies differ significantly from our results. And, Wilke et al. (2000) find a value of $\mathcal{R}$=1.1, for the galaxy UGC 12343, which is in good agreement with our result. The average value of this ratio, when all galaxies listed in Table 7 are taken into account, is found to be 1.35±0.36, which is in agreement with the dynamical simulations of Athanassoula (1992), who predicted a value for $\mathcal{R}$ of 1.2±0.2. If we classify the galaxies according to the morphological type, we find that the ratio $\mathcal{R}$ is 1.15±0.28 for the early types (Sa-Sab), 1.30±0.30 for the intermediate types (Sb-Sbc) and 1.35±0.28 (Sc-Sd) for the late types. Our results are in agreement, within the error bars, with those of Rautiainen et al. (2008), who find $\mathcal{R}$ to be equal to 1.15±0.25, 1.44±0.29 and 1.82±0.63, for early, intermediate and late type galaxies respectively. Both results reveal that the corotation radius to the bar radius ratio shows a dependence on the morphology, with a trend to increasing values from early to late type galaxies.

Table 7. Ratio of Corotation Radius to Bar Length for Selected Barred Galaxies

| Name | | Bar Corotation | Bar length | $\mathcal{R}$ | Reference |
|---|---|---|---|---|---|
| UGC | NGC | (arcsec) | (arcsec) | | |
| (1) | (2) | (3) | (4) | (5) | (6) |
| 508 | 266 | 51.8±2.3 | 44.2 | 1.17±0.05 | De Jong 1996 |
| 763 | 428 | 51.9±1.9 | 53.5 | 0.97±0.04 | Cabrera-Lavers & Garzón 2004 |
| 1736 | 864 | 56.1±2.1 | 44.5 | 1.26±0.05 | Cabrera-Lavers & Garzón 2004 |
| 1913 | 925 | 118.5±3.1 | 119.8 | 0.99±0.03 | Elemegreen et al. 1998 |
| 3013 | 1530 | 76.5±3.2 | 67.6 | 1.13±0.05 | Reynaud & Downes 1998 |
| 3685 | -- | 27.7±1.7 | 27 | 1.03±0.06 | Erwin et al. 2008 |
| 3809 | 2336 | 35.4±1.8 | 26.5 | 1.34±0.07 | Wilke et al. 1999 |
| 4284 | 2541 | 52.9±1.9 | 38.15 | 1.79±0.05 | Cabrera-Lavers & Garzón 2004 |
| 4422 | 2595 | 36.5±2.4 | 38.9 | 0.94±0.08 | Comerón et al. 2010 |
| 5532 | 3147 | 16.3±3.3 | 7.5 | 2.1±0.5 | Casasola et al. 2008 |
| 5786 | 3310 | 43.6±8.7 | 24.9 | 1.7±0.3 | Comerón et al. 2010 |
| 5840 | 3344 | 31.6±4.2 | 25 | 1.3±0.2 | Menéndez-Delmestre et al. 2007 |
| 6118 | 3504 | 37.6±2.2 | 34 | 1.11±0.07 | Erwin et al. 2008 |
| 6537 | 3726 | 59.8±2.5 | 46.1 | 1.30±0.05 | Marinova & Jogee 2007 |
| 7021 | 4045 | 39.4±1.9 | 29 | 1.36±0.07 | Erwin et al. 2008 |
| 7154 | 4145 | 18.1±2.1 | 15 | 1.38±0.08 | Speltincx et al. 2008 |
| 7323 | 4212 | 44.7±2.8 | 34.4 | 1.30±0.08 | Marinova & Jogee 2007 |
| 7420 | 4303 | 31.1±2.0 | 30.3 | 1.03±0.07 | Erwin 2004 |
| 7853 | 4618 | 54.5±1.7 | 47 | 1.16±0.04 | Cabrera-Lavers & Garzón 2004 |
| 7901 | 4651 | 15.6±2.1 | 14 | 1.11±0.08 | Speltincx et al. 2008 |
| 8937 | 5430 | 15.2±3.0 | 11.6 | 1.3±0.2 | Hoyle et al. 2011 |
| 9363 | 5668 | 36.7±1.2 | 20.0 | 1.8±0.4 | Marino et al. 2012 |
| 9358 | 5678 | 39.3±3.1 | 25 | 1.6±0.1 | Laine et al. 2002 |
| 9943 | 5970 | 19.1±2.8 | 12 | 1.6±0.2 | Laine et al. 2002 |
| 9969 | 5985 | 57.3±5.8 | 48 | 1.2±0.1 | Menéndez-Delmestre et al. 2007 |
| 10470 | 6217 | 57.8±3.4 | 48 | 1.28±0.07 | Cabrera-Lavers & Garzón 2004 |
| 11012 | 6503 | 21.8±1.8 | 12.5 | 1.7±0.1 | Freeland et al. 2010 |
| 11218 | 6643 | 48.6±4.0 | 38.7 | 1.3±0.1 | Cabrera-Lavers & Garzón 2004 |
| 11407 | 6764 | 64.9±2.3 | 40 | 1.62±0.06 | Leon et al. 2007 |
| 11872 | 7177 | 19.9±2.8 | 17 | 1.2±0.2 | Erwin et al. 2008 |
| 12343 | 7479 | 52.4±3.3 | 50.9 | 1.03±0.07 | Marinova & Jogee 2007 |
| 12754 | 7741 | 69.6±7.3 | 53.7 | 1.3±0.1 | Marinova & Jogee 2007 |

**Notes.** Columns (1) and (2) show the name of galaxies with measured bar lengths available in the literature. The bar corotation radius found in this work is given in Column (3), while Column (4) shows the deprojected radial bar length, which can be found in the articles given in Column (6). The coefficient between corotation radius and bar length is shown in Column (5).

DISCUSSION AND CONCLUSIONS

We have developed a method for finding resonance radii in spiral galaxies, using their emission in Hα, which implies that the method in this form can be applied only to galaxies with widely distributed star formation, i.e. not to early types. The technique relies on the prediction that as we move outwards the residual velocity field (after the subtraction of the rotational velocity under gravitational equilibrium) should undergo a phase change in its radial component of $180°$, which can be detected directly using a two-dimensional velocity map with sufficiently high spatial and velocity resolution. The method, using Hα emission as presented in the article could profitably be extended to maps in the CO molecular emission lines and to maps in the 21 cm line of HI. At the present time there are limited numbers of maps of these kinds with angular resolution better than 5 arcsec, but these numbers will grow steadily in the near future. In particular line maps of galaxies with ALMA should give especially good results. Our method takes advantage of the fact that the gas response to a density wave pattern is much more rapid than the response of a stellar population, so the resonance radii are correspondingly better defined and detected. We have shown using a representative model that these phase reversals should indeed be detectable at corotation, and that a similar effect but with significantly reduced amplitude, can be expected at an ILR and also with reduced amplitude at an OLR. These predictions have enabled us to make the appropriate observational distinctions, and to detect not only corotations but also ILR´s and OLR´s in specific cases. Special care was needed in our interpretation of resonances defined by only one resolved transition point; we have explained in the text how we determined which of these "N=1" transitions can be used as resonances, and the reader can judge their degree of reliability. We have applied our observational method to 100 galaxies from the GHASP data base of Hα data cubes, and complemented the results with four galaxies observed with the GHαFaS interferometer, which have higher S:N and angular resolution.

We wish to pick out two general results from the present study. Firstly we have found at least one corotation radius, defined using the peaks in our histograms of the radial distribution of resolved points in the galaxies showing clear, statistically significant phase changes, in each and every one of the galaxies measured. The number of peaks detected in our sample varies between a minimum of a single peak and a maximum of seven, with a mode of three, a median value of 4, and a mean of 4.2 peaks per galaxy. The galaxies with the highest number of detected peaks appear to be those in which the dust pattern reveals both radially and azimuthally structured gas distribution, a sort of "organized pseudo-flocculence". In barred galaxies we consistently find that there is a peak whose galactocentric radius is consistent with the range of values of corotation expected from general theoretical models. The remaining peaks in those galaxies, and the peaks in the unbarred galaxies, cannot be easily identified with specific morphological parameters. It is particularly interesting to note, however, that the existence of multiple pattern speeds which dominate over restricted concentric galactocentric radial ranges has been claimed by observers and in recent times even hinted at by theorists. Meidt et al. (2008) applying a "Radial Tremaine-Weinberg Method" predicted the presence of multiple pattern speeds in galactic disks, while in Meidt et al. (2009) they used "high quality HI and CO data cubes" to demonstrate the presence of multiple pattern speeds in four nearby galaxies where data of high enough spatial resolution could be obtained. Buta & Zhang, using a potential-density phase shift method, found multiple corotation radii using the stellar components of 153 galaxies, via H band images. Uncertainties in the corotation radii are quoted, however, for only a small fraction of the galaxies they analyzed, and not for all corotations of a given galaxy. It is also pertinent to cite Foyle et al. (2011) at this point. They find, using a wide selection of stellar and gaseous spectral and photometric measurements, that the systematic angular offsets between indicators, predicted for spiral arms produced within a stable density wave system, are not detected, and conclude that this is evidence for the absence of such a system. They use the prediction that the peak of the cross-correlation functions between pairs of indicators should be shifted in opposite directions inside and outside corotation, and find that this prediction is not satisfied by their data. It is interesting to see in their article, however, that there do appear to be such shifts within limited radial ranges, and not simply shifts in opposite senses at a single galactocentric radius. We suggest, tentatively, that if the results were analyzed on the basis of models with multiple corotations, they could well satisfy the predictions, and that, perhaps ironically, their work could well be evidence in favor of density wave produced stellar patterns, albeit rather complex ones, of the kind we have detected in our work here.

Our second general result is also of considerable interest. We made the initial assumption (borne out in practice as we have seen) that most, if not all, of the peaks we detect in our histograms do in fact correspond to corotation radii. We then went on to use the rotation curves derived from our observations to compute the ILR, the OLR and the two ultraharmonic resonances (I4:1 and O4:1) for each corotation. This led us to discover a pattern which has not been predicted (perhaps it would be more accurate to say that only half of the pattern has been previously predicted) or observed, and which is found, at least once, in a total of 70% of the galaxies studied here. The pattern is as follows: we will call two of the corotations found in a given galaxy CR:1 and CR:2, the corresponding angular rate are $\Omega_1$ and $\Omega_2$, respectively; the pattern of resonance between them is such that the OLR of $\Omega_1$ (OLR:1) falls at CR:2, and the I4:1 resonance of $\Omega_2$ falls on CR:1. This second relation is suggested in the simulations by Rautiainen & Salo (2002) but the double coincidence is not, as far as we know, referred to anywhere in the literature. The remarkable point is that this pattern is found in 70% of the galaxies observed; in 42 of them it occurs once, in 26 we find it twice, between two different pairs of corotations, in five galaxies we find the pattern three times, and in one we find it as many as four times. We plot the distribution of coupling patterns in Figure 10 right panel; while in left panel we plot the distribution of the number of peaks identified in the histogram of phase-reversals. Furthermore, of the galaxies in which we fail to detect the pattern, eight are close to edge-on (with inclination angle over 70º) and four are close to face-on (with inclinations lower than 25º). In the edge-on galaxies the combination of dust and foreshortening make it quite possible for any pattern of this sort to be present but not detected. In the galaxies which are face-on we are not able to measure accurately the in-plane residual velocities, so here again the coupling pattern might be present but undetected. Neglecting face-on and edge-on galaxies from the list, the fraction of objects showing at least one coupling pattern is increased from 70% up to 80%. So we are left with 18 galaxies of the sample in which we might have been able to detect the pattern but did not do so. For three of these, which are highly flocculent, we might in fact expect not to find a pattern which reflects resonant structure. The remaining 15 galaxies are a rather mixed bag of lopsided and disturbed objects (two of them are Arp objects), and galaxies with quasi-rigid body rotation curves, where it is not possible to detect resonant structure from the kinematics. To summarize, the pattern we have detected occurs in at least two thirds and up to four fifths of the galaxies presented here, and in over one third of those with the pattern it occurs more than once. This is very clear evidence in favor of persistent resonant structure in galaxies. However we need help from our theorist colleagues to provide a complete explanation of the phenomena observed.

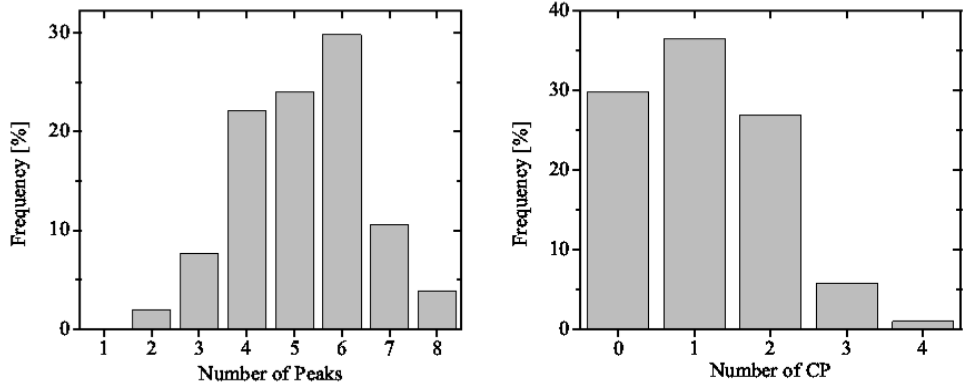

**Figure 10. Left panel**. Histogram showing the number of peaks in the plot of the observed frequency of the phase changes in radial velocity measured on the residual velocity maps of the galaxies in the sample. We assign almost all of these to corotation radii detected in the gas component, apart from a specific few which mark Inner Lindblad Resonances (see the text for details). **Right panel**. Histogram showing the frequency with which we detect the pattern of interaction between two of the resonances, also termed a Coupling Pattern (CP, the Outer Lindblad resonance of the internal corotation coincides with the external corotation, while the Inner 4:1 resonance of the external corotation coincides with the internal corotation) in a given galaxy of the sample. Note that the number of non-detections is certainly an upper limit, since we cannot measure the residual velocities adequately if the galaxy is close to edge-on or close to face-on. Note that the pattern occurs more than once in almost one third of the galaxies in the sample.

ACKNOWLEDGMENTS

We would like to thank Sharon Meidt and Javier Zaragoza Cardiel for very useful discussions. We are most grateful to the anonymous referee for comments, which substantially improved the paper during review. Some of the observations for this article were taken at the William Herschel Telescope of the Isaac Newton Group, in the Spanish Observatorio del Roque de los Muchachos, La Palma, Canary Islands. The research was supported by the grant AYA2007-6762-C02-02, of the Spanish Ministry of Science and Innovation, MINECO, and by project P3/86 of the Instituto de Astrofísica de Canarias. We also acknowledge financial support from the DAGAL network from the People Programme (Marie Curie Actions) of the European Union's Seventh Framework Programme FP7/2007-2013/ under REA grant agreement number PITN-GA-2011-289313.

REFERENCES

Aguerri, J.A.L., Beckman, J.E. & Prieto, M. 1988, AJ, 116, 2136.
Aguerri, J.A.L., Muñoz-Tuñon, C., Varela, A.M. & Prieto, M. 2000, A&A, 361, 841.
Aguerri, J.A.L., Debattista, V.P. & Corsini, E.M. 2003, MNRAS, 338, 465.
Athanassoula, E. 1992, MNRAS, 259, 345.
Athanassoula, E. 2002, ApJ Letters, 569 L83.
Beckman, J.E. & Cepa, J. 1990, A&A, 229, 37.
Broeils, A. 1995, AIP Conference proceedings, 336, 125.

Buta,R. & Zhang, X. 2009, ApJ Suppl. Ser., 182, 559.
Cabrera-Lavers, A. & Garzon, F. 2004, ApJ, 127, 1386.
Canzian, B. 1993, ApJ 414, 487.
Canzian, B. & Allen, R.J. 1997, ApJ 479, 723.
Casasola, V., Combes, F., Garcia-Burillo, S. et al. 2008, A&A, 490, 61.
Comerón, S., Knapen, J.H., Beckman, J.E. et al. 2010, A&A, 402, 2462.
Contopoulos, G. 1980, A&A 81, 198.
Contopoulos, G. & Mertzanides, C. 1977, A&A, 61, 477.
Contopoulos, G. & Papayannopoulos, T. 1980, A&A, 92, 33.
Contopoulos, G. 1981, A&A, 102, 265
Contopoulos,G., Gottesman, S.T., Hunter J.H. Jr. & England, M.N. 1989, ApJ 343, 608.
Corsini, E.M. 2011, MSAIS, 18, 23.
Debattista, V.P. & Sellwood, J.A. 2000, ApJ, 543, 704.
Debattista, V.P. & Shen, J. 2007, ApJL, 654, 127.
de Jong, R.S. 1996, A&AS, 118, 557.
Del Rio, M.S. & Cepa, J. 1998, A&A, 340, 1.
Downes, D., Reynaud, D., Solomon, P.M. & Radford, S.J.E. 1996, ApJ, 461, 186.
Duval, M.F. & Monnet, G. 1985, A&AS, 61, 141.
Duval, M.F. & Monnet, G., Boulesteix, J. et al. 1991, A&A, 241,375.
Egusa, F., Kohno, K., Sofue, Y., Nakanishi, H. & Komugi, S. 2009, ApJ, 697, 1870.
Elmegren, B. G. & Elmegren, D.M. 1985, ApJ, 288, 438.
Elmegren, B. G., Wilcots, E. & Pisano, C.J. 1998, ApJ, 494,L37.
England, M., Gottesman, S.T. & Hunter, J.H. Jr. 1990, ApJ, 348, 456.
Englmeier, P. & Shlosman, I: 2004, ApJL, 617, L115.
Epinat, B., Amram, P., Marcelin, M. et al. 2008, MNRAS, 388, 500.
Erwin, P. 2004, A&A, 415, 941.
Erwin, P., Pohlen, M. & Beckman, J.E. 2008, AJ, 135, 20.
Fathi, K., Beckman, J.E:, Piñol-Ferrer, N., et al. 2009, ApJ, 704, 1657.
Font,J., Beckman, J.E., Epinat, B., et al. 2011, ApJ Letters, 741, L14.
Foyle, K., Rix, H.-W., Dobbs, C.L., Leroy, A.K. & Walter, F. 2011, ApJ 735, 101.
Freeland, E., Chomiuk, L., Keenan, R. & Nelson, T. 2010, ApJ, 139, 865.
García-Burillo, S., Combes F. & Gerin, M. 1993, A&A, 274, 148.
Garrido O., Marcelin, M., Amram, P. & Boulesteix, J. 2002, A&A, 387,821.
Gerola, H. & Seiden, P.E:, 1978, ApJ 223, 129.
Hernandez, O., Wozniak, H., Carignan, C., et al. 2005, ApJ. 632, 253.
Hernandez, O., Fathi, K., Carignan, C., et al. 2008, PASP. 120, 665.
Hernquist, L. & Weinberg, M. 1992, ApJ 400, 80.
Hoyle, B., Masters, K.L., Nuchol, R.C. et al. 2011, MNRAS, 415, 362.
Huntley, J.M., Sanders, R.H. & Roberts, W.W., Jr. 1978, ApJ, 221, 521.
Huntley, J.M. 1980, ApJ 238, 524.
Kalnajs, A.J. 1978, in IAU symp. No. 77, Structure and properties of nearby galaxies. Ed. E.M. Berkhuijsen & R. Wielebinski (Boston: IAU), 113.
Koda, J. & Sufue, Y. 2006, PASJ, 58, 299.
Kranz, T., Slyz, A. & Rix, H-W. 2003, ApJ, 586,143.
Krajnovic, D., Cappellari, M., de Zeeuw, P.T. & Copin, Y. 2006, MNRAS 366, 1126.
Laine, S., Shlosman, I. & Heller, C.H. 1998, MNRAS, 297, 1052.
Laine, S., Shlosman, I., Knapen, J.H. & Peletier, R.F. 2002, ApJ, 567, 97.
Leon, S., Eckart, A. & Laine, S. 2007, A&A, 473, 747.
Lindblad, B. 1961, Stockholms Obs. Ann: 1961, vol. 21. No. 8.
Lin, C.C. & Shu, F.H. 1964 ApJ Letters, 140, 646.
Marino, R.A., Gil de Paz, A., Castillo-Morales, A., et al. 2012, ApJ, 754,61.
Marinova, I. & Jogee, S. 2007, ApJ, 659, 1176.
Martinez-Valpuesta, I., Shlosman & I., Heller, C. 2006, ApJ 637, 214.

Masset, F. & Tagger, M. 1997 A&A, 322, 442.
Meidt, S.E., Rand, R.J., Merrifield, M.R., Shetty, R., Vogel, S., 2008a ApJ 688. 224.
Meidt, S.E., Rand, R.J., Merrifield, M.R., Debattista, V.P., Shen, J. 2008b, ApJ 676, 899.
Meidt, S.E., Rand, R. & Merrifield, M.R. 2009, ApJ, 702, 277.
Menendez-Delemestre, K., Sheth, K., Schinnerer, E., Jarrett, T. H. & Scoville, N. Z. 2007, ApJ, 657, 790.
Merrifield, M. & Kuijken, K. 1995, MNRAS, 274, 933.
Patsis, P., Hiotelis, N. Contopoulos, G., & Grosbol, P. 1994, A&A, 286,46.
Pérez, I. 2008, A&A, 478, 717.
Puerari, I. & Dottori, H. 1997, ApJ, 476, L73.
Quillen, A.C., Frogel, J.A., Kenney, J.P.D., Pogge, R.W. & Depoy, D.L. 1995, ApJ, 441, 549.
Rand, R.J. & Wallin, J.F. 2004, ApJ, 614, 142.
Rautianinen,P. & Salo, H. 1999, A&A, 348, 737.
Rautianinen,P. & Salo, H. 2002, A&A, 362, 465.
Rautiainen, P., Salo, H. & Laurikainen, E. 2005, ApJ Letters, 631, L129.
Rautiainen, P., Salo, H. & Laurikainen, E. 2008, MNRAS, 338, 1803.
Regan, M.W., Teuben, P.J., Vogel, S.N. & van der Hults, T. 1996, AJ, 112, 2549.
Reynaud, D. & Downes, D. 1998, A&A, 337, 671.
Sanders, R.H. & Huntley, J.M. 1978, ApJ, 209, 53.
Sanders, R.H. & Tubbs, A.D. 1980, ApJ 235, 803.
Schinnerer, E., Maciejewski, W., Scoville, N. & Moustakas, L.A. 2002, 575, 826.
Sempere, M.J., Combes, F. & Casoli, F. 1995, A&A, 299, 371.
Sellwood, J.A. 1981, A&A, 99, 362.
Sellwood, J.A. & Sparke, L.S. 1988, MNRAS, 231, 25.
Sellwood, J.A. & Debattista, V.P. 2006, ApJ, 639, 868.
Sempere, M.J., García-Burillo, S., Combes, F. & Knapen, J.H. 1995 A&A, 296, 45.
Shen, J. & Debattista, V.P. 2009 ApJ, 690, 758.
Sheth, K., Regan, M., Hinz, J.L. et al. 2010, PASP, 122, 1397.
Speltincx, T, Laurikainen, E. & Salo, H. 2008, MNRAS, 383, 317.
Sygnet, J.F., Tagger, M., Athanassoula, E., & Pellat, R. 1988, MNRAS, 232, 733.
Tagger, M., Sygnet, J.F., Athanassoula, E., & Pellat, R. 1987, ApJ Letters, 318, L43.
Teuben, P.J. & Sanders, R.H. 1985, MNRAS, 212, 257.
Tinsley, B.M. 1981, MNRAS, 194,63.
Toomre, A. 1969, ApJ 158, 899.
Tremaine, S. & Ostriker, J.P. 1999, MNRAS, 306, 662.
Tremaine, S. & Weinberg, M. 1984, ApJ Letters, 282 L5.
van Albada, T.S. & Sanders, R.H. 1982, MNRAS, 201, 303.
Vera-Villamizar, N., Puerari, I. & Dottori, H. 2003, RevMexAA, 17, 201.
Villa-Vargas, J., Shlosman, I. & Heller, C. 2010, ApJ 719, 1470.
Wada, K. 1994, PASJ, 46, 165.
Weinberg, M. 1985, MNRAS 213, 451.
Weiner, B.J., Williams, T.B., van Gorkom, J.H. & Sellwood, J.A. 2001, ApJ. 546, 916.
Weiner, B.J., Sellwood, J.A. & Williams, T.B. 2001, ApJ 546, 931.
Wilke, K., Mollenhoff, C. & Mathias, M. 1999, A&A, 344, 787.
Wilke, K., Mollenhoff, C. & Mathias, M. 2000, A&A, 361, 507.
Zimmer, P., Rand, R.J. & McGraw, J.T. 2004, ApJ, 607, 285.

ONLINE MATERIAL

Figure 9

**UGC508**

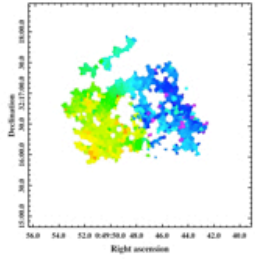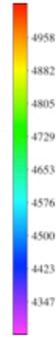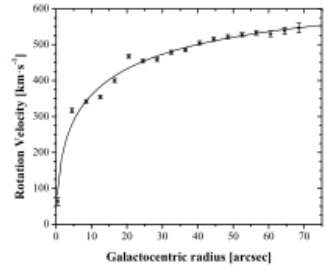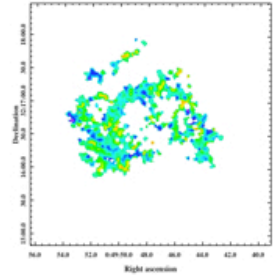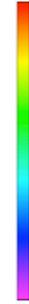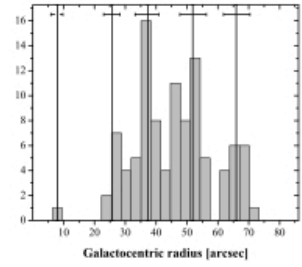

**UGC763**

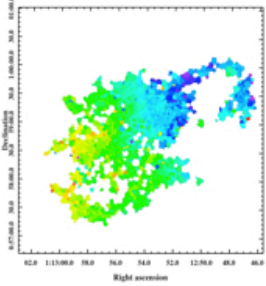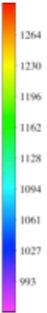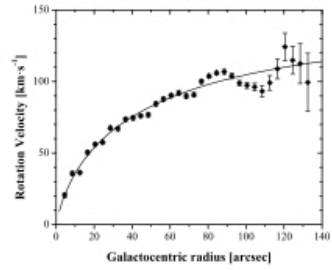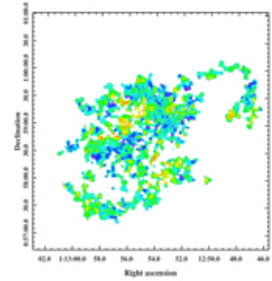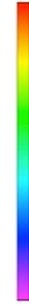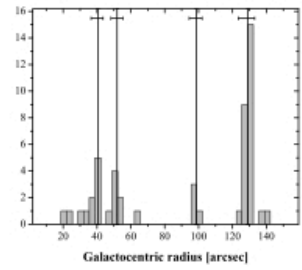

**UGC1256**

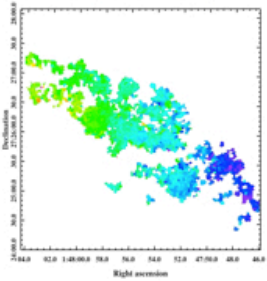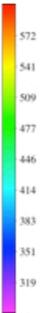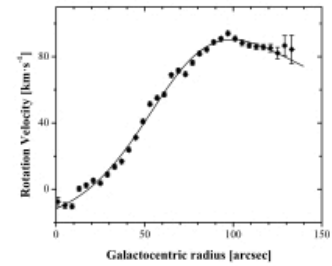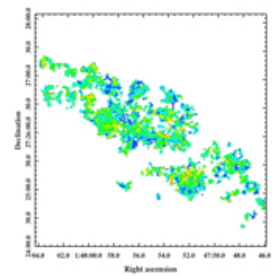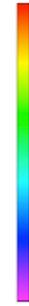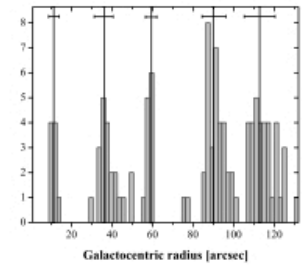

**UGC1317**

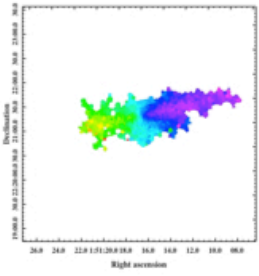 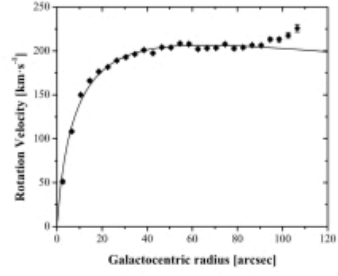 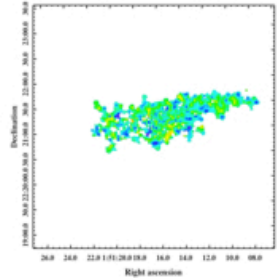 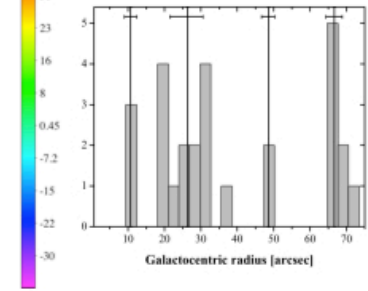

**UGC1437**

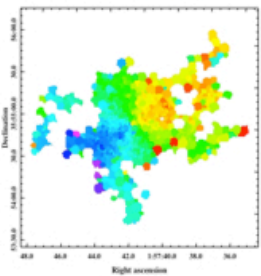 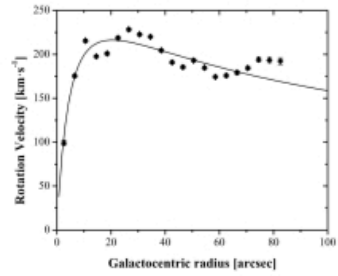 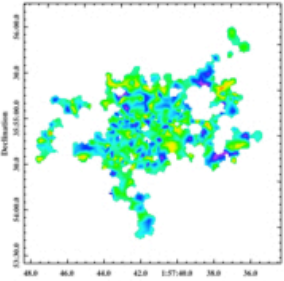 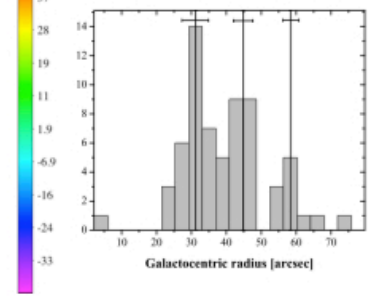

**UGC1736**

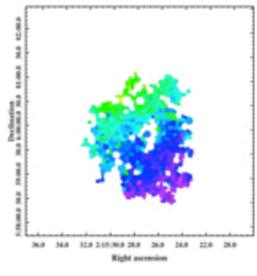 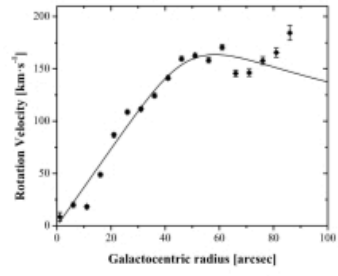 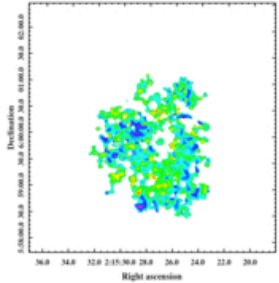 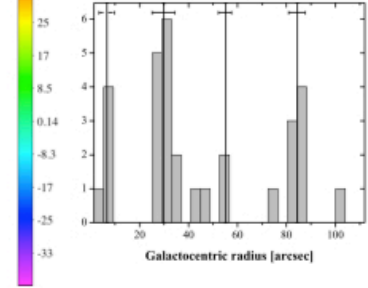

## UGC1913

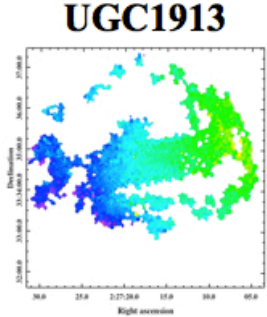 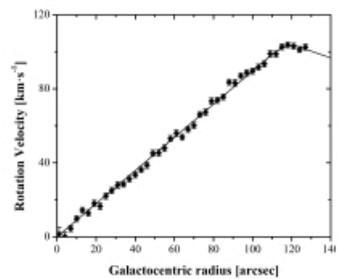 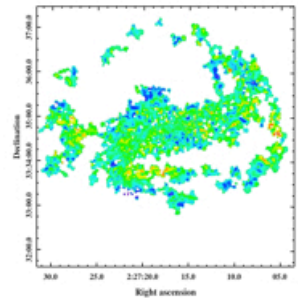 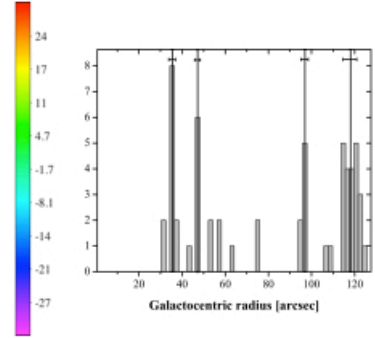

## UGC2080

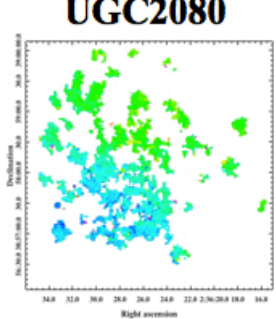 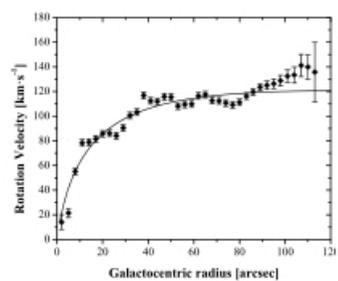 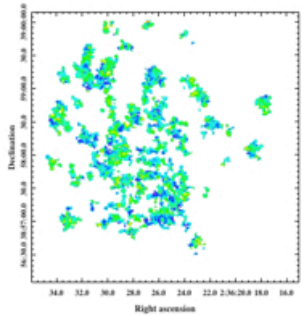 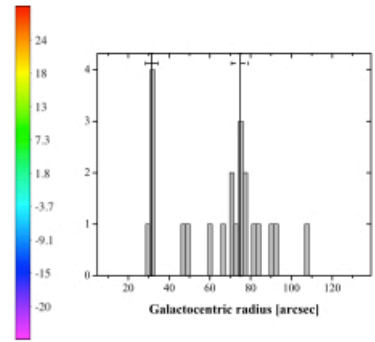

## UGC2141

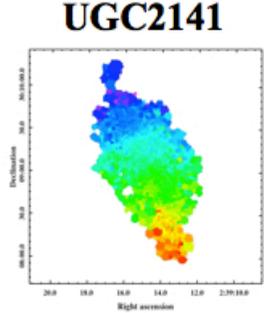 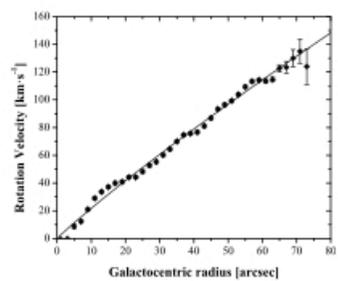 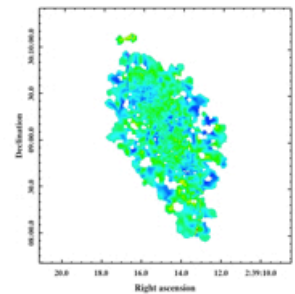 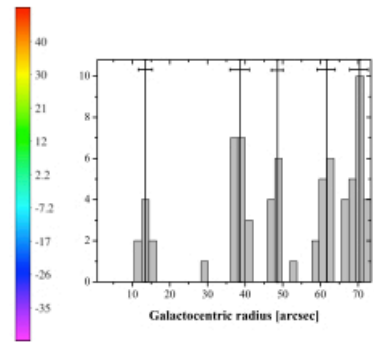

**UGC2193**

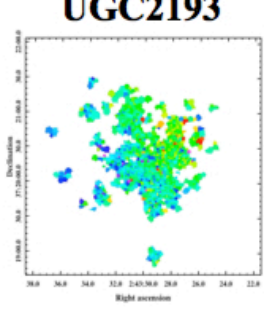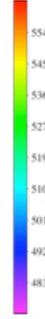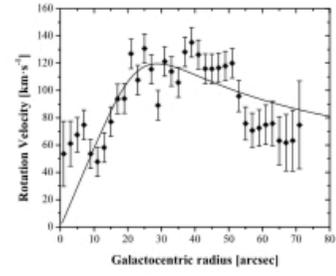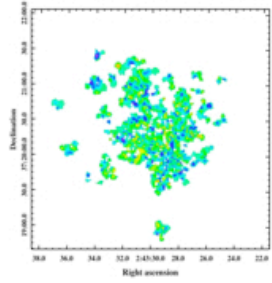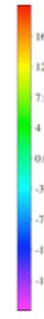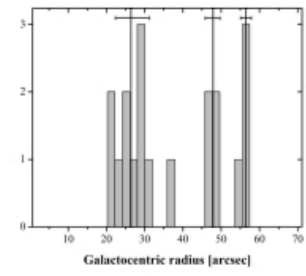

**UGC2855**

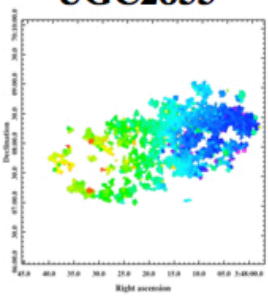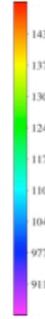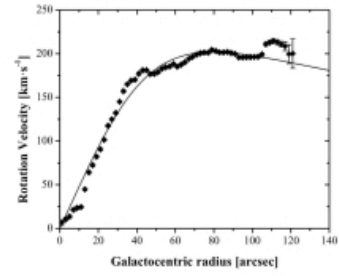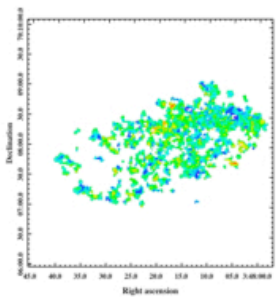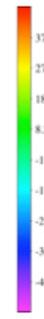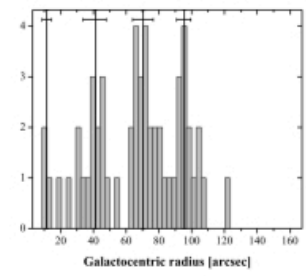

**UGC3013**

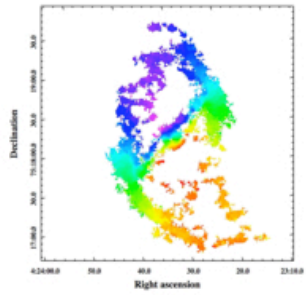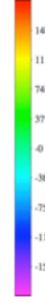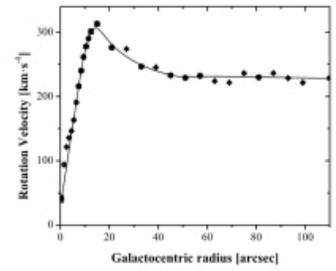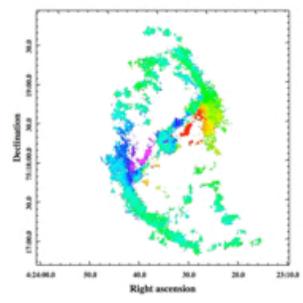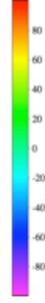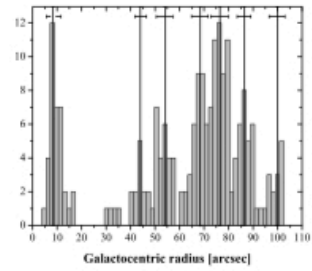

**UGC3273**

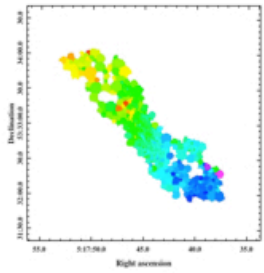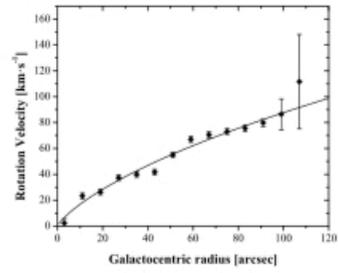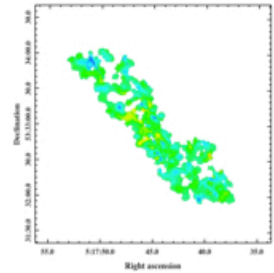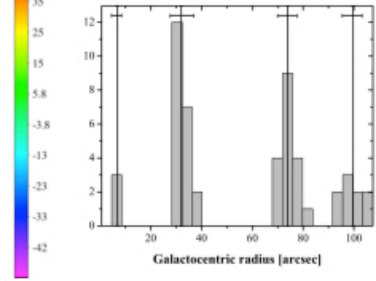

**UGC3463**

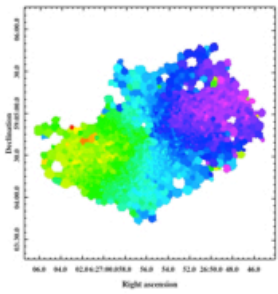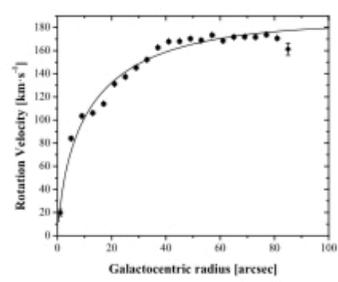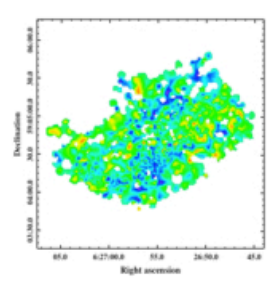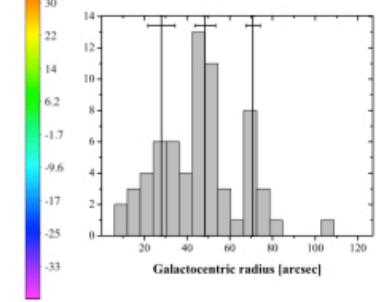

**UGC3574**

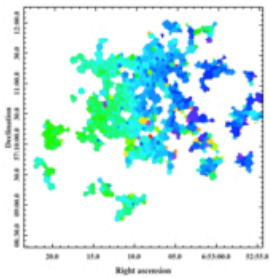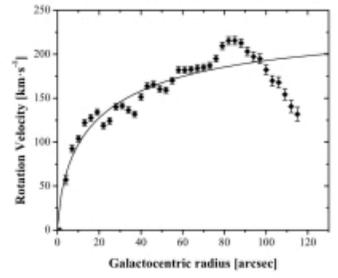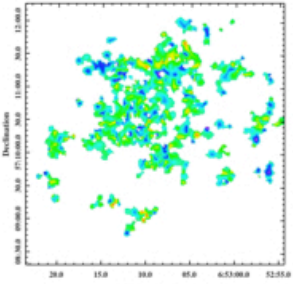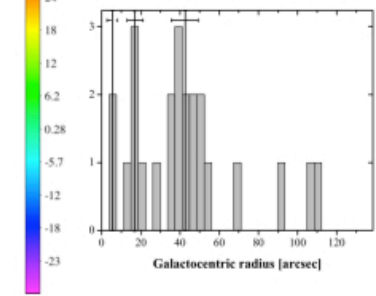

**UGC3685**

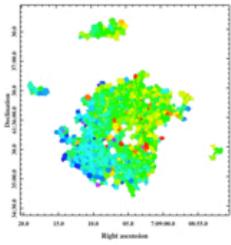 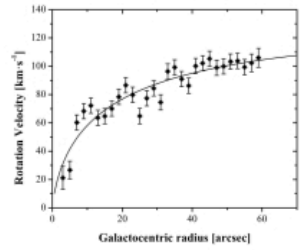 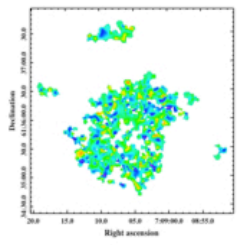 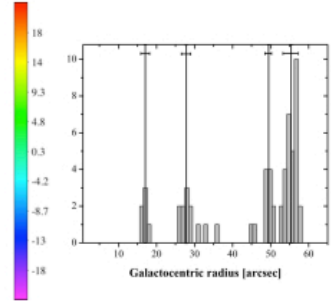

**UGC3691**

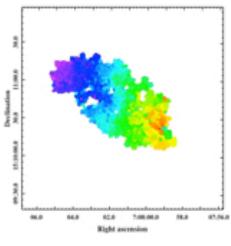 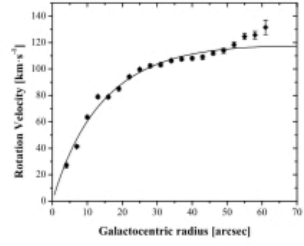 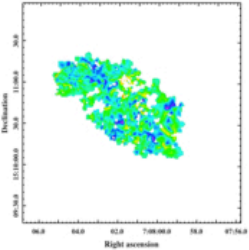 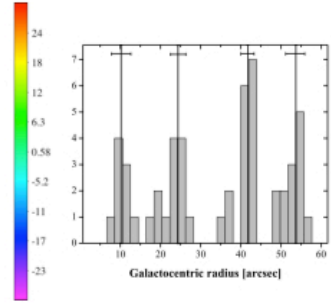

**UGC3709**

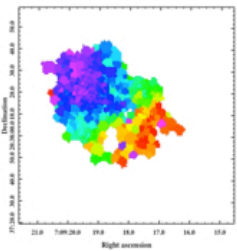 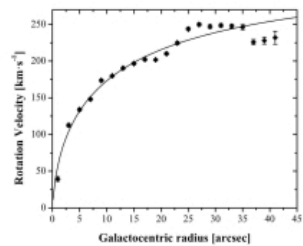 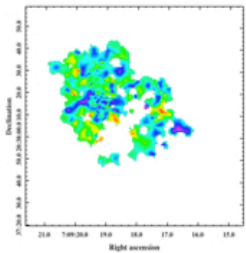 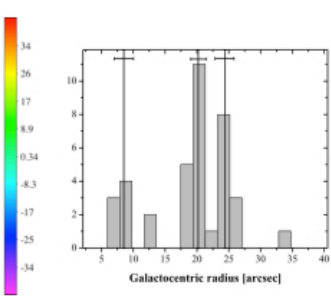

**UGC3734**

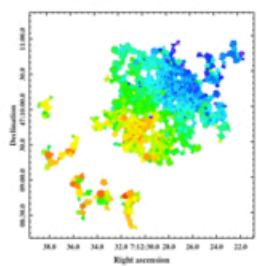 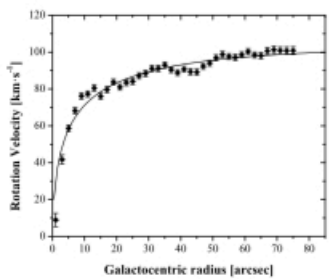 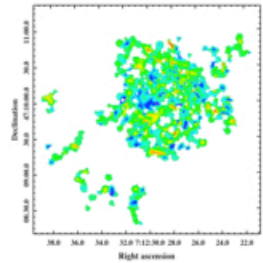 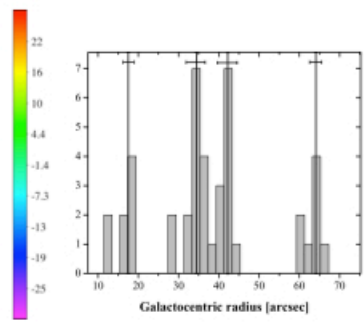

**UGC3740**

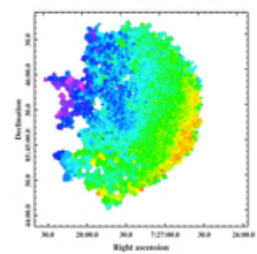 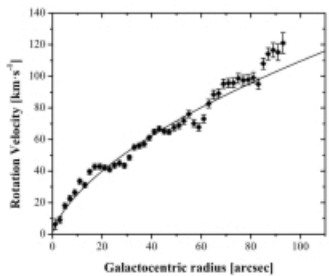 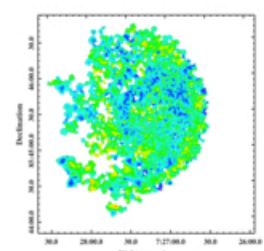 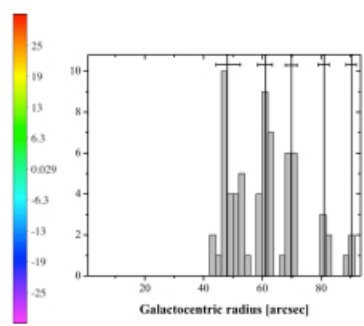

**UGC3809**

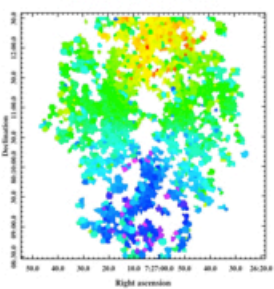 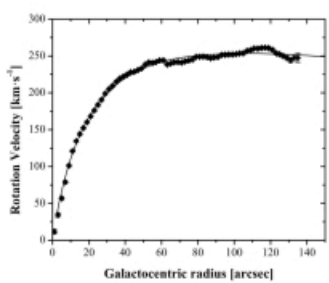 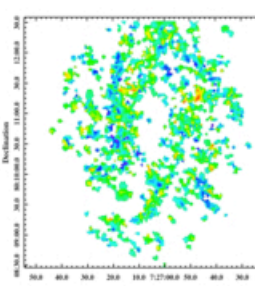 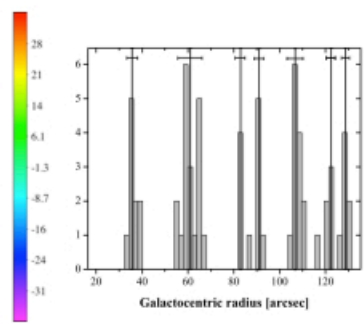

**UGC3826**

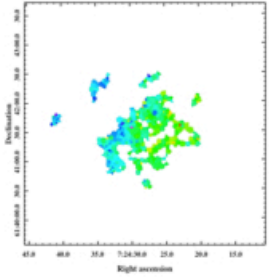 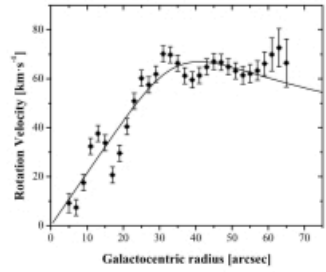 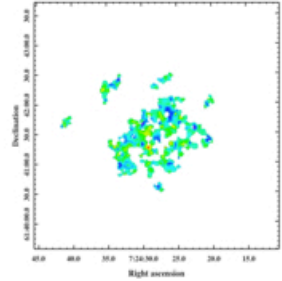 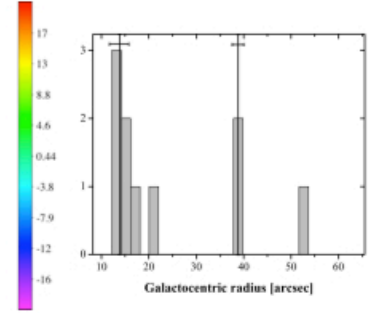

**UGC3876**

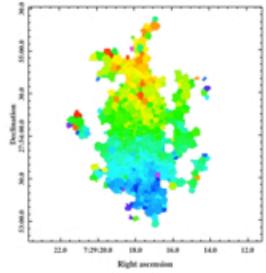 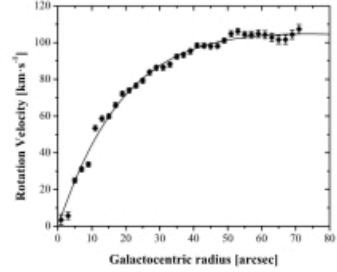 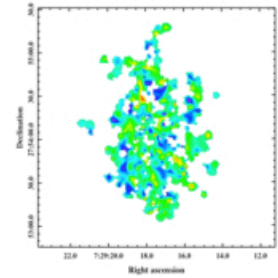 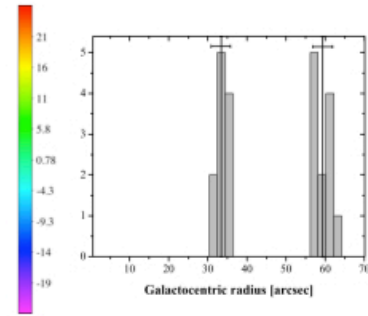

**UGC3915**

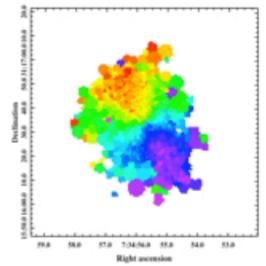 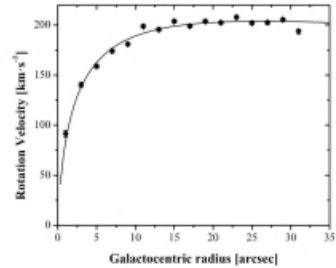 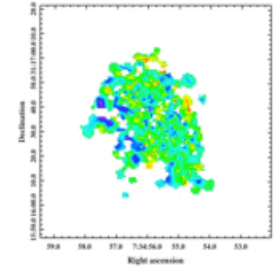 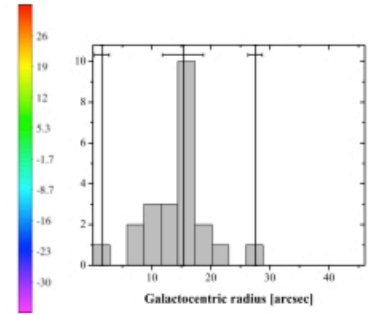

**UGC4165**

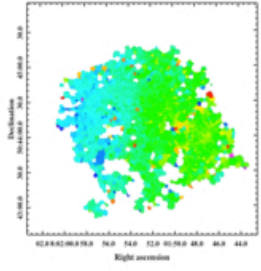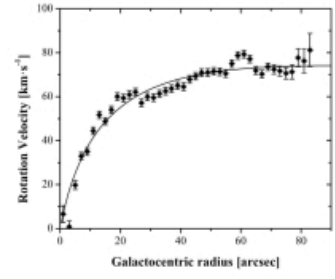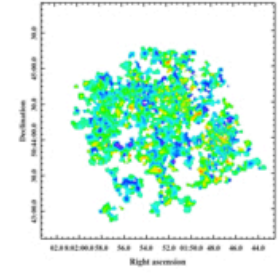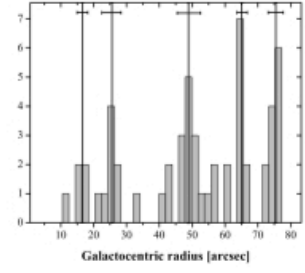

**UGC4273**

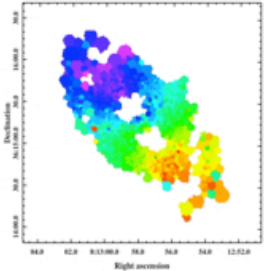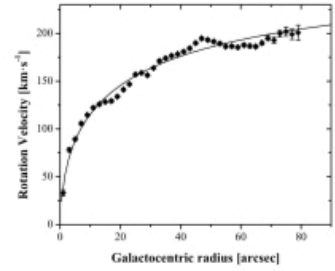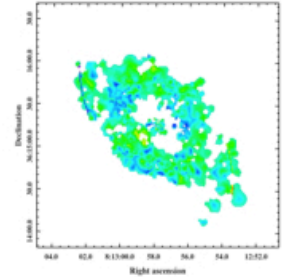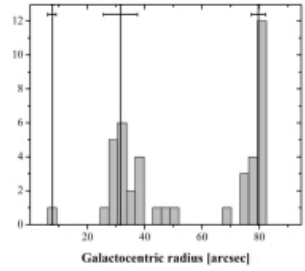

**UGC4284**

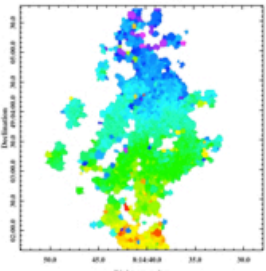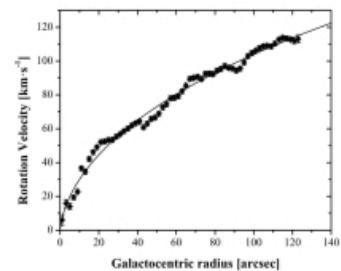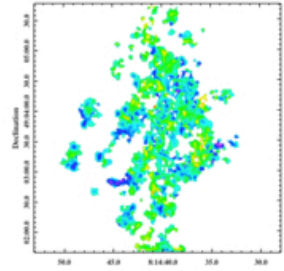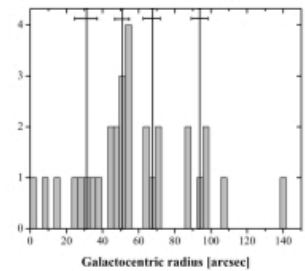

**UGC4325**

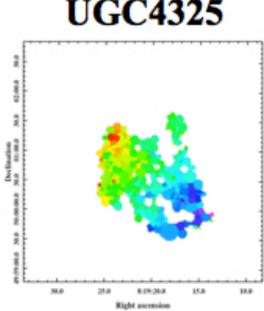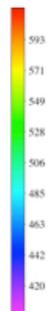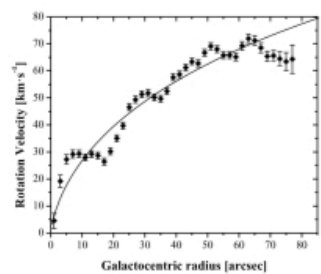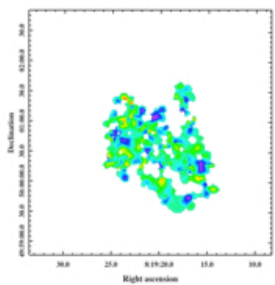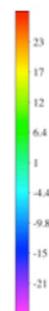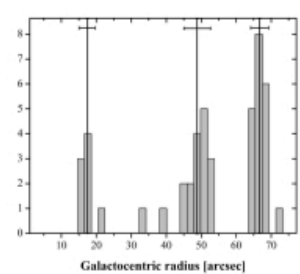

**UGC4422**

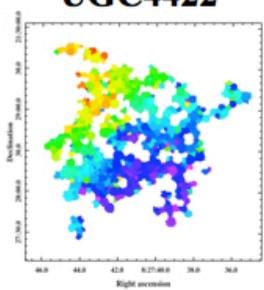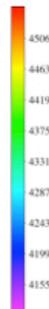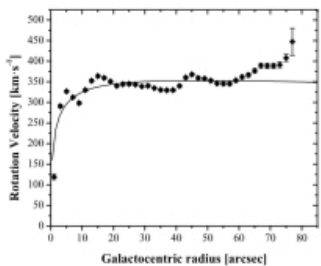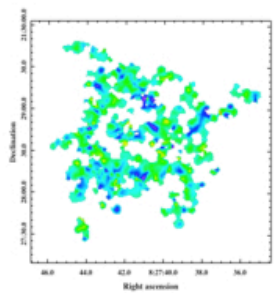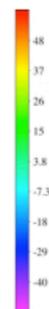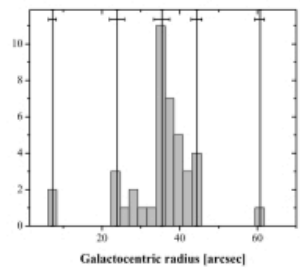

**UGC4555**

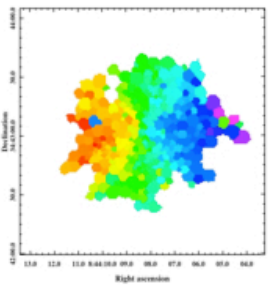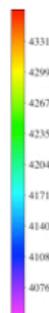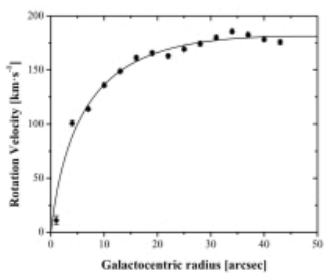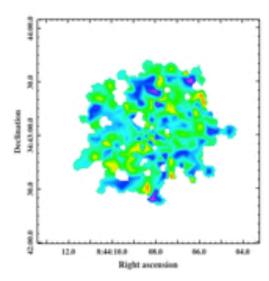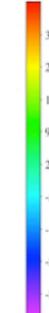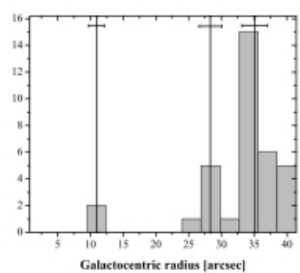

## UGC4936

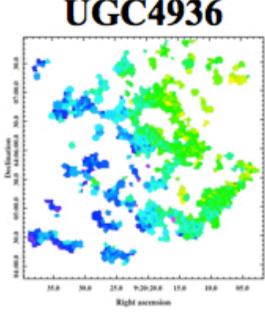 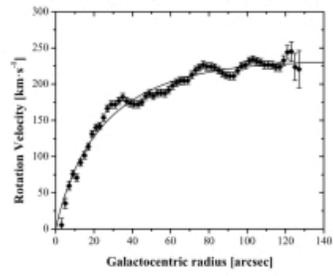 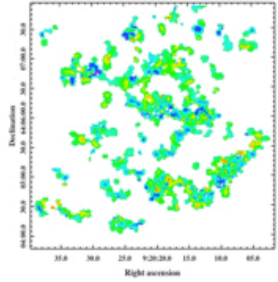 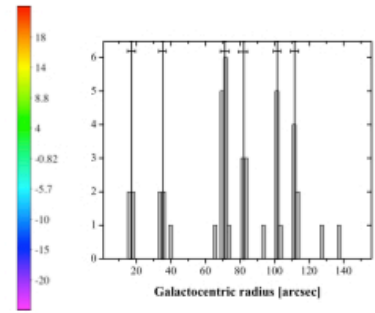

## UGC5175

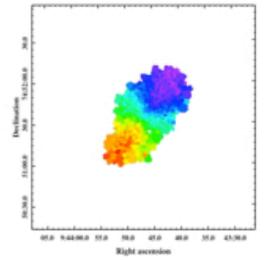 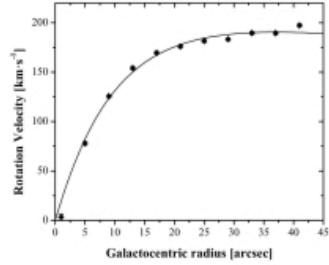 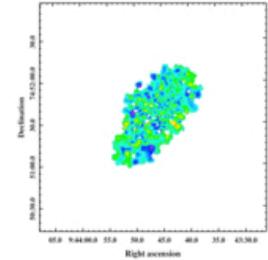 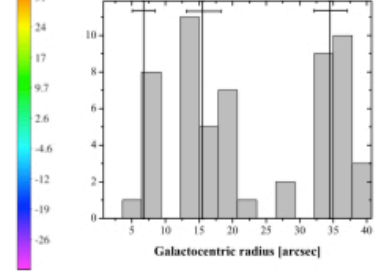

## UGC5228

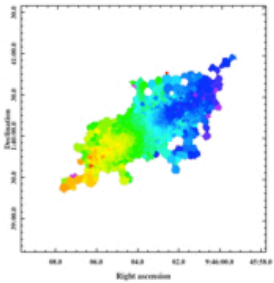 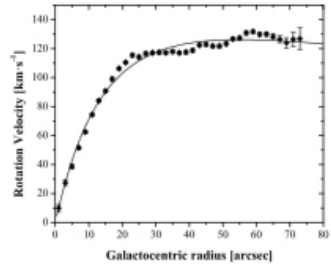 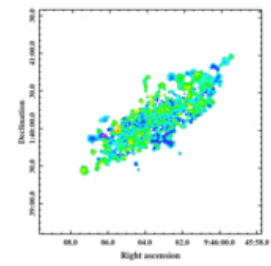 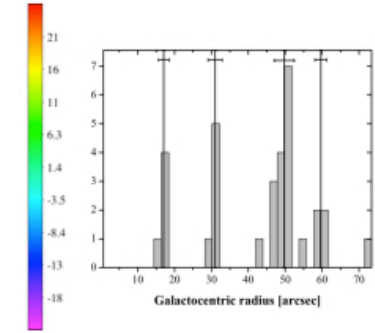

**UGC5251**

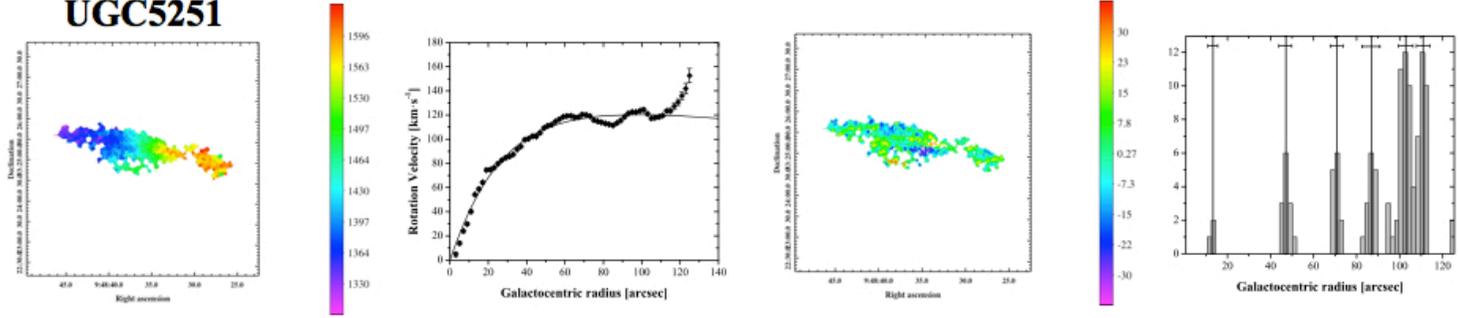

**UGC5253**

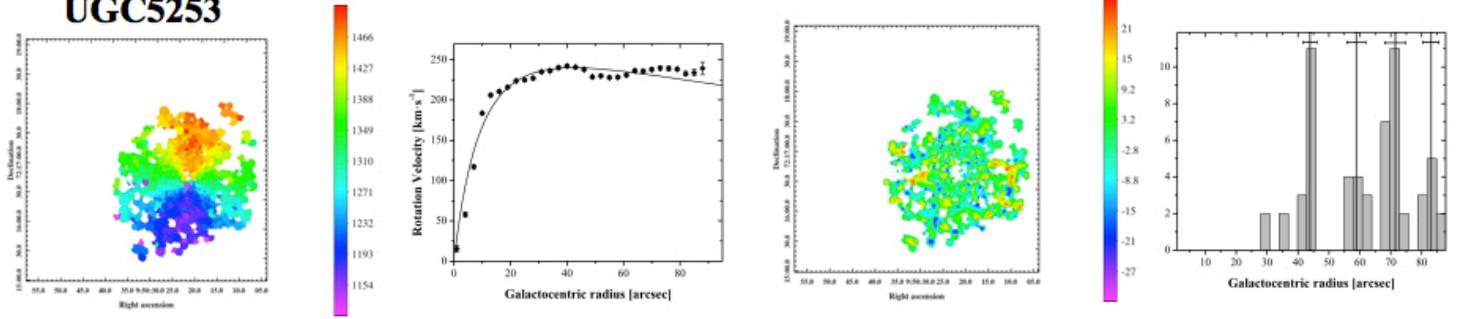

**UGC5303**

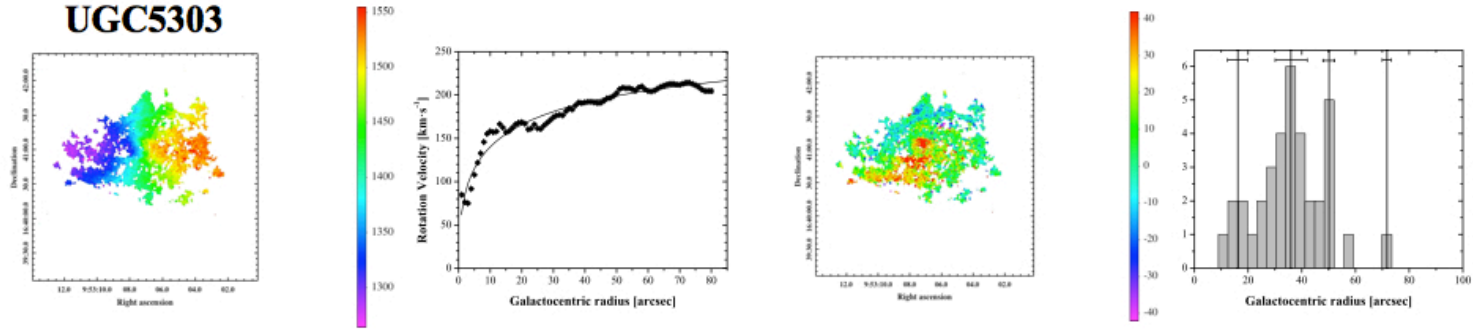

**UGC5319**

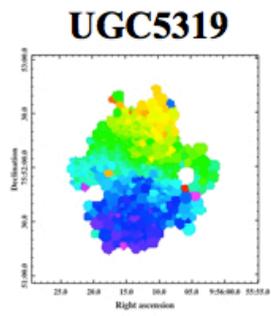 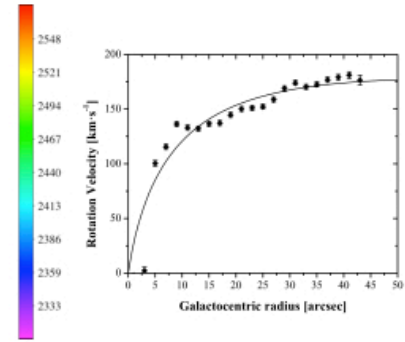 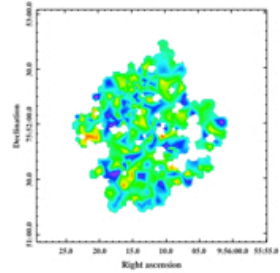 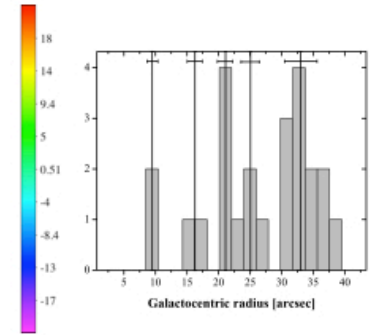

**UGC5414**

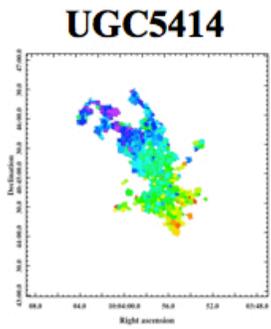 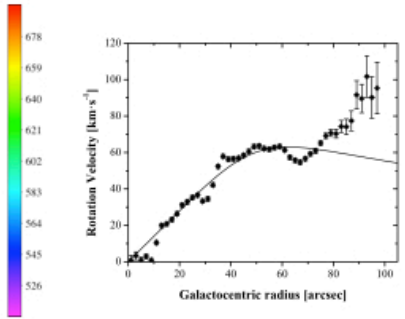 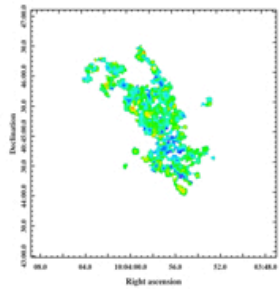 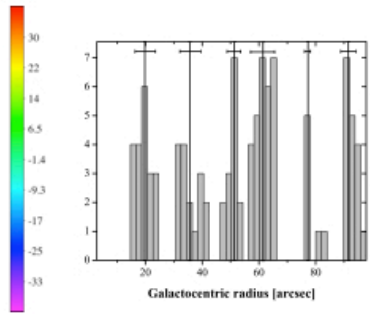

**UGC5510**

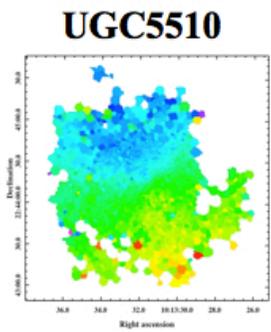 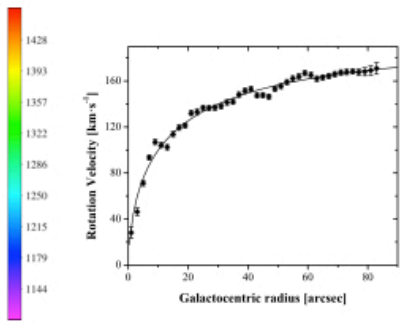 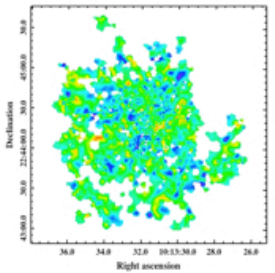 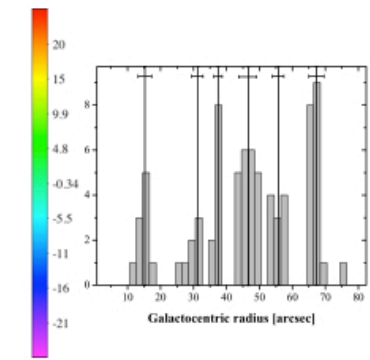

**UGC5532**

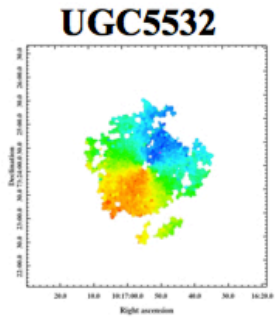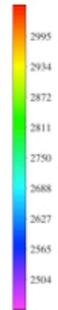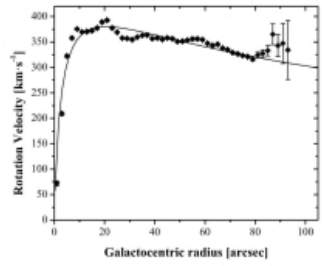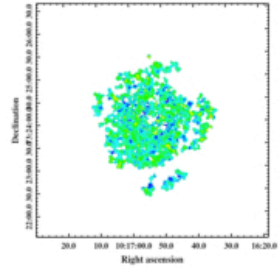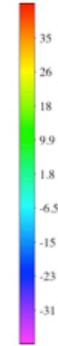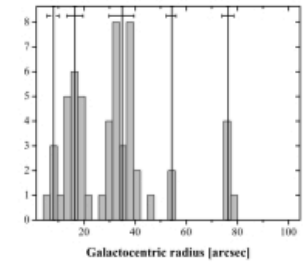

**UGC5786**

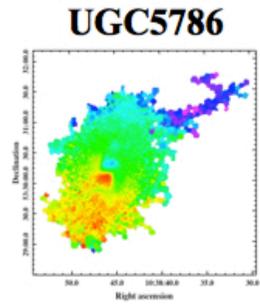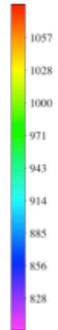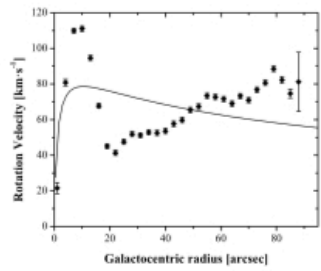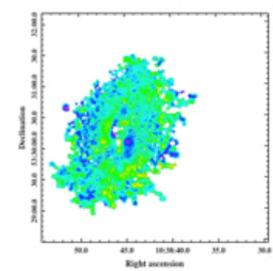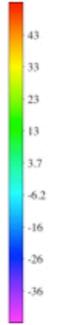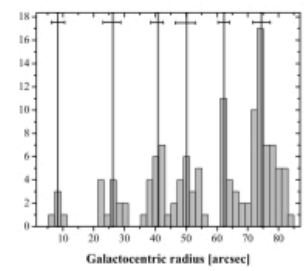

**UGC5840**

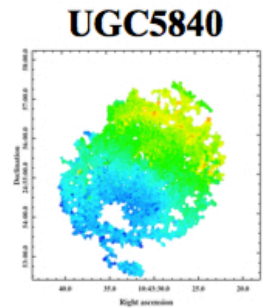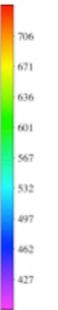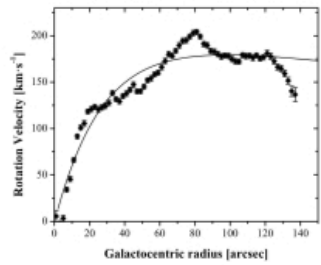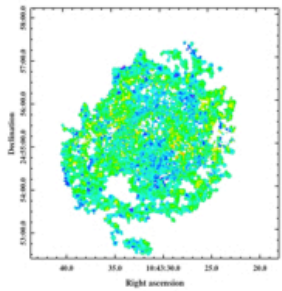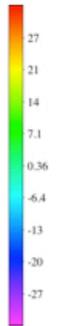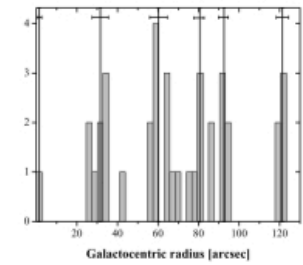

## UGC5842

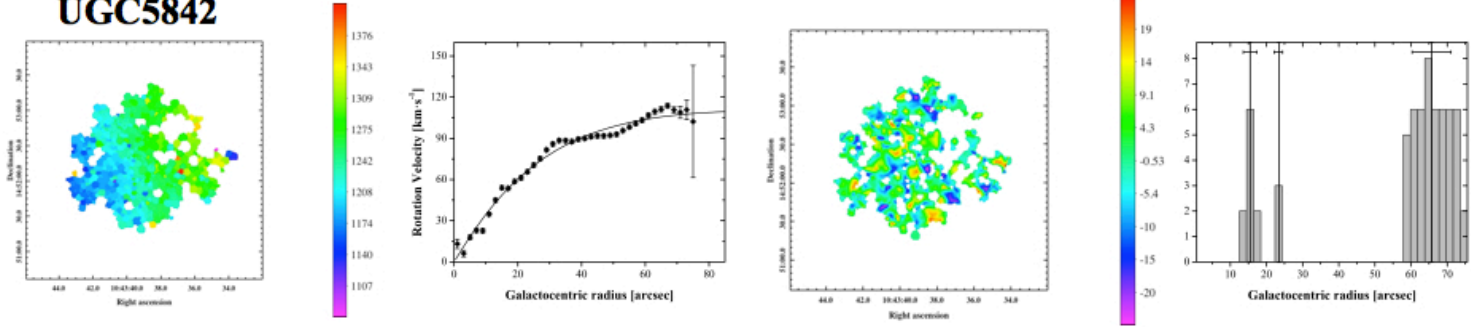

## UGC5982

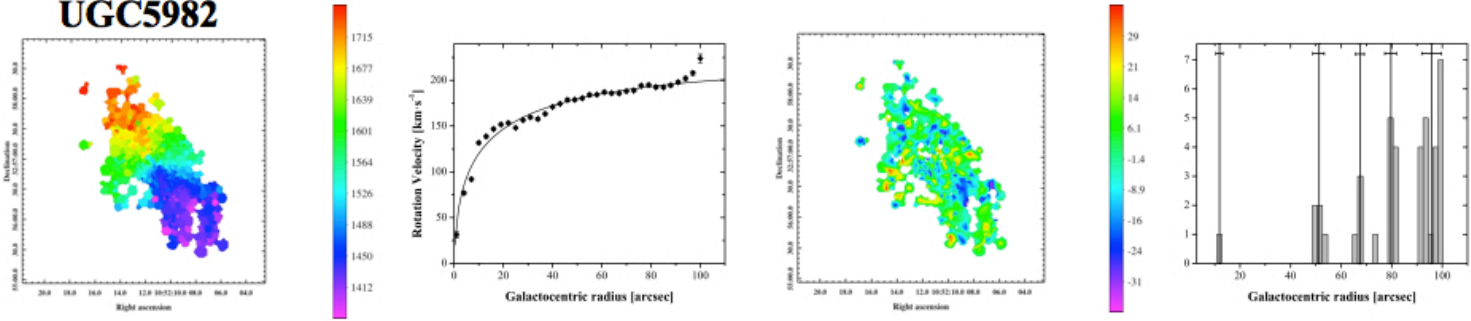

## UGC6118

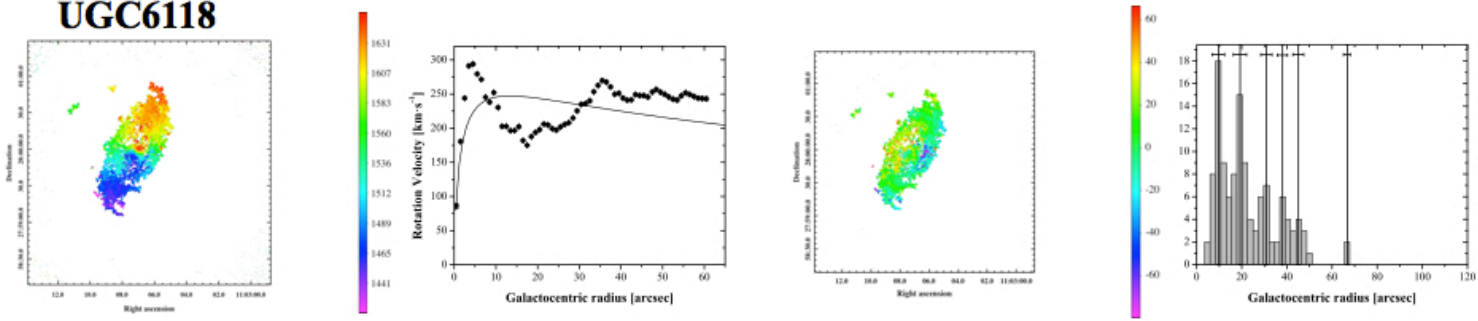

## UGC6277

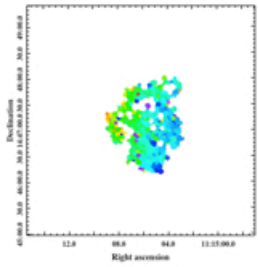 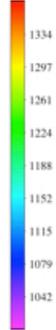 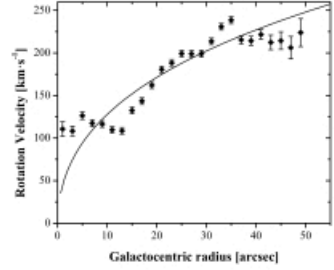 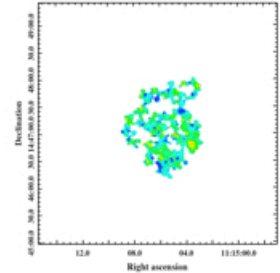 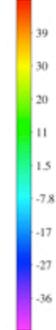 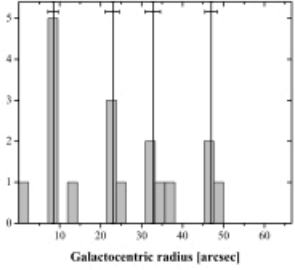

## UGC6521

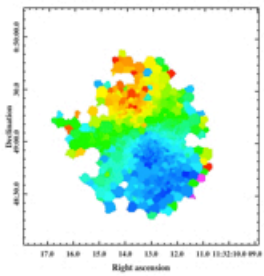 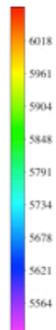 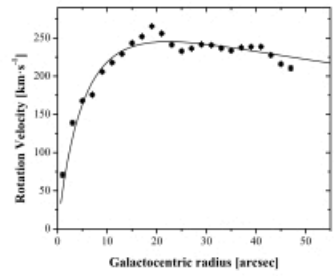 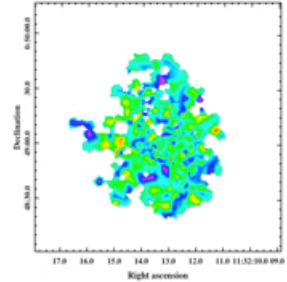 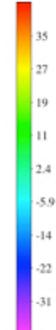 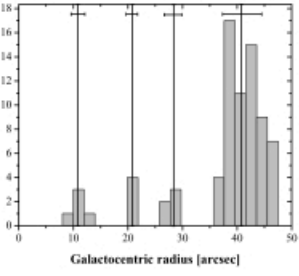

## UGC6523

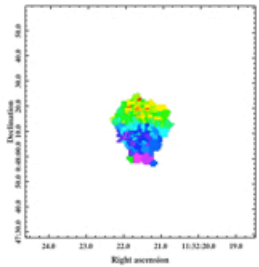 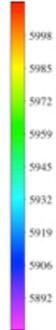 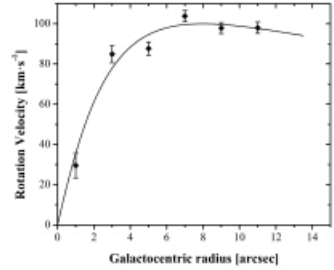 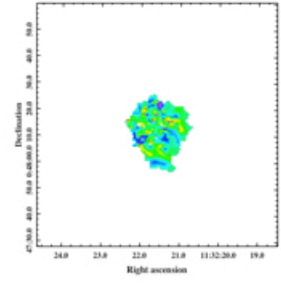 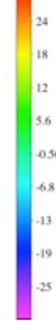 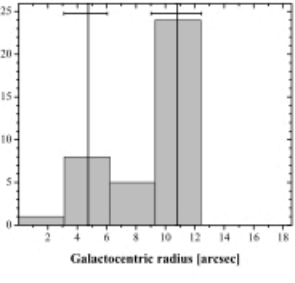

**UGC6537**

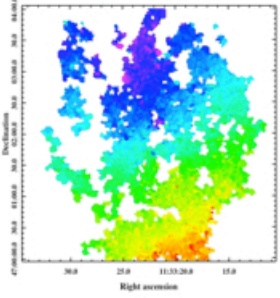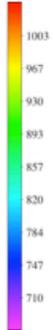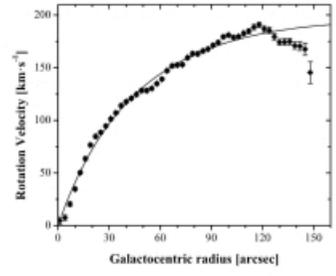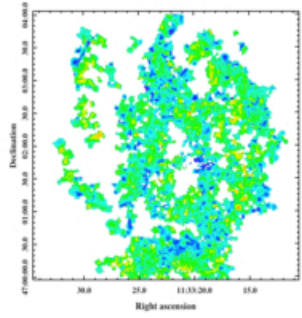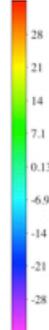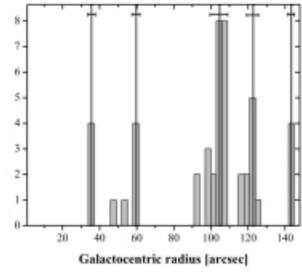

**UGC6702**

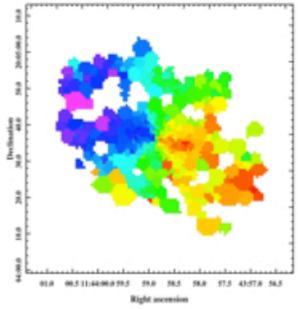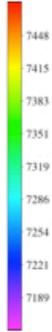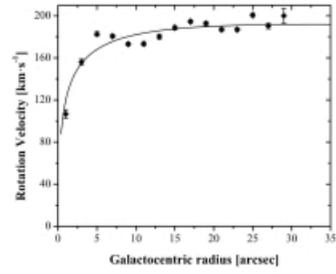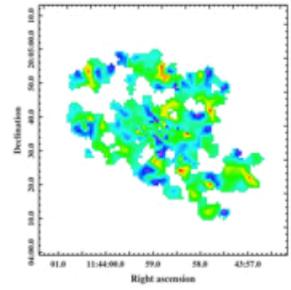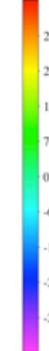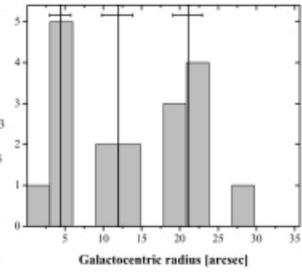

**UGC6778**

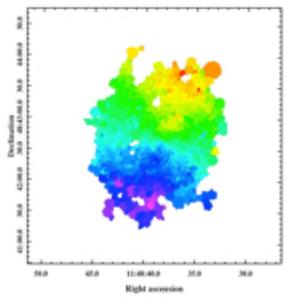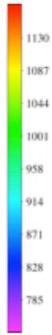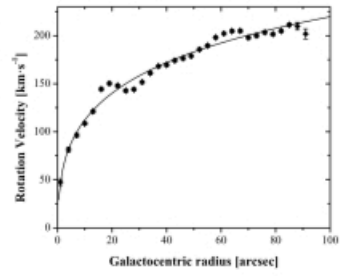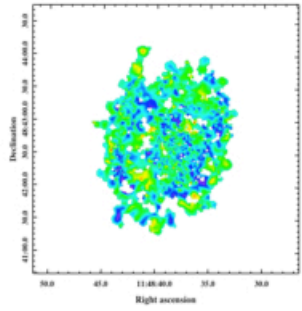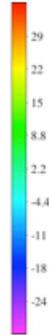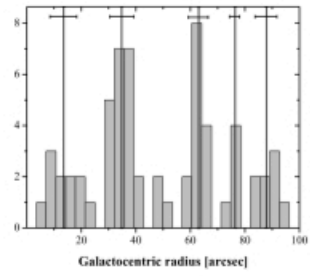

**UGC7021**

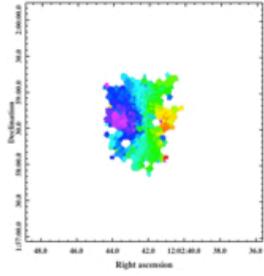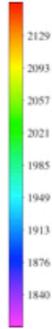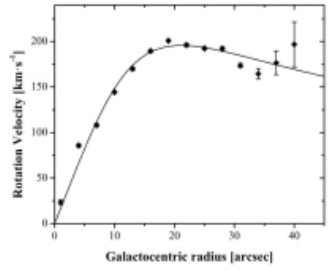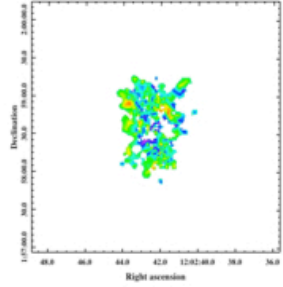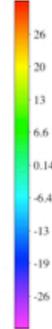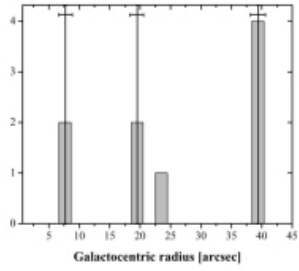

**UGC7045**

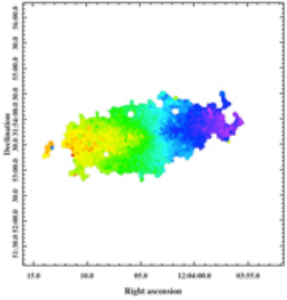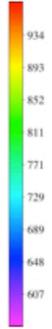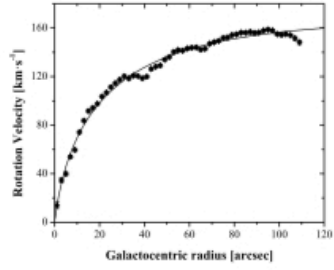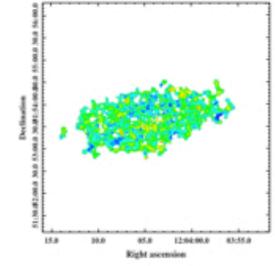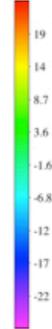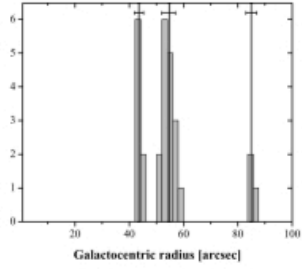

**UGC7154**

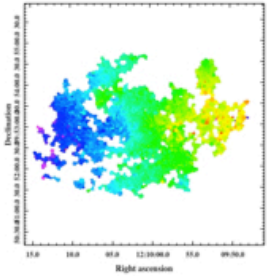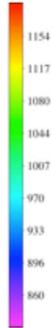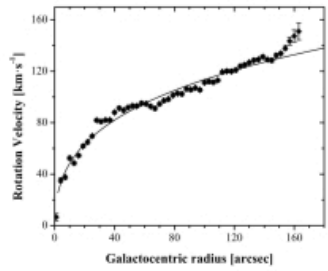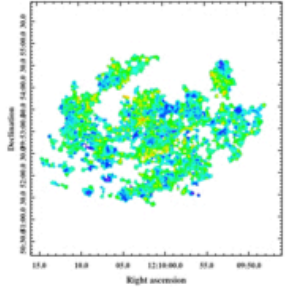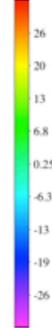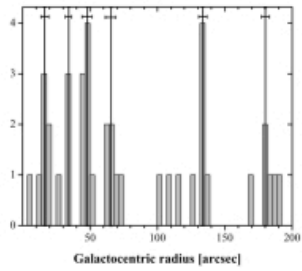

**UGC7323**

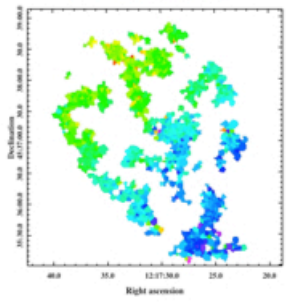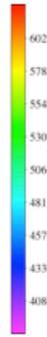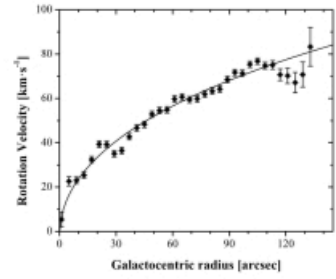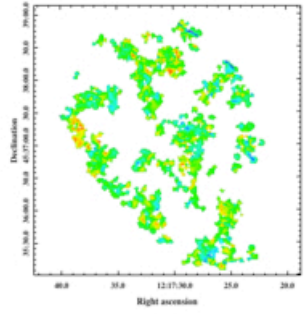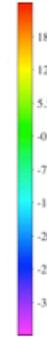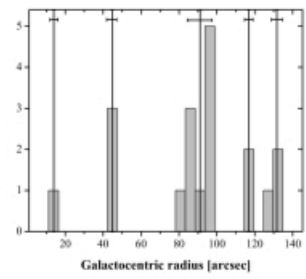

**UGC7420**

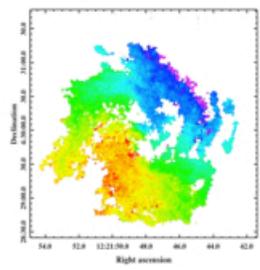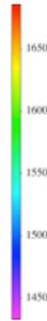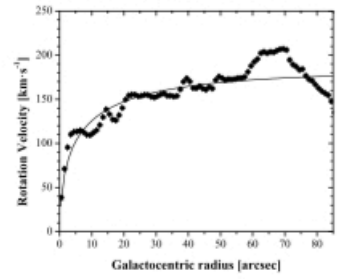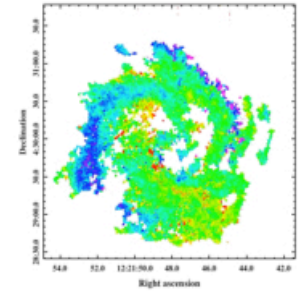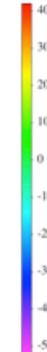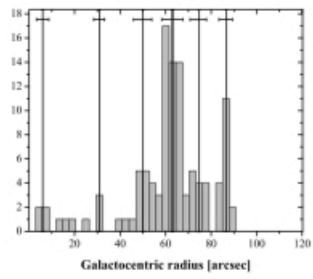

**UGC7766**

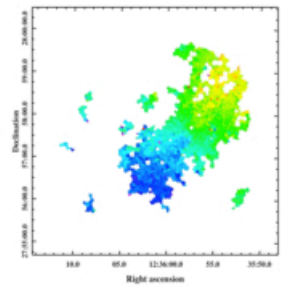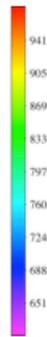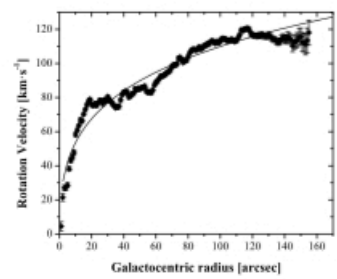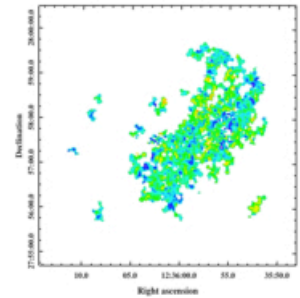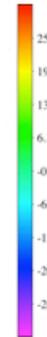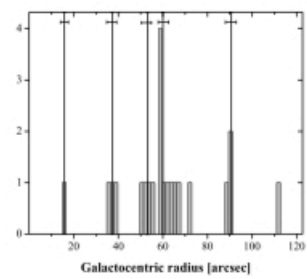

## UGC7831

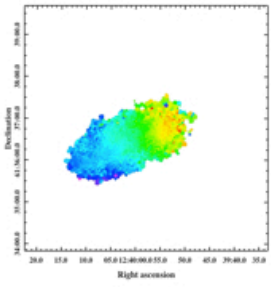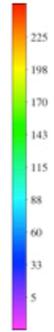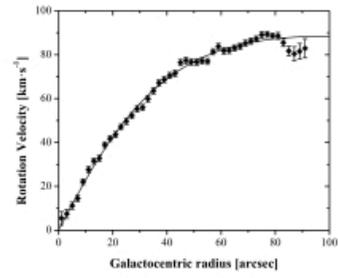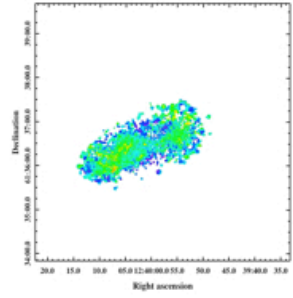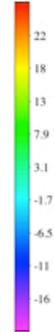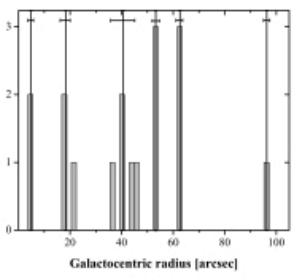

## UGC7853

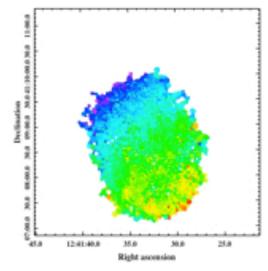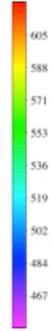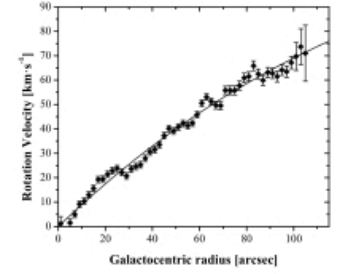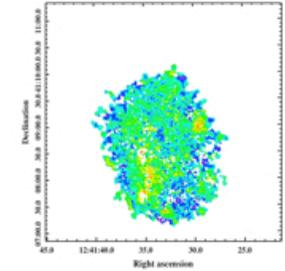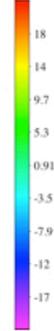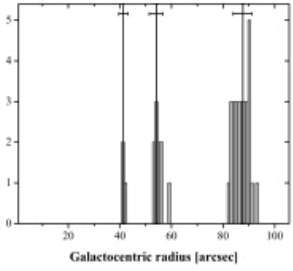

## UGC7861

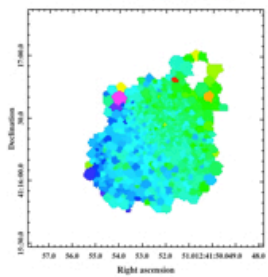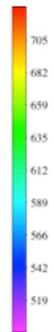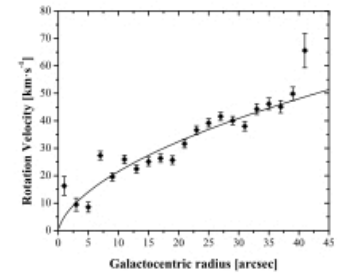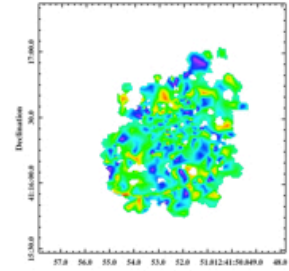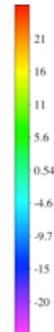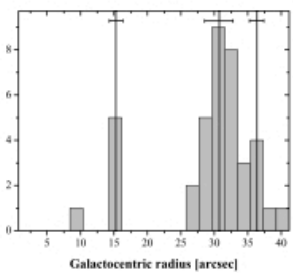

**UGC7876**

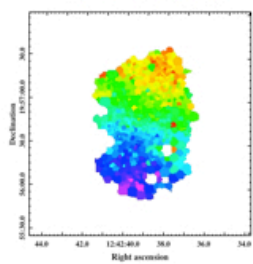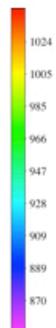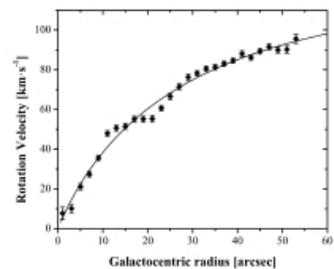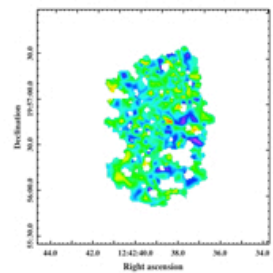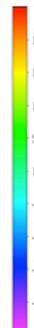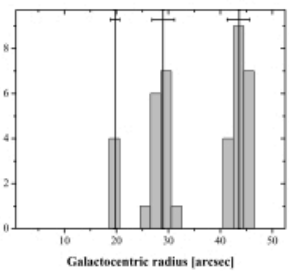

**UGC7901**

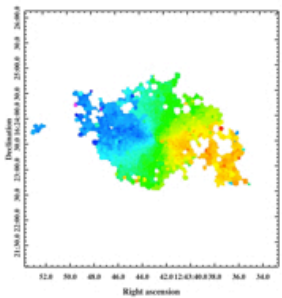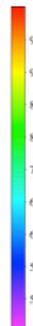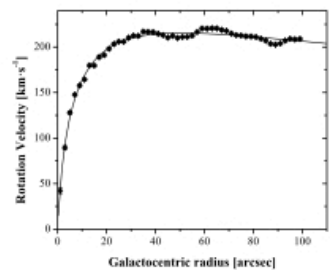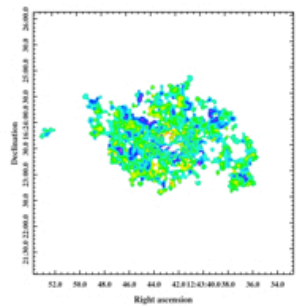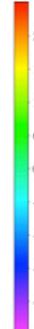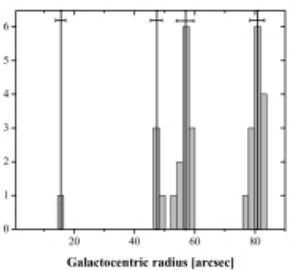

**UGC7985**

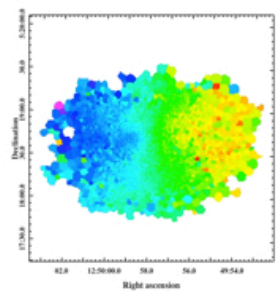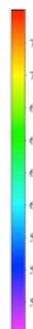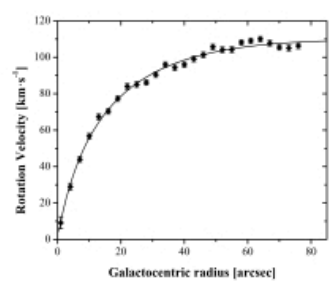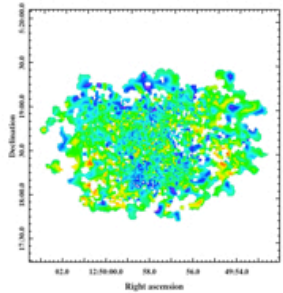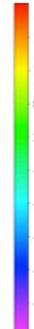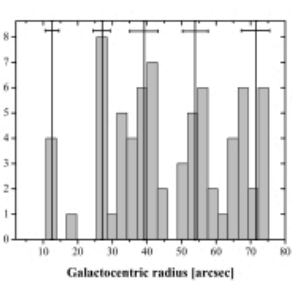

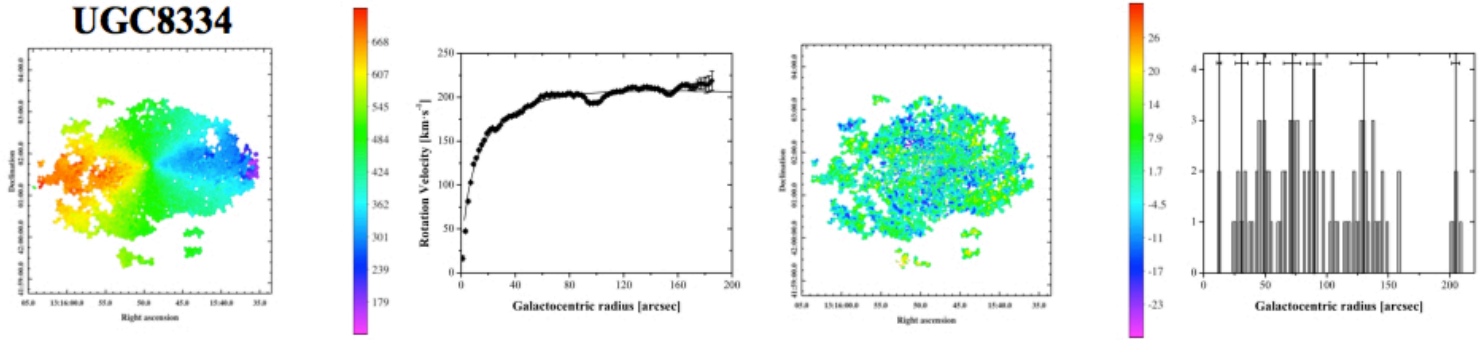
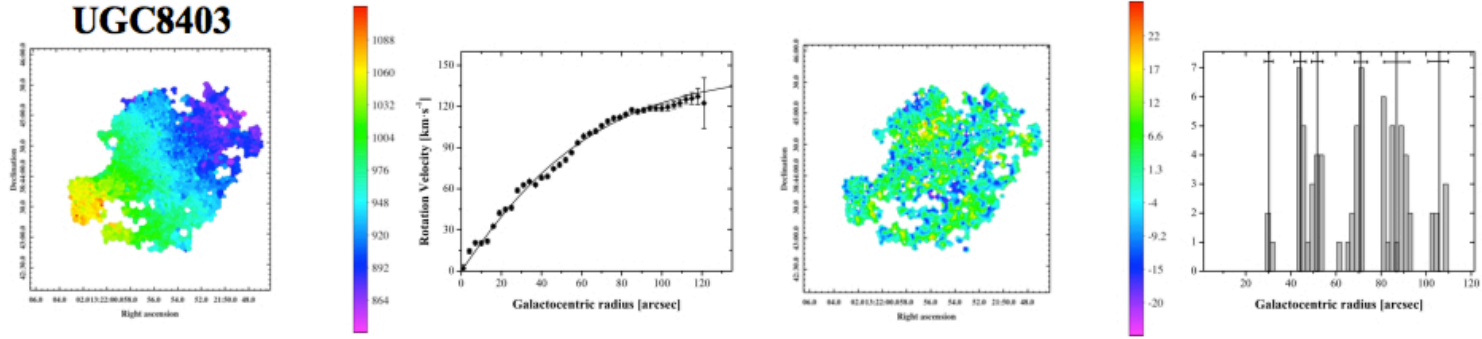
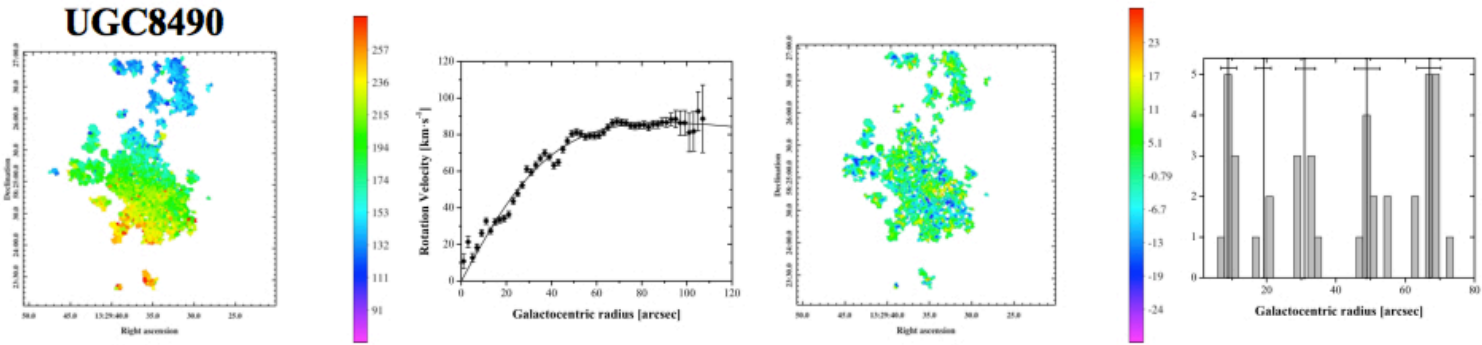

**UGC8709**

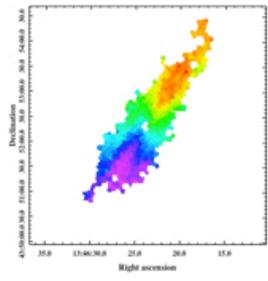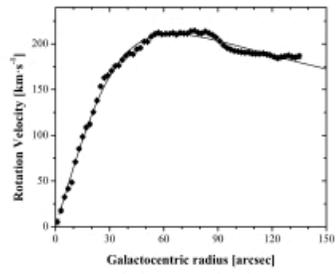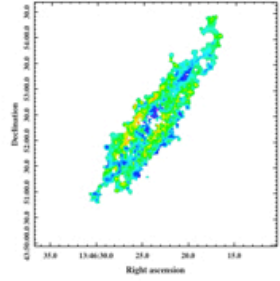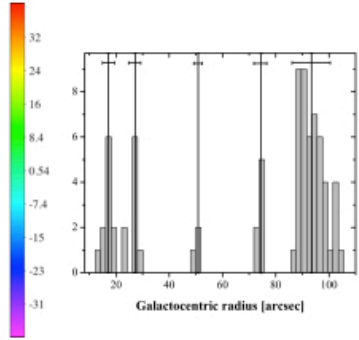

**UGC8852**

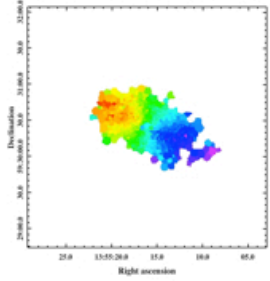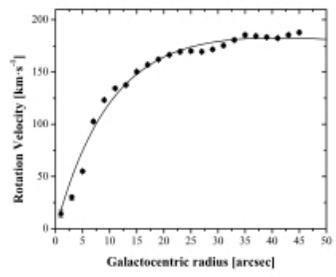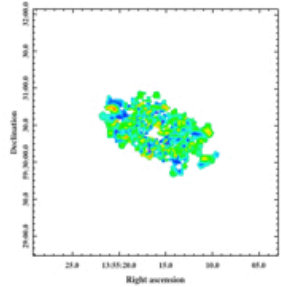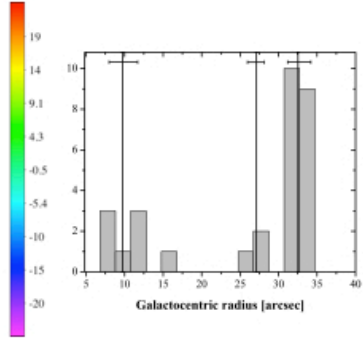

**UGC8937**

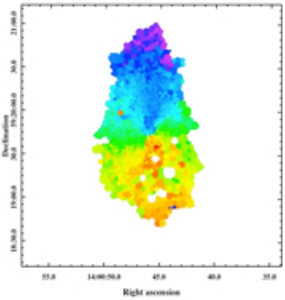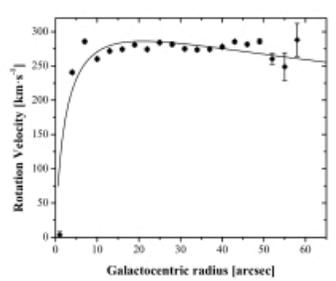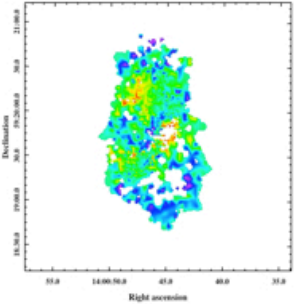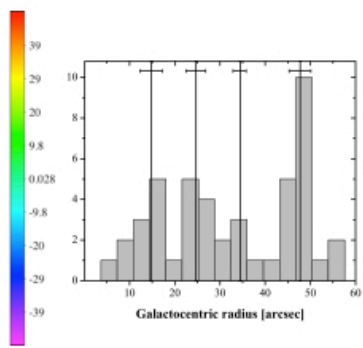

**UGC9179**

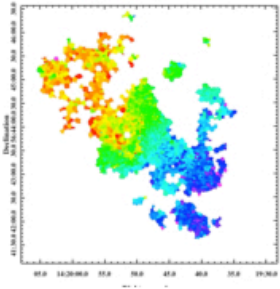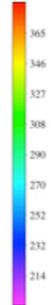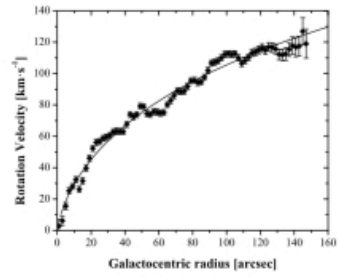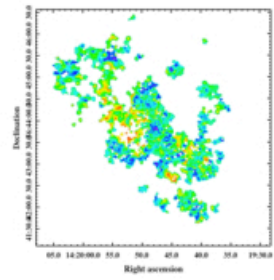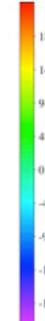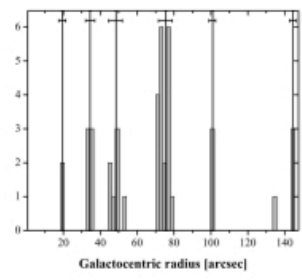

**UGC9248**

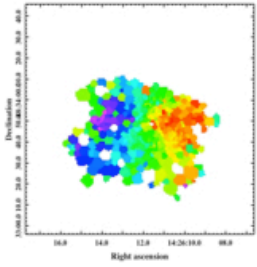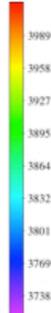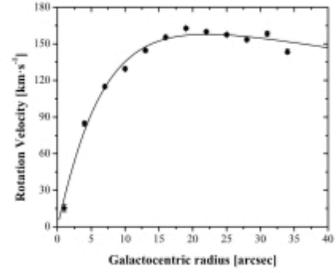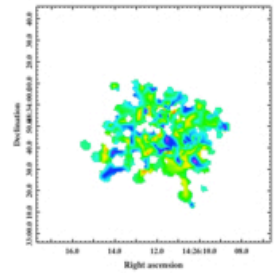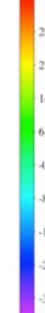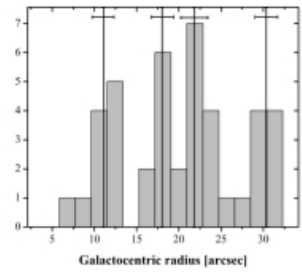

**UGC9358**

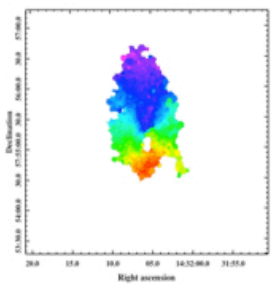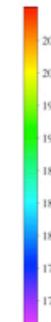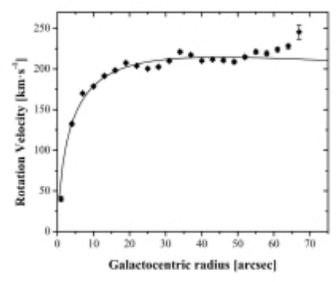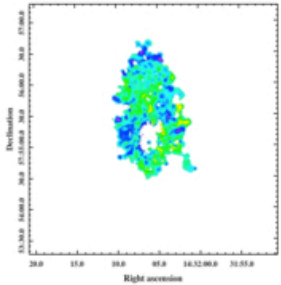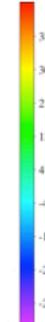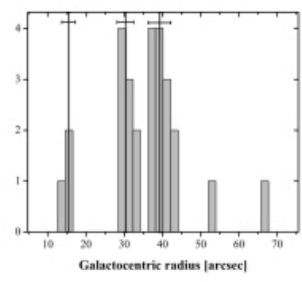

**UGC9363**

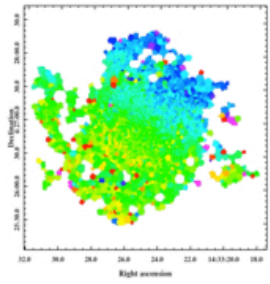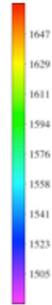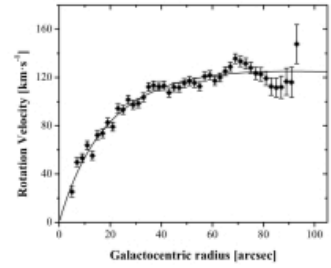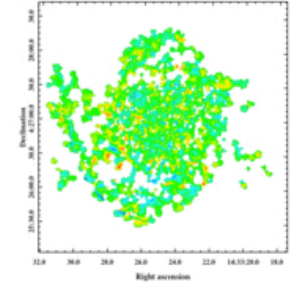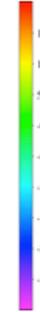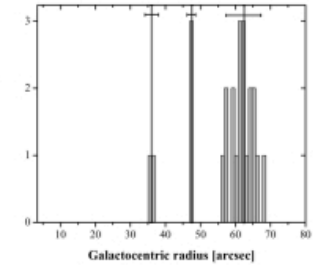

**UGC9366**

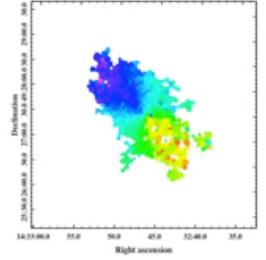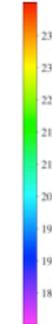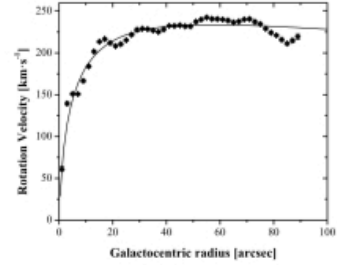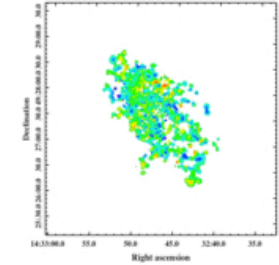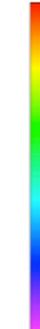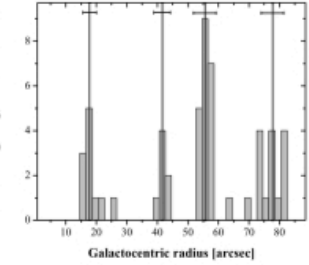

**UGC9465**

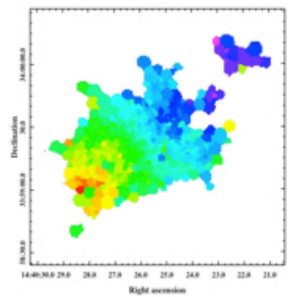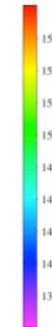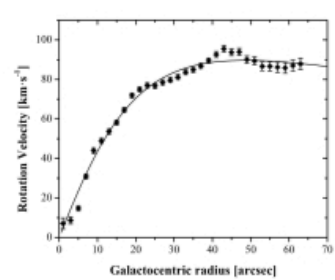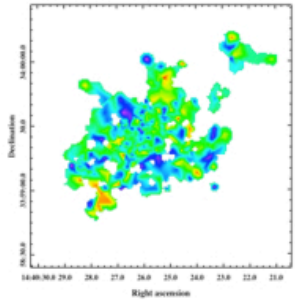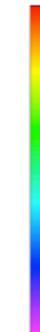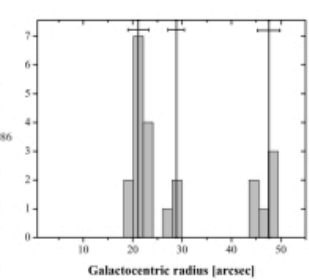

**UGC9736**

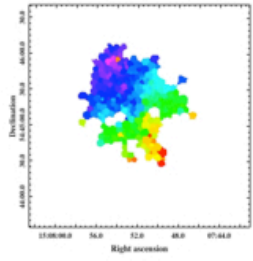 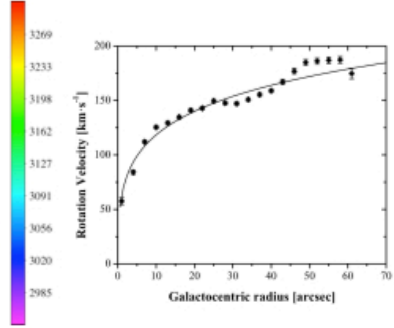 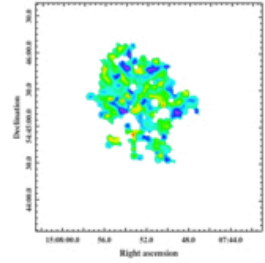 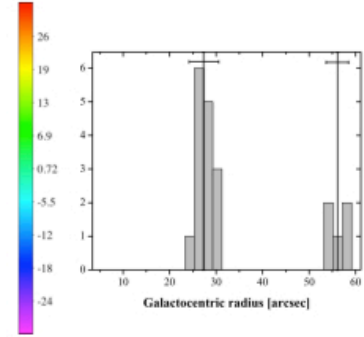

**UGC9753**

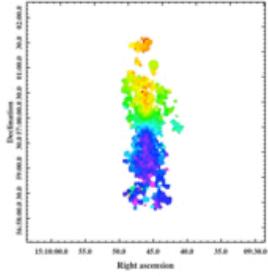 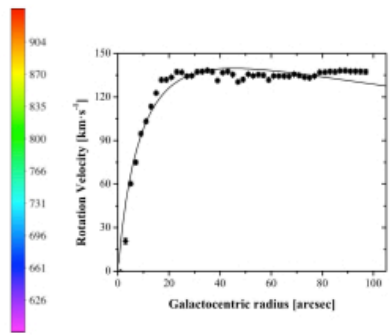 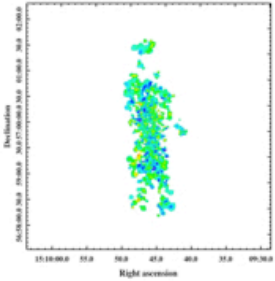 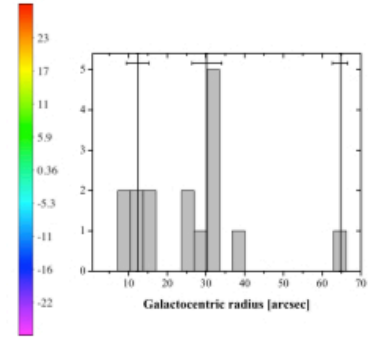

**UGC9866**

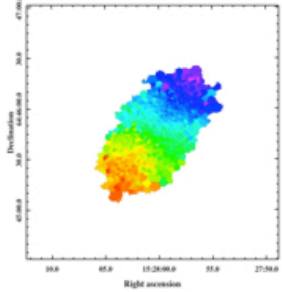 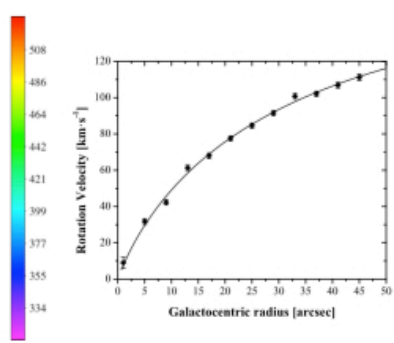 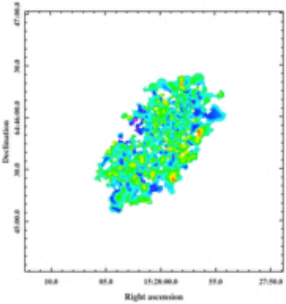 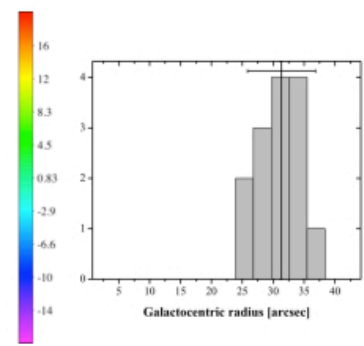

**UGC9943**

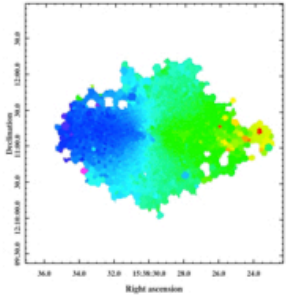 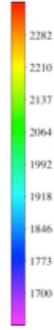 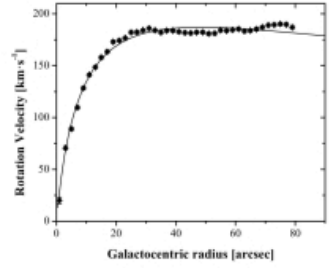 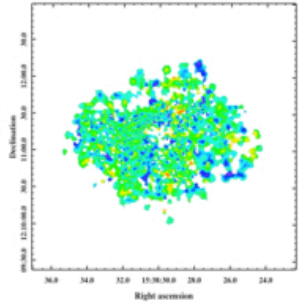 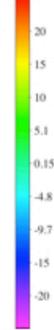 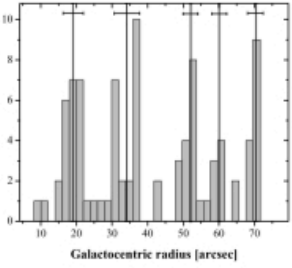

**UGC9969**

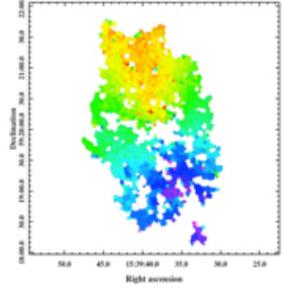 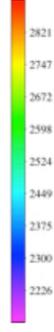 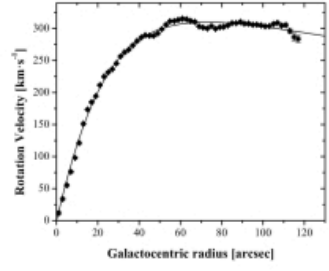 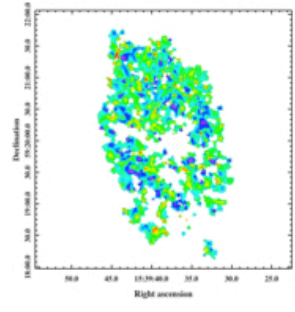 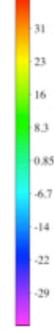 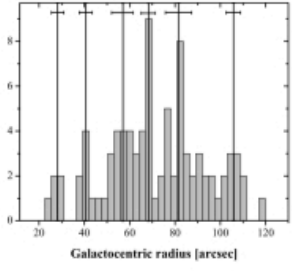

**UGC10075**

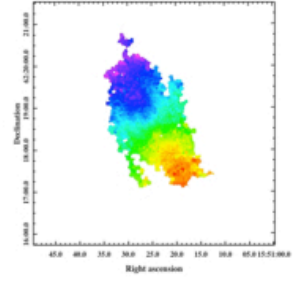 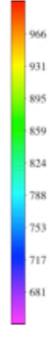 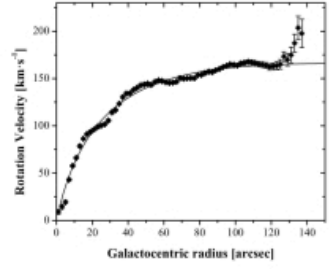 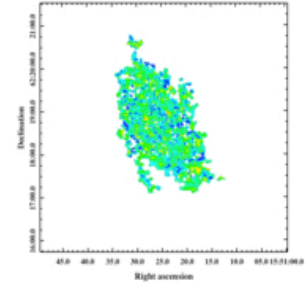 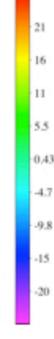 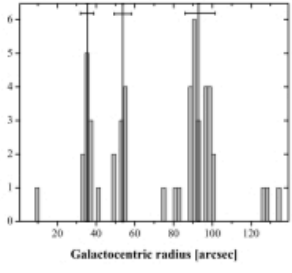

**UGC10359**

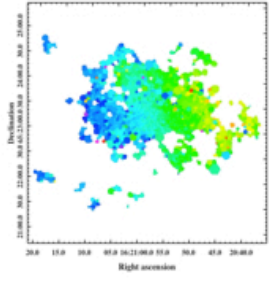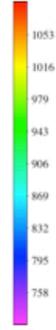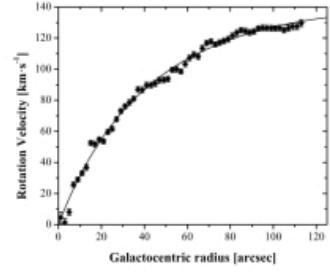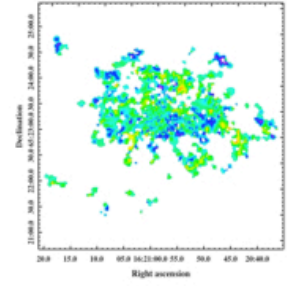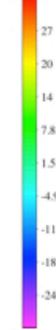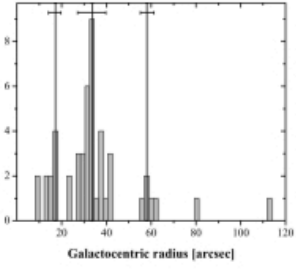

**UGC10445**

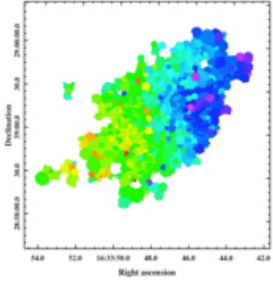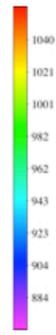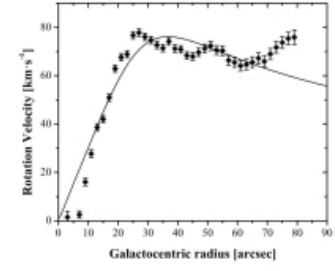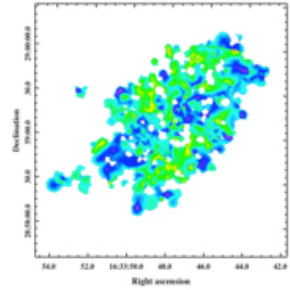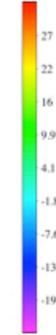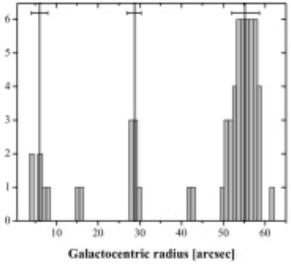

**UGC10470**

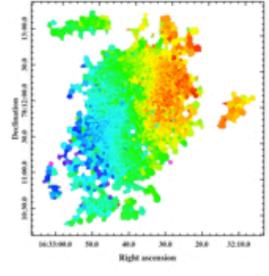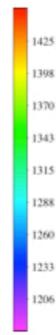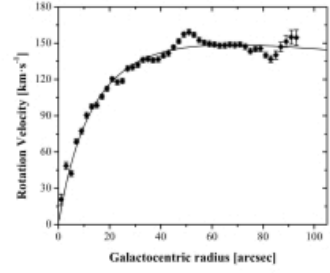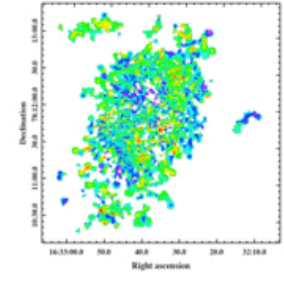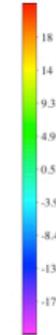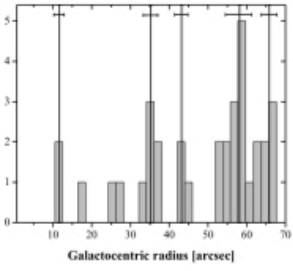

## UGC10502

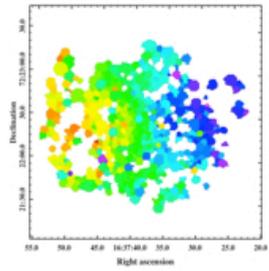 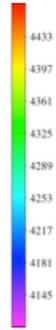 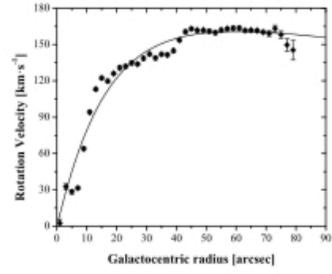 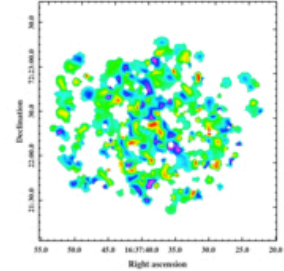 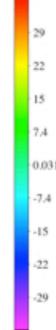 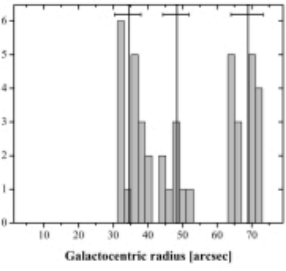

## UGC10521

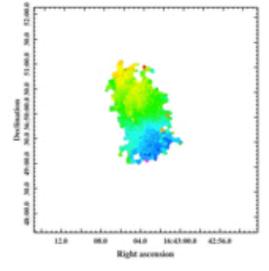 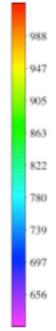 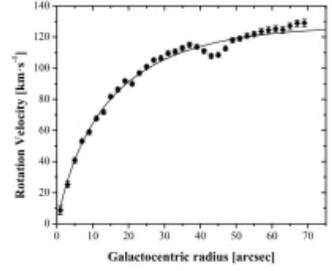 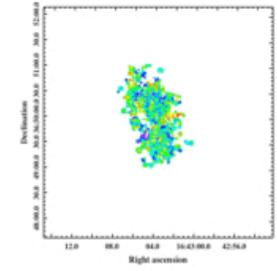 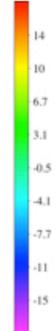 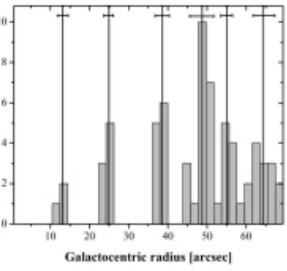

## UGC10546

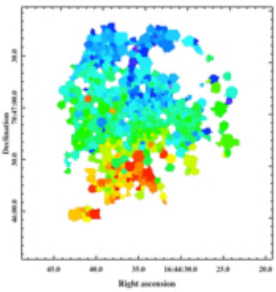 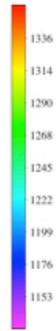 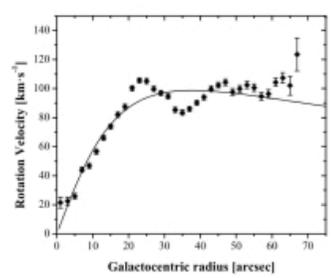 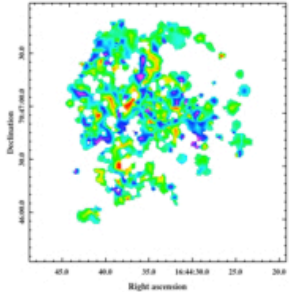 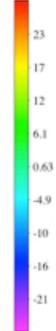 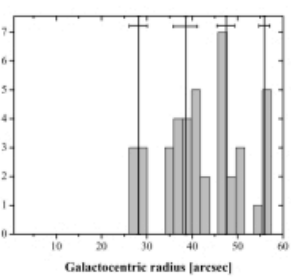

**UGC10564**

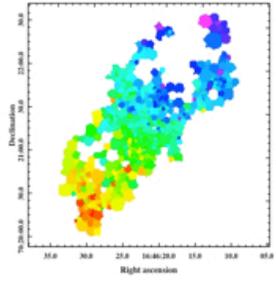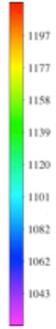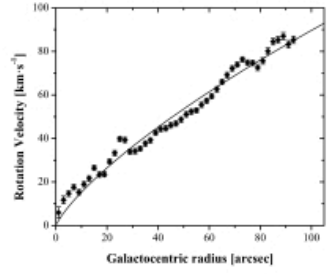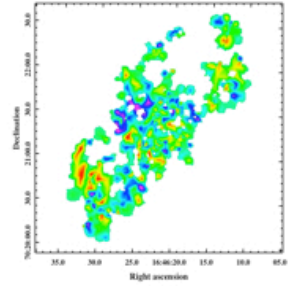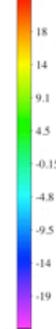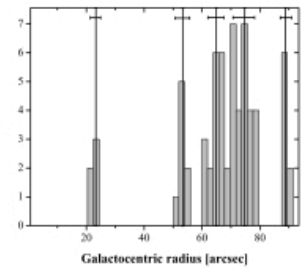

**UGC10652**

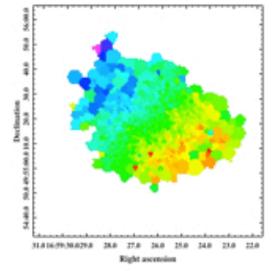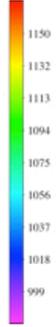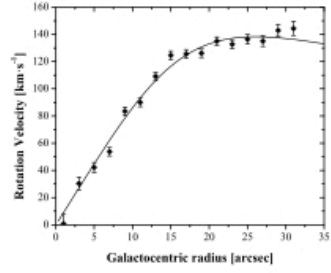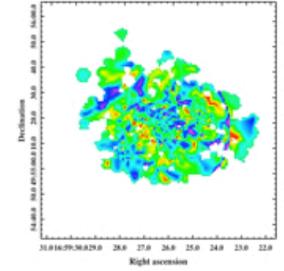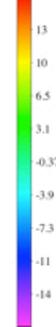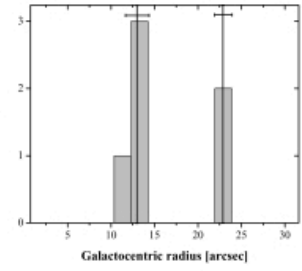

**UGC10757**

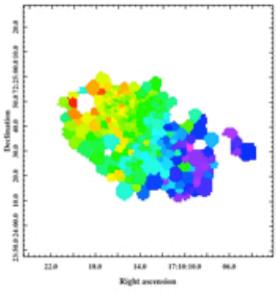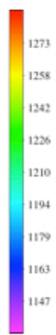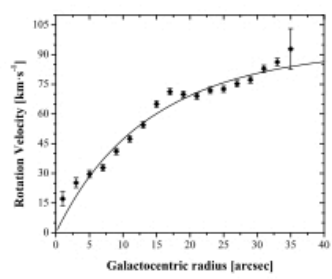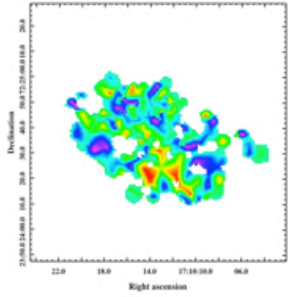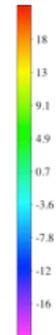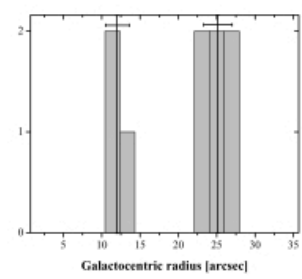

**UGC10897**

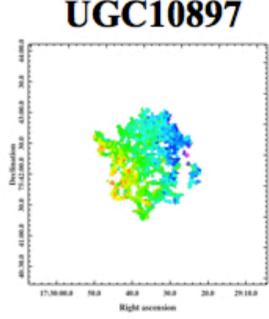 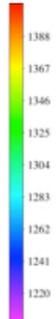 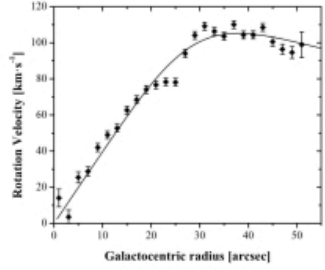 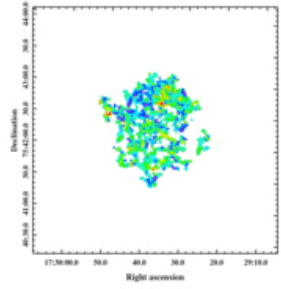 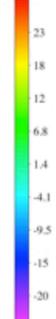 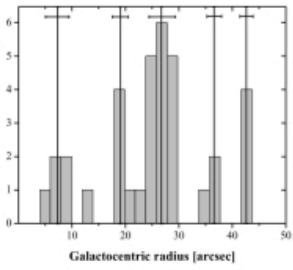

**UGC11012**

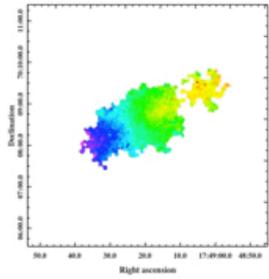 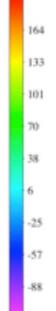 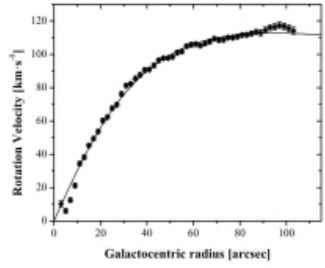 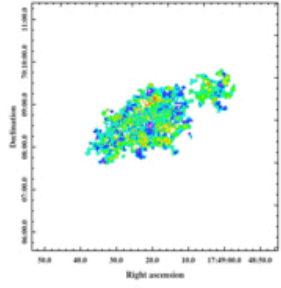 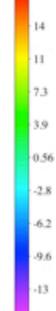 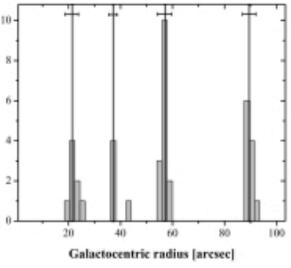

**UGC11124**

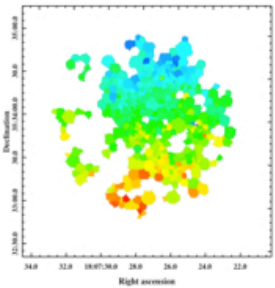 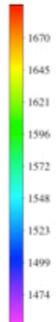 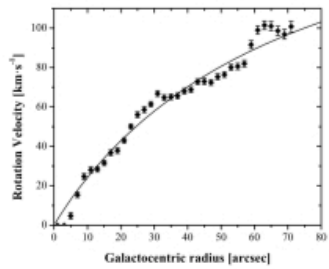 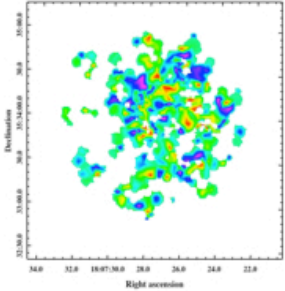 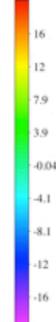 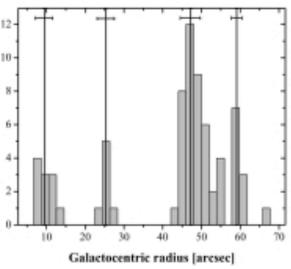

**UGC11218**

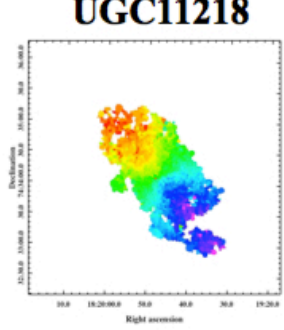 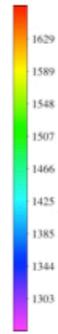 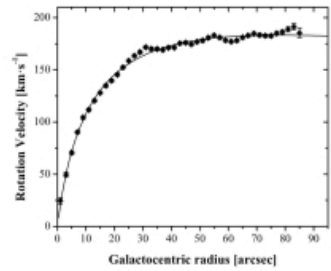 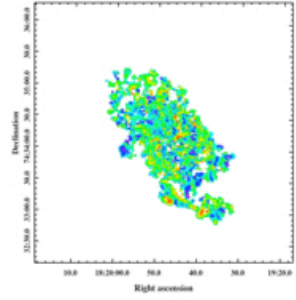 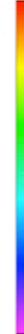 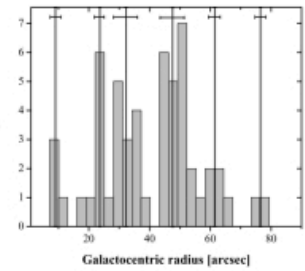

**UGC11283**

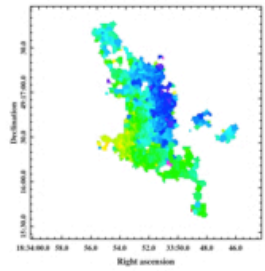 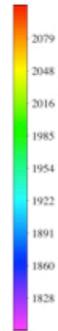 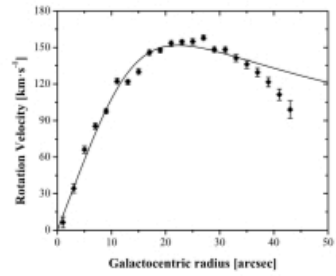 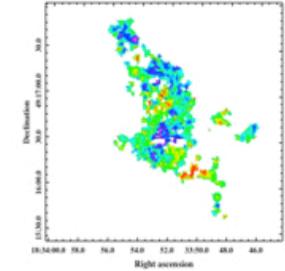 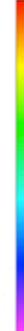 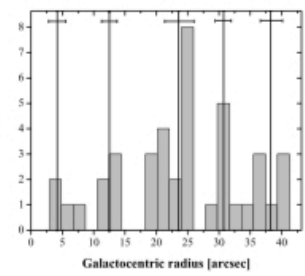

**UGC11407**

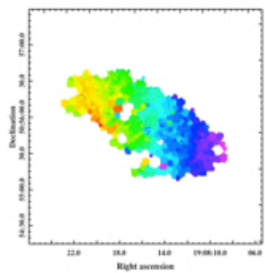 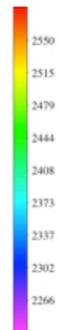 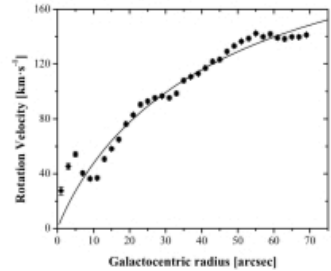 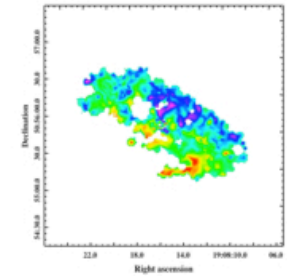 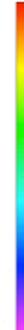 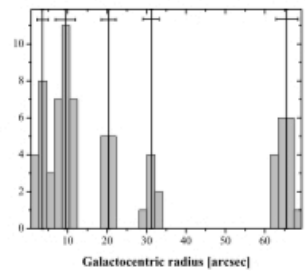

**UGC11466**

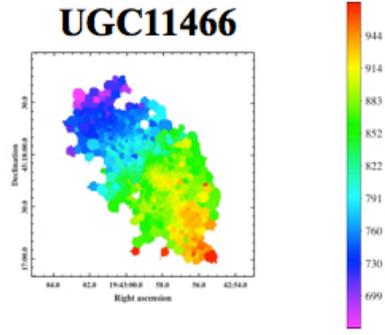 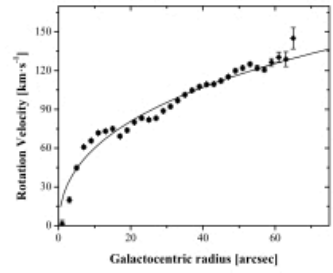 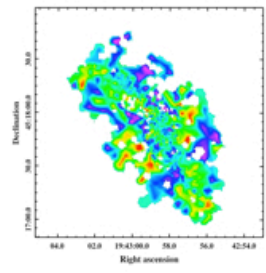 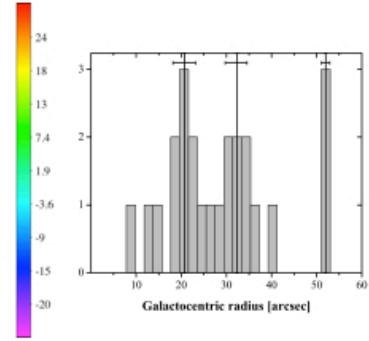

**UGC11557**

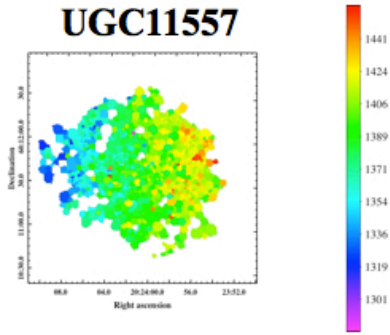 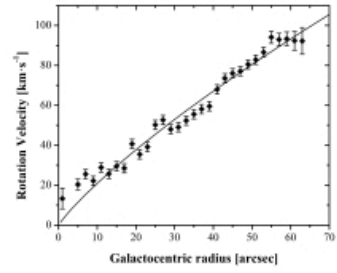 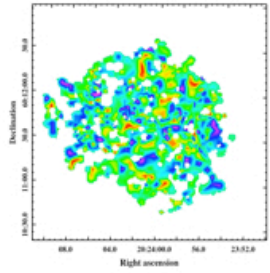 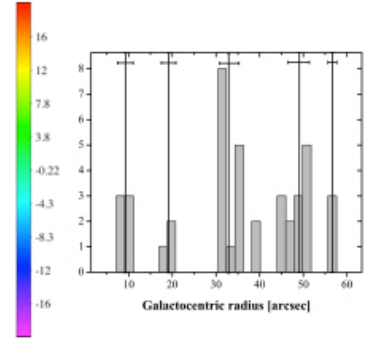

**UGC11861**

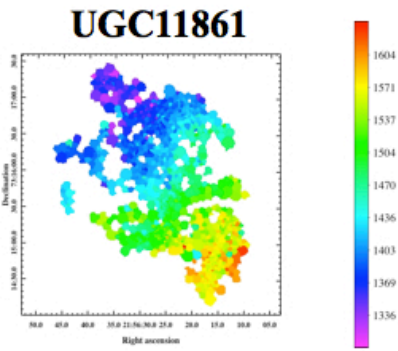 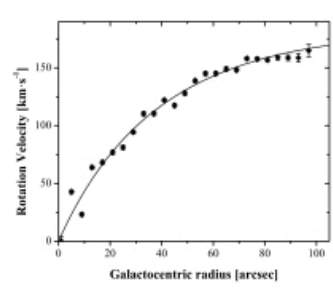 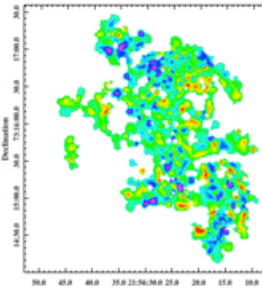 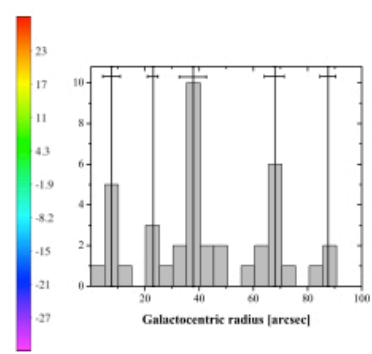

## UGC11872

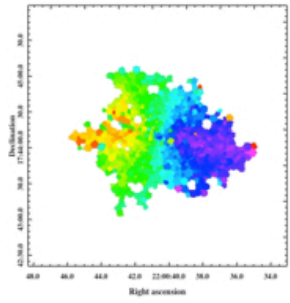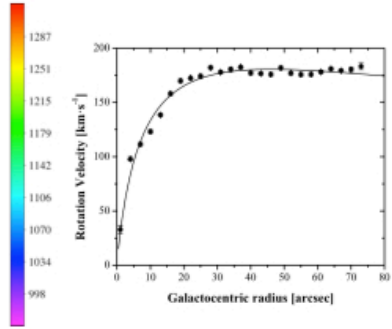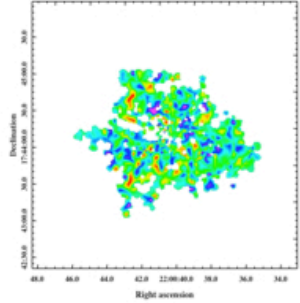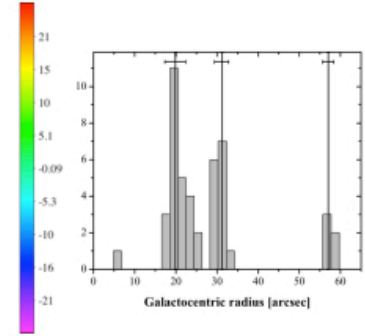

## UGC11914

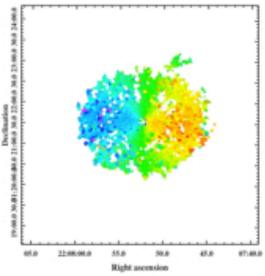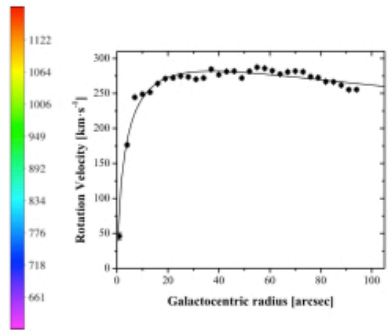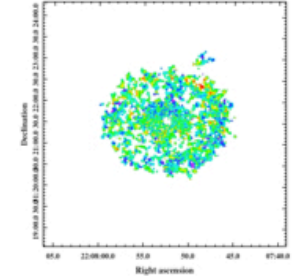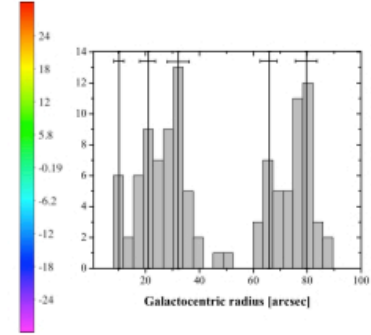

## UGC12276

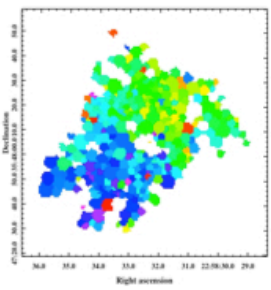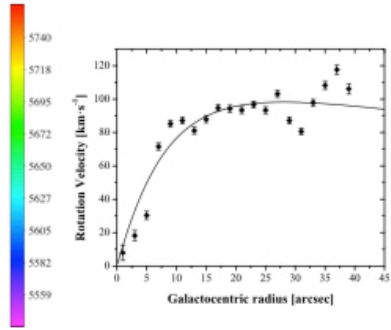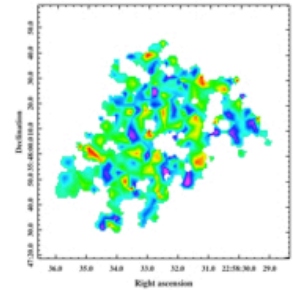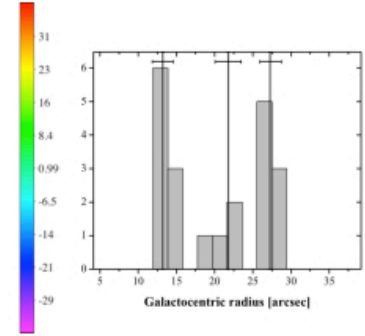

**UGC12343**

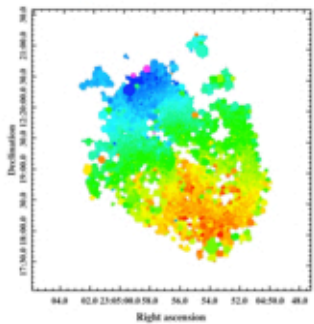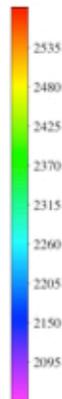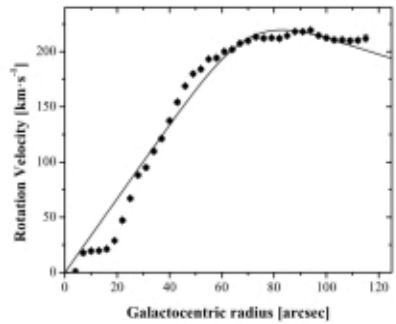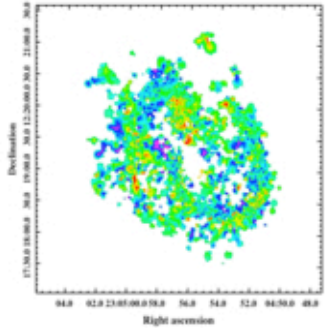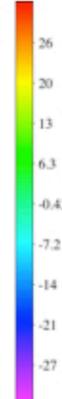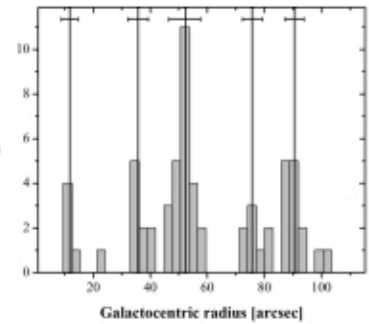

**UGC12754**

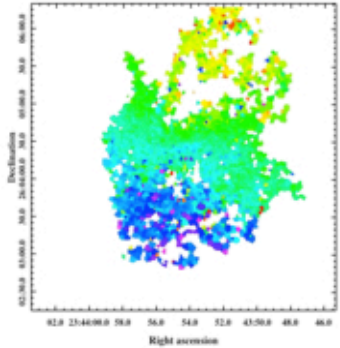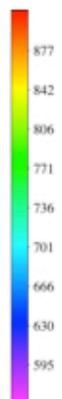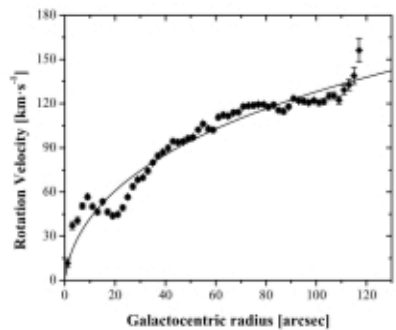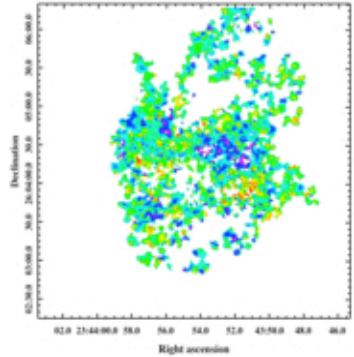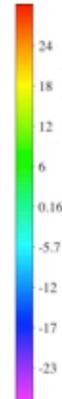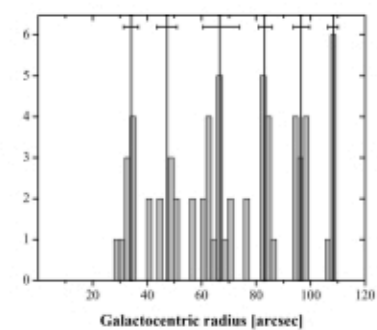